\documentclass[11pt]{article}
\pdfoutput=1
\usepackage{amsmath,amssymb,amsbsy,amstext, amsthm, simplewick, amsfonts}
\usepackage{hyperref}
\usepackage{graphicx}
\usepackage{upgreek}
\usepackage{framed}
\usepackage{bbm}
\usepackage{textcomp}
\usepackage{adjustbox}
\usepackage{makecell}
\usepackage{tcolorbox}
\usepackage{physics}
\usepackage{empheq}
\usepackage[normalem]{ulem}

\numberwithin{equation}{section}

\def\be {\begin{equation}}
\def\ee {\end{equation}}

\def\e{\epsilon}

\def\ba{\begin{eqnarray}}
\def\ea{\end{eqnarray}}

\def \bfx {\textbf{x}}

\def\vpi {\varphi}
\def \bfk {\textbf{k}}
\def \bfq {\textbf{q}}
\def \del {\partial}
\def\cU {{\cal U}}
\def \blangle {\Big\langle}
\def \brangle {\Big\rangle}
\def \bl {\Big |}
\usepackage[framemethod=default]{mdframed}
\newmdenv[skipabove=7pt,
skipbelow=7pt,
rightline=false,
leftline=false,
topline=false,
bottomline=false,
backgroundcolor=gray!10,
linecolor=gray,
innerleftmargin=5pt,
innerrightmargin=5pt,
innertopmargin=5pt,
innerbottommargin=5pt,
leftmargin=0cm,
rightmargin=0cm,
linewidth=4pt]{eBox}
\newmdenv[skipabove=7pt,
skipbelow=7pt,
rightline=false,
leftline=false,
topline=false,
bottomline=false,
backgroundcolor=gray!10,
linecolor=gray,
innerleftmargin=5pt,
innerrightmargin=5pt,
innertopmargin=-5pt,
innerbottommargin=5pt,
leftmargin=0cm,
rightmargin=0cm,
linewidth=4pt]{eBox2}

\usepackage{array}

\definecolor{blue3}{RGB}{31, 119, 180}
\definecolor{red3}{RGB}{	214, 39, 40}
\definecolor{orange3}{RGB}{255, 127, 14}
\definecolor{green3}{RGB}{44, 160, 44}

\newcommand{\then}{\quad \Rightarrow\quad}

\usepackage{colortbl}
\definecolor{lightgreen}{cmyk}{0.2, 0, 0.2, 0.2}
\definecolor{lightgray}{cmyk}{0.1,0.2,0,0.1}
\definecolor{lightgray2}{cmyk}{0.1,0.1,0,0.1}

\makeatletter
\newlength{\apb@width}
\newcommand{\autoparbox}[2][c]{\settowidth{\apb@width}{#2}\parbox[#1]{\apb@width}{#2}}

\makeatother


\def\Mpl{M_{\text{Pl}}}

\def \bfp {\textbf{p}}

\def\beq{\begin{equation}}
\def\eeq{\end{equation}}

\allowdisplaybreaks[1]
\begin{document}



\begin{titlepage}
\setcounter{page}{1} \baselineskip=15.5pt 
\thispagestyle{empty}

\begin{center}
{\fontsize{18}{18} \bf The Cosmological Optical Theorem \vspace{0.1cm}
\;}\\
\end{center}

\vskip 18pt
\begin{center}
\noindent
{\fontsize{12}{18}\selectfont Harry Goodhew\footnote{\tt
			hfg23@damtp.cam.ac.uk}, Sadra Jazayeri\footnote{\tt
			sj571@damtp.cam.ac.uk} and Enrico Pajer\footnote{\tt
			ep551@damtp.cam.ac.uk}}
\end{center}

\begin{center}
  \vskip 8pt
\textit{Department of Applied Mathematics and Theoretical Physics, University of Cambridge, Wilberforce Road, Cambridge, CB3 0WA, UK} 

\end{center}

\vspace{0.4cm}
The unitarity of time evolution, or colloquially the conservation of probability, sits at the heart of our descriptions of fundamental interactions via quantum field theory. The implications of unitarity for scattering amplitudes are well understood, for example through the optical theorem and cutting rules. In contrast, the implications for in-in correlators in curved spacetime and the associated wavefunction of the universe, which are measured by cosmological surveys, are much less transparent. For fields of any mass in de Sitter spacetime with a Bunch-Davies vacuum and general local interactions, which need not be invariant under de Sitter isometries, we show that unitarity implies an infinite set of relations among the coefficients $  \psi_{n} $ of the wavefunction of the universe with $  n $ fields, which we name Cosmological Optical Theorem. For contact diagrams, our result dictates the analytic structure of $  \psi_{n} $ and strongly constrains its form. For example, any correlator with an odd number of conformally-coupled scalar fields and any number of massless scalar fields must vanish. For four-point exchange diagrams, the Cosmological Optical Theorem yields a simple and powerful relation between $  \psi_{3} $ and $  \psi_{4} $, or equivalently between the bispectrum and trispectrum. As explicit checks of this relation, we discuss the trispectrum in single-field inflation from graviton exchange and self-interactions. Moreover, we provide a detailed derivation of the relation between the total-energy pole of cosmological correlators and flat-space amplitudes. We provide analogous formulae for sub-diagram singularities. Our results constitute a new, powerful tool to bootstrap cosmological correlators.

 
 \noindent


\end{titlepage}


\newpage
\setcounter{tocdepth}{2}
\tableofcontents

\newpage




\section{Introduction}

%

Cosmological observations provide us with a unique opportunity to probe the laws of physics at the highest energies as well as the perturbative regime of quantum gravity through the study of correlators of primordial initial conditions. While we have so far only measured the two-point correlator of scalar perturbations, one of the main goals of 21st century cosmological surveys is to detect higher-point scalar correlators and correlators involving tensor perturbations, i.e. primordial gravitational waves. After four decades of intense research activity and inflationary model building, it has become increasingly clear that there is often a vast degeneracy of models that lead to observationally indistinguishable predictions, especially in classes of models that are most symmetric and display a minimal set of ingredients. As a result, much recent effort has been devoted to derive model-independent predictions, which are based on general principles and symmetries and which can be tested without detailed assumptions about the inflationary dynamics.  \\

In this paper, we derive some of the consequences for correlators of unitary evolution during the early universe. Unitarity, or colloquially the conservation of probability, is one of the pillars of quantum physics and of our understanding of fundamental interactions. The implications of unitarity for the natural observables of particle physics, namely scattering amplitudes, are well understood and typically discussed in most textbooks on quantum field theory (QFT), see e.g. \cite{Schwartz:2013pla,WeinbergBook1}. In particular, the optical theorem, cutting rules and spectral decomposition are a very useful part of the standard toolkit to tackle QFT problems in flat space. Conversely, the implications of unitarity for the natural observables of cosmology, namely the asymptotic future of equal-time correlators of the product of fields, are far from clear. In virtually all instances in the literature, unitary evolution is ensured by performing explicit calculations that start from Hermitian Hamiltonians. While this procedure ensures that whatever correlators we compute are consistent with unitarity, it is ultimately unsatisfactory from various points of view. First, to carve out the space of all possible correlators compatible with unitarity one would need to compute these correlators for all possible models, which is both impractical and unnecessary. Second, when analyzing a future observational detection, it is hard to learn anything about unitarity because of the degeneracy with various model building assumptions. Third, many, often lengthy, calculations are actually unnecessary because the final results are fixed by unitarity in terms of other, simpler, calculations. This is precisely what happens when applying the (generalized) optical theorem to amplitudes: some higher-order amplitudes are fixed in terms of a specific combination of lower-order ones. Fourth, when combined with symmetry, unitarity can impose such strong constraints that many quantities of interest are uniquely determined without the need of explicit calculations. This is the case for example with on-shell methods for amplitudes: Lorentz invariance, unitarity and locality uniquely fix all possible three-particle amplitudes, which in turn can be used to derive higher-particle amplitudes using recursion relations in many interesting classes of theories (for reviews see \cite{BenincasaReview,ElvangHuang,CheungReview}).\\

So we would like to formulate explicitly the consequences of unitarity for correlators. Actually, we can already make a baby step in this direction using results in the literature, by proceeding as follows. It has been observed \cite{Maldacena:2011nz,Raju:2012zr} and later discussed e.g. in \cite{Arkani-Hamed:2017fdk,Arkani-Hamed:2018kmz,Benincasa:2018ssx}, that correlators as function of the so-called ``total energy'', namely the sum of the norms of the momenta in the correlator, possess poles, the residue of which is the (UV-limit of the) flat-space amplitude. The (generalized) optical theorem applied to the amplitude then imposes one constraint on the correlator, as a consequence of unitarity. But one can do much better. In this work, by mimicking the derivation of the optical theorem in flat space, we will derive a relation (Section \ref{sec:COT}) that, when evaluated on the residue of the total-energy pole, reduces to the standard optical theorem, but that is highly non-trivial also on the rest of the correlator, away from this pole. We name this result the \textit{Cosmological Optical Theorem}, and summarize it in the next subsection.\\

Given that we currently understand amplitudes much better than correlators, the relation between amplitudes and correlators on the total-energy pole can be extremely useful to uplift our flat-space knowledge to curved spacetime and hence cosmology. Because of its importance, we provide in this work a detailed and self-contained analysis of this relation and derive additional results for the residues of partial-energy poles, arising when the total-energy of sub-diagrams vanishes (see Section \ref{sec:polo} and previous work \cite{Benincasa:2018ssx,Benincasa:2019vqr}). These additional partial-energy singularities can also be extracted from the Cosmological Optical Theorem, which relates lower-order correlators to higher-order ones.\\

An important theme in our discussion is the focus on observables, which in a quantum theory including gravity necessarily live at the boundary of spacetime, namely at the spacelike asymptotic future of the quasi-de Sitter inflationary universe. This perspective has a long history. In this context, the idea to focus on observables goes back to the S-matrix program of the 60's \cite{Eden:1966dnq}. Much progress has been made since then on the amplitude front, and modern on-shell methods have blossomed into a very rich and powerful network of new results and ideas, which bring to the surface the rigidity of general principles such as unitarity, causality and symmetry without relying on fictitious degrees of freedom such as gauge fields or ``virtual'' states \cite{BenincasaReview,ElvangHuang,CheungReview}.  The boundary perspective has also been at the core of the gauge-gravity correspondence, which has revolutionized our understanding of strongly coupled systems both with and without gravity \cite{Aharony:1999ti}. \\

A boundary perspective on cosmology is therefore not only perfectly natural, since after all cosmological observations do measure the future asymptotic of primordial correlators, but also extremely powerful. Many steps forward have been made over the years on this front \cite{maldacena2003non,McFadden:2009fg,McFadden:2010vh,Mata:2012bx,Ghosh:2014kba,Arkani-Hamed:2015bza,Maldacena:2011nz, Baumann:2017jvh} and much interest and progress has appeared recently in the literature \cite{Arkani-Hamed:2018kmz,Baumann:2019oyu,Sleight:2019mgd,Sleight:2019hfp,baumann2020cosmological,Sleight:2020obc}. The most powerful results so far rely heavily on the highly constraining isometries of de Sitter spacetime. However, only seven of the de Sitter isometries have been confirmed by cosmological observations, namely rotations, translations and dilations. There is currently no evidence that the three de Sitter boosts (acting as special conformal transformations on the boundary) are even approximate symmetries at play in the early universe. In fact, we already know that these boosts are broken in many models of inflation, e.g. those featuring a non-luminal speed of propagation $  c_{s} $ and other effects induced by the inflaton background that permeates the universe. In contrast to what happens for dilations, which lead to (approximate) scale invariance, the breaking of de Sitter boosts is in general \textit{not} slow-roll suppressed (see e.g. \cite{Baumann:2019ghk} for a recent discussion). Moreover, we also know that in models with a minimal set of ingredients, non-Gaussian correlators are large, and hence of phenomenological interest, precisely when de Sitter boosts are strongly broken, as evident in the well-known relation $  f^{\text{eq,ort}}_{NL}\sim c_{s}^{-2} $ \cite{Alishahiha:2004eh,Chen:2006nt,Cheung:2007st}. This simple observation has recently been promoted to a precise theorem in \cite{EnricoDan}, where it was proven that the only theory of single-clock inflation where scalar curvature perturbations are fully de Sitter invariant is the free theory. In this spirit, it is important both to keep exploring the powerful results that can be derived by a cosmological bootstrap based on de Sitter isometries, as advocated in \cite{Arkani-Hamed:2018kmz}, and also to pay special attention to all results that do not rely on an enlarged set of symmetries and in particular on de Sitter boosts, as recently proposed in \cite{Enrico}. The Cosmological Optical Theorem presented here makes some progress in this direction. In particular, while our most useful results do assume de Sitter mode functions, we do allow and account for different speeds of sound as well as for non-boost invariant interactions, as arising from the ubiquitous coupling to the inflaton foliation of time.\\

Our exact formulae for the residue of the total-energy pole of correlators, \eqref{eq:kTlim} and \eqref{finalBtoA}, provide an additional opportunity to move the bootstrap idea away from full de Sitter isometries. In particular, when de Sitter boosts are broken by the inflaton foliation, the flat-space limit of the correlator is an amplitude that breaks Lorentz boosts. These amplitudes have been recently systematically studied in \cite{Boostless} (at tree-level) using on-shell methods and the spinor helicity formalism for all massless relativistic spinning particles. This provides a first step towards a boundary understanding of boost-breaking spinning correlators. A more detailed discussion can be found in \cite{Enrico}, where a large number of bispectra containing scalar and tensor perturbations have been derived from a set of ``bootstrap rules'', which do not rely on de Sitter boosts.\\


The rest of the paper is organized as follows. In the final parts of this section, we summarize our main results and conventions. In Section \ref{sec:COT}, we derive the general form of the Cosmological Optical Theorem as a consequence of unitary evolution. We then study the implications of this theorem for cosmological correlators and the coefficients of the wavefunction of the universe, discussing separately contact diagrams, resulting in \eqref{unicont}, and exchange four-point diagrams, resulting in \eqref{exchCOPT}. In Section \ref{NonTC}, we perform a series of non-trivial checks of our general results in various systems of practical interest, including inflationary correlators from gravitational interactions and self-interactions. We provide a detailed and self-contained derivation of the relation between the total-energy pole of correlators and flat-space amplitudes in Section \ref{sec:polo}, where we also derive exact expressions for the partial-energy poles. Finally we conclude in Section \ref{sec:conc} with a discussion of future directions. Some technical details are collected in the appendices. In particular, in Appendix \ref{wfu} we provide a pedagogical review of the wavefunction of the universe and its application to the computation of cosmological correlators.

 
\subsection*{Summary of the main results}\label{ssec:}

For the convenience of the reader, we summarize below our main results:
\begin{itemize}
\item We derive a formal, non-perturbative expression for the consequences of unitary evolution in \eqref{UNIT}, which we name the generalized Cosmological Optical Theorem. This becomes of practical use when phrased in terms of the coefficients $  \psi_{n} $ of the wavefunction of the universe with $  n $ fields, defined in \eqref{WFu}. We assume a Bunch-Davies vacuum and de Sitter mode functions with arbitrary speed of sound and allow for derivative interactions that break de Sitter boosts. Taking advantage of the analytical properties of de Sitter mode functions for arbitrary masses, we derive the following \textit{Cosmological Optical Theorem for contact diagrams} (see \eqref{unicont})
\begin{align}
\qquad \psi'_n(k_a,\hat{k}_a \cdot \hat{k}_b)+\left[\psi'_n(-k_a-i\epsilon,\hat{k}_a \cdot \hat{k}_b)\right]^*=0\,, \qquad k_a\in \mathbb{C}^{n -}\,,
\end{align}
where $\psi_n$ is the wavefunction coefficient, the prime indicates that we have stripped off the momentum conserving delta function, $a=1,\dots,n $ enumerates the momenta, $  k\equiv |\bfk| $ and $  \hat k\equiv \bfk / k$. This result sheds light on the analytical structure of the wavefunction coefficients and is the cosmological equivalent of the Hermitian analyticity for amplitudes \cite{olive1962unitarity}. Further specifying to massless fields $  \phi $ and conformally-coupled fields $  \varphi $ with IR finite interactions\footnote{By IR-finite interactions we mean those that lead to (tree-level) correlators that are regular at future infinity, $\eta_0\to 0$.}
, this tells us that the wavefunction coefficient for an even (odd) number of $  \varphi $'s must be real (imaginary). In particular, massless fields such as curvature perturbations, $  \psi_{n} $ must be real when they are IR finite. As a lemma, IR-finite correlators involving an odd number of conformally-coupled fields and an arbitrary number of massless fields must vanish. We also discuss IR-divergent contributions and fields of general scaling dimensions around \eqref{phicube} and \eqref{fnstar}, respectively.
\item When specializing to four-point wavefunction coefficients $  \psi'^{s}_4 $ of a massless scalar $  \phi $ arising from the $ s  $-channel exchange of some potentially different scalar $  \sigma $, we find the \textit{Cosmological Optical Theorem for exchange diagrams} (see \eqref{exchCOPT})
\begin{align}
& \psi'^s_4(k_1,k_2,k_3,k_4,s)+[\psi'^s_4(-k_1,-k_2,-k_3,-k_4,s)]^*= \\ \nonumber
& P_\sigma(s)\, \left[\psi'^{\phi\phi\sigma}_3(k_1,k_2,s)- \psi'^{\phi\phi\sigma}_3(k_1,k_2,-s) \right]\,\left[\psi'^{\phi\phi\sigma}_3(k_3,k_4,s)- \psi'^{\phi\phi\sigma}_3(k_3,k_4,-s) \right]\,.
\end{align}
where $  s=|\bfk_{1}+\bfk_{2}| $ and $  P_{\sigma} $ is the power spectrum of $  \sigma $. This expression implies a similar relation for the full $ \psi'_4  $ upon summing over all channels, \eqref{sum}. As a lemma, one finds the following \textit{Cosmological Optical Theorem for exchange correlators} (see \eqref{cotcor})
\begin{align}
& B_4^s(k_1,k_2,k_3,k_4,s)+B_4^s(-k_1,-k_2,-k_3,-k_4,s)= \nonumber \\ &\frac{1}{2 P_\sigma(s)} \left[B_{\phi\phi\sigma}(k_1,k_2,s)-B_{\phi\phi\sigma}(k_1,k_2,-s)\right]\left[B_{\phi\phi\sigma}(k_3,k_4,s)-B_{\phi\phi\sigma}(k_3,k_4,-s)\right]\,,
\end{align}
which relates the $  \phi $-trispectrum $  B_{4} $ to the $  \phi\phi\sigma $-bispectrum $ B_{\phi\phi\sigma}  $. Similar expressions from the exchange of a graviton can be found in \eqref{eq:internalg} and \eqref{eq:externalg}.
\item We provide checks of all the above results in explicit calculations, including the inflationary trispectrum from graviton exchange and from $  \dot \phi^{3} $ interactions in $  P(X) $-theories or the Effective Field Theory of inflation (see Section \ref{NonTC}).
\item We provide an explicit and detailed derivation of the relation between the residue of the total-energy pole of $  n $-point correlators $  B_{n} $, i.e. in the limit $  k_{T}\equiv \sum_{a=1}^{n} |\bfk_{a}|\to 0 $, and the flat space $ n  $-particle amplitude $  A_{n} $ including several details that had not been made previously explicit in the literature. For contact interactions involving a total of $  N $ derivatives, counting equally time and space derivatives, we find (see \eqref{eq:kTlim})
\begin{equation}
\lim_{k_T\rightarrow 0} B_n=2(-1)^{n}H^{2n+N-4}(n+N-4)!\Re\left\lbrace \frac{iA_n'}{(-ik_T)^{N+n-3}}\prod_a^n\frac{i+k_a\eta_0}{2k_a^2}\right\rbrace,
\end{equation}
where $  H $ is the Hubble parameter and $  \eta_{0} $ is the late time at which the correlator is evaluated, which can be taken to zero for IR-finite contributions. This agrees with the scaling independently derived in \cite{Enrico} using dimensional analysis and scale invariance. For more general diagrams we find (see \eqref{finalBtoA})
\begin{align}
\lim_{k_T\rightarrow 0} B_n=2\,(-1)^{n} H^{2I+2n+\sum\limits_\alpha^V(N_\alpha-4)} \left(\sum\limits_\alpha^VD_\alpha-4\right)! \times \Re\left\lbrace  \frac{iA_n'}{(-ik_T)^{p} }\prod_a^n\frac{i+k_a\eta_0}{2k_a^2}\right\rbrace,
\end{align}
where $  \alpha $ runs over all $  V $ vertices with $  n_{\alpha} $ fields and $  N_{\alpha} $ derivatives, which correspond to operators of dimension $  D_{\alpha}=n_{\alpha}+N_{\alpha} $, $  I $ is the number of internal legs, and the order $  p $ of the total-energy pole is
\begin{align}
p\equiv 1+\sum\limits_\alpha^V \left( D_\alpha -4\right)\,.
\end{align}
Additional poles arising when a subset of norms of the momenta vanish are discussed in Section \ref{sec:op}.
\end{itemize}


\subsection*{Notation and conventions}

We use natural units throughout and define $ \Mpl^{2}=1/(8\pi G_{N})$. The 3d Fourier transformation is defined through, 
\begin{align}
f(\bfx)&=\int \dfrac{d^3\bfk}{(2\pi)^3}{f}(\bfk)\exp(i\bfk\cdot\bfx)\equiv\int_{\bfk}{f}(\bfk)\exp(i\bfk\cdot\bfx) \,,\\
{f}(\bfk)&=\int d^3\bfx f(\bfx)\exp(-i\bfk\cdot\bfx)\equiv \int_{\bfx}  f(\bfx)\exp(-i\bfk\cdot\bfx)\,.
\end{align}
Where we have dropped the usual tilde to avoid cluttered notation as it will always be clear when we are in Fourier Space. 
We use bold letters to refer to vectors, i.e. $\bfk$, and write their magnitude as $k\equiv |\bfk|$. A prime on a wave function coefficient or correlator is defined to mean that we drop the momentum conserving delta function, 
\begin{align}
\psi_n(\bfk_1,\dots ,\bfk_n)& \equiv \psi_n'(\bfk_1,\dots ,\bfk_n)(2\pi)^3 \delta^3(\sum \bfk_a)\,,\\ \nonumber
\langle {\cal O}(\bfk_1)\dots {\cal O}(\bfk_n)\rangle &\equiv \langle {\cal O}(\bfk_1)\dots {\cal O}(\bfk_n)\rangle' (2\pi)^3\delta^3(\sum \bfk_a)\,.
\end{align}
Otherwise, a prime refers to the derivative with respect to the conformal time, i.e. $\phi'=\del_\eta \phi$. 
When computing four-point exchange correlators, we use the following variables
\be
s\equiv |\bfk_1+\bfk_2|\,,\qquad t\equiv |\bfk_1+\bfk_3|\,,\qquad u\equiv |\bfk_1+\bfk_4|\,. 
\ee
They satisfy the relation, 
\be
s^2+t^2+u^2=\sum_{a=1}^4\,k_a^2\,.
\ee
We define the ``total energy'' $  k_{T} $ as
\be\label{kT}
k_T^{(n)} \equiv \sum_{a=1}^{n} c_{a} k_a\,,
\ee
where $  c_{a} $ is the speed of sound of the particle with momentum $  \bfk_{a} $ and we will drop the superscript ``$  (n) $'' in $  k_{T} $ when it is clear from the context.
The symbols $a_\bfk$ and $a_{\bfk}^\dagger$ refer to annihilation and creation operators, respectively. 
The flat space amplitude will be written as 
\be
{\cal S}-\mathbf{1}\equiv i\, A_n(p_1^\mu\,,\dots\, p_n^\mu) \,(2\pi)^4\,\delta^4\left( \sum\,p_a^{\mu} \right)\,.
\ee
with all the four-momenta defined to be ingoing. The correlators are sometimes written in terms of $B_n$ (bispectrum $B_3$, trispectrum $B_4$, etc), 
\be
\langle \phi(\bfk_1)\,,\dots \,\phi(\bfk_n)\rangle \equiv B_n(\bfk_1,...,\bfk_n)\,  (2\pi)^3\delta^3\left(\sum \bfk_a\right)\,.
\ee

\section{The Cosmological Optical Theorem}\label{sec:COT}

In this section, after building the preliminaries of our set-up, we provide a detailed derivation of the Cosmological Optical Theorem separately for contact terms and exchange diagrams of massless scalar fields. 

\subsection{Generalities} 

In this section, we introduce our setup and review the formalism of the wavefunction of the universe. The reader who is already familiar with this formalism should move on directly to Section \ref{s:deriv}. 

In this work, we will assume that a cosmological background is well approximated by a de Sitter (dS) spacetime,
\be
ds^2=a^2(\eta)\left(-d\eta^2+d\bfx^2\right)\,, \qquad a(\eta)=-\dfrac{1}{\eta H}\,,
\ee
where $H$ is the Hubble constant. Although de Sitter boosts may be broken by the (non-gravitational) background, we demand invariance under dilatations, spatial translations and rotations. 
In most models of inflation, dilations are only an approximate symmetry. However, for simplicity we neglect the presumably small deviations from scale invariance in the following. Diffeomorphism invariance will be irrelevant for most of our discussions and we will be freely inserting local interactions that may not come from covariant Lagrangians. 

When discussing scalar fields, we will assume the following second-order Lagrangian, 
\be
\label{quad}
S_2=\int  d\eta d^3\bfx\, a^2(\eta)\,\left[\dfrac{1}{2}\sigma'^2-\dfrac{1}{2}c_\sigma ^2(\partial_i \sigma)^2-\dfrac{1}{2}a^2(\eta) m_\sigma^2 \sigma^2 \right]\,.
\ee
Above, the constants $c_\sigma$ and $m_\sigma$ indicate the speed of sound and the mass of the scalar field $\sigma$, respectively. The mode function appearing in the free field operators 
\be
\hat \sigma(\bfk,\eta)=a_{\bfk} ( \sigma)\,\sigma^-(k,\eta)+a^{\dagger}_{-\bfk}(\sigma)\,\sigma^{+}(k,\eta)\,,
\ee
follows from  \eqref{quad}, 
\begin{align}\label{due}
\sigma^+(k,\eta)=i\,\dfrac{\sqrt{\pi} H}{2 } e^{-i\frac{\pi}{2}(\nu+\frac{1}{2})}\, \left( \frac{ -\eta}{c_\sigma}  \right)^{\frac{3}{2}} H^{(2)}_{\nu}(-c_\sigma  k\eta)\,,\qquad \sigma^-(k,\eta)=(\sigma^+(k,\eta))^*\,,
\end{align}
where $\nu=\sqrt{\frac{9}{4}-\frac{m_\sigma^2}{H^2}}$ and in the last equation we assumed $  k \in \mathbb{R}^+ $. Henceforth, we set $c_\sigma=1$ except for one massless scalar field ($c_s\neq 1$), merely for simplicity in our notation. We will be especially interested in massless and conformally coupled scalars denoted by $\phi$ ($m_\phi=0$) and $\vpi$ ($m_\vpi=\sqrt{2}H$), respectively. Their mode functions are given by\footnote{Notice that the normalization of the mode function of the curvature perturbation $\zeta$ is different from that of $\phi$.}
\begin{align}
\phi^\pm(k,\eta) &= \dfrac{H}{\sqrt{2 c_s^3 k^3}}(1 \mp i c_s k \eta)\exp(\pm i c_s k \eta)\,, \\ \nonumber
\vpi^\pm (k,\eta)&=\mp \dfrac{ i H}{\sqrt{2 k}} \eta \exp( \pm i  k \eta)\,.
\end{align}
We consider interactions of the following type, 
\be
\label{inte}
S_{\text{int}}=\int d\eta d^3 \bfx \,\mathcal{L}_{\text{int}}=g_n \int d\eta d^3\bfx \, a^{4-\sum N_a} \partial^{N_1} \sigma_{\alpha_1} \dots   \partial^{N_n} \sigma_{\alpha_n}\,,
\ee
where $N_a$ denotes the number of spatial or time derivatives acting on the field, $\partial$'s are either $\partial_\eta$ or $\partial_i$, and the indices $\lbrace \alpha_a, a=1,..,n\rbrace$ denote the field types. Notice that the powers of the scale factor $  a $ are fixed by scale invariance. Sometimes, to avoid unnecessary technical difficulties, we restrict our attention to interactions with at most one temporal derivative per field. It is often convenient to write  \eqref{inte} in Fourier space, 
\begin{align}
\label{contact}
S_{\text{int}} =& g_n \int d\eta\, a(\eta)^{4-\sum s_a}\int \left(\prod_{a=1}^n \dfrac{d^3\bfk_a}{(2\pi)^3}\right) (2\pi)^3 \delta^3\left( \sum \limits_{i=1}^n\bfk_a  \right)\,\\ \nonumber
&\times \,\left[ \prod_{a=1}^{n} \partial_\eta^{s_a}\sigma_{\alpha_a}(\bfk_a,\eta) \right]\,F_n\left[\dfrac{\bfk_a \cdot\bfk_b}{a^2(\eta)}\right]\,.
\end{align}
Where $F_n[..]$ is the product $\prod \bfk_a \cdot \bfk_b / a^2$, with $a$ and $b$ running over the fields on which the spatial derivatives act.
The objects of ultimate interest are correlators that can be derived from the coefficients appearing in the wavefunction of the universe $  \Psi $ (see Appendix \ref{wfu} and \cite{Anninos:2014lwa, Goon:2018fyu} for more details\footnote{We thank Garrett Goon for sharing an unpublished  manuscript on the wavefunction of the universe.}). Given a set of fields $\sigma_{\alpha_{a}}$, which are weakly interacting with each other, the wavefunction can be expanded in $\bar{\sigma}_\alpha(\bfk)\equiv \sigma_\alpha (\bfk,\eta_0)$ (the value of the scalar field in momentum space and on the boundary $\eta=\eta_0$), 
\begin{align}\label{WFu}
\Psi[ \bar{\sigma}_\alpha(\bfk),\eta_0 ] &= \exp \Big[ -\dfrac{1}{2!}\sum_\alpha \int_{\bfk_1,\bfk_2} \psi_2^\alpha(\bfk_1,\bfk_2)\bar{\sigma}_\alpha(\bfk_1)\bar{\sigma}_\alpha(\bfk_2) \\ \nonumber
&-\sum_{n=3}^{\infty}\sum_{\alpha_1,..,\alpha_n}\int_{\bfk_1,..,\bfk_n}\frac{1}{r_{n}[\alpha_1\dots \alpha_n]}\psi_n^{\alpha_1\dots  \alpha_n}(\bfk_1,\dots ,\bfk_n)\bar{\sigma}^{\alpha_1}(\bfk_1)\dots  \bar{\sigma}^{\alpha_n}(\bfk_n) \Big]\,,
\end{align}
where we have introduced a symmetry factor $r_{n}[\alpha_1\dots  \alpha_n]$ that accounts for the number of trivial permutations of $\alpha_1,\dots ,\alpha_n$ (e.g. $n!$ when $\alpha_1=\alpha_2=\dots =\alpha_n$). By definition, $\psi_n^{\alpha_1\dots  \alpha_n}(\bfk_1,\dots ,\bfk_n)$ has the same symmetry under the permutation of momenta as the symmetries of $\alpha_1,\dots ,\alpha_n$ (e.g. $\psi_n$ is totally symmetric when all the fields are the same). To simplify our notation, we have dropped a potential $  \eta_{0} $ dependence in $  \psi_{n} $ and in the following we will drop the $\alpha_1,\dots ,\alpha_n$ indices where no ambiguity arises. With a small misuse of terminology, sometimes we refer to $\psi_n$ as the $ n  $-point function. 

Due to spatial translations, all the wavefunction coefficients $\psi_n$ have to be proportional to the Dirac delta function of the sum of spatial momenta, i.e. 
\be
\psi_n(\bfk_1,\dots ,\bfk_n)=\psi_n'(\bfk_1,\dots ,\bfk_n)(2\pi)^3\,\delta^3(\bfk_1+\dots +\bfk_n)\,.
\ee
It can be shown that in the tree-level approximation and assuming the Bunch-Davies vacuum in the far past, the wavefunction of the system is given by 
\be
 \Psi[ \bar{\sigma}_\alpha(\bfk),\eta_0 ]=e^{i S_{\text{cl}}[\sigma^{\text{cl}}_\alpha(\eta, \bfk)]}\,,
\ee
where $S_{\text{cl}}$ is the classical action evaluated on the field configuration $\sigma^{\text{cl}}_\alpha(\eta,\bfk)$ that (i) solves the \textit{classical equations of motion},
\begin{align}
\mathcal{O}_\alpha (\eta,k)\sigma_\alpha=-\frac{\partial }{\partial t}\frac{\delta \mathcal{L}_{\text{int}}}{\delta\dot{\sigma}_\alpha}+\frac{\delta\mathcal{L}_{\text{int}}}{\delta\sigma_\alpha},
\end{align}
where $  \mathcal{O}_\alpha (\eta,k)\sigma_\alpha  $ is a shorthand notation for the linearized equations of motion which depend only on the magnitude of the momentum,
\begin{align}
\mathcal{O}_\alpha (\eta,k)\sigma_\alpha\equiv \frac{\partial }{\partial t}\frac{\delta \mathcal{L}_2}{\delta\dot{\sigma}_\alpha}-\frac{\delta\mathcal{L}_2}{\delta\sigma_\alpha}\,.
\end{align}
And (ii) $\sigma^{\text{cl}}_\alpha(\eta,\bfk)$ satisfies the following \textit{boundary conditions}, 
\be
\lim_{\eta\to -\infty(1-i\epsilon)}\sigma^{\text{cl}}_\alpha(\eta, \bfk)=0\,,\qquad \sigma^\text{cl}_\alpha(\eta_0,\bfk)=\bar{\sigma}_\alpha(\bfk)\,.
\ee
Assuming interactions are weak, we can solve this perturbatively using the bulk-to-boundary propagator $K_\alpha$ and the bulk-to-bulk propagator $G_\alpha$. They satisfy
\begin{equation}
\mathcal{O}_\alpha(\eta,k)K_\alpha(k,\eta)=0,\quad 
\end{equation}
\begin{equation}
\mathcal{O}_\alpha (\eta,k)G_\alpha (k,\eta,\eta')=\delta (\eta-\eta')\,,
\end{equation}
and are subject to the following boundary conditions, 
\begin{align}
 &\, \; \lim_{\eta\rightarrow\eta_0}K_\alpha(k,\eta)=1,&  \, \; \lim_{\eta\to -\infty(1-i\epsilon)}K_\alpha(k,\eta)=0\,,\\ \nonumber
& \lim_{\eta,\eta'\rightarrow\eta_0}G_\alpha (k,\eta,\eta')=0,&  \lim_{\eta,\eta'\to -\infty(1-i\epsilon)}G_\alpha(k,\eta,\eta')=0\,.
\end{align}
Then, up to linear order in ${\cal L}_{\text{int}}$, the classical solution is given by,
\begin{equation}
\label{class}
\sigma_\alpha^{\text{cl}}(\eta,\textbf{k})=K_{\alpha}(k,\eta)\bar{\sigma}_\alpha(\textbf{k})+\int d\eta' \left.\left(-\frac{\partial }{\partial t}\frac{\delta \mathcal{L}_{\text{int}}}{\delta\dot{\sigma}_\alpha}+\frac{\delta\mathcal{L}_{\text{int}}}{\delta\sigma_\alpha}\right)\right\rvert_{K_{\alpha}(k,\eta)\bar{\sigma}_\alpha(\textbf{k}) }G_\alpha(k,\eta,\eta').+{\cal O}({\cal L}_{\text{int}}^2).
\end{equation}
Where the bulk-to-boundary and bulk-to-bulk propagators are given by
\begin{align}\label{uno}
K_\alpha(k,\eta) &= \dfrac{\sigma_\alpha^+(k,\eta)}{\sigma^+_\alpha(k,\eta_0)}\,,\\ \nonumber
G_\alpha(k,\eta,\eta')&= i\Big[\sigma_\alpha^-(k,\eta)\sigma^+_\alpha(k,\eta')\theta(\eta-\eta')\\ \nonumber
&+\sigma^-_\alpha(k,\eta')\sigma^+_\alpha(k,\eta)\theta(\eta'-\eta)-\sigma^+_\alpha(k,\eta)\sigma^+_\alpha(k,\eta')\dfrac{\sigma^-_\alpha(k,\eta_0)}{\sigma^+_\alpha(k,\eta_0)}\Big]\,.
\end{align}
The classical solution \eqref{class} can be extended to arbitrary order in ${\cal L}_{\text{int}}$ and thereby one can compute $S_{\text{cl}}[\sigma^{\text{cl}}_\alpha(\eta, \bfk)]$ in a diagrammatic fashion.

\subsection{The Cosmological Optical Theorem} \label{s:deriv}

In this section, we will derive some interesting implications of unitarity for the wavefunction coefficients. Let's start by considering the evolution operator in the interaction picture, which can be computed perturbatively in terms of the couplings of the theory, 
\begin{align}\label{eq:timeevo}
{\cal U}(\eta_0) &= {\cal T}\exp\left(-i\int_{-\infty}^{\eta_0} d\eta \, H_{\text{int}}(\eta)\right)\\ \nonumber
&= \mathbbm{1}-i\int_{-\infty}^{\eta_0}d\eta\, H_{\text{int}}(\eta)-\int_{-\infty}^{\eta_0} d\eta d\eta' \,H_{\text{int}}(\eta)\,H_{\text{int}}(\eta')\,\theta(\eta-\eta')+\dots \,. 
\end{align}
Above, $H_{\text{int}}(\eta)$ is the time-dependent interaction Hamiltonian of the system, and for the moment we leave implicit the Wick rotation that projects onto the vacuum of the free theory at past infinity. In flat space this property gives the well known (generalised) optical theorem (see eg. \cite{Schwartz:2013pla})
\begin{equation}
A_{i\rightarrow f}-A_{f\rightarrow i}^*=i\sum_X\int d\Pi_X(2\pi)^4\delta^4(p_i-p_X)A_{i\rightarrow X}A^*_{f\rightarrow X},
\end{equation}
where $A_{i\rightarrow f}$ is the amplitude from some initial state, $i$ to a final state $f$, the sum is over all possible states $X$ and $d\Pi_X$ represents an integral over all the momenta in the state $X$. Paralleling this derivation, we decompose ${\cal U}$ as
\be
\cU=\mathbbm{1}+\delta \cU\,,
\ee
Then $\,\cU^\dagger \cU =\mathbbm{1}\,$ implies 
\be
\label{opt}
\delta \cU +\delta\cU ^\dagger=-\delta \cU \, \delta \cU ^\dagger\,.
\ee
This operator equation can be sandwiched between any two $ n  $-particle states of the free theory, defined via 
\be
\ket{\{\bfk,\alpha\}_{n}}\equiv\big | \bfk_1, \alpha_1; \dots ;\bfk_n,\alpha_n \big\rangle=a^\dagger(\bfk_1,\alpha_1)\dots  a^\dagger (\bfk_n,\alpha_n)\big |0\big\rangle\,,
\ee
(where the $\alpha_a$'s characterize the particle type and $|0 \rangle$ is the vacuum of the free theory.) Consider now sandwiching  \eqref{opt} between $\langle \{\bfk,\alpha\}_{n} |$ and $|0\rangle$ for arbitrary $n$ and $\alpha_a$'s. Inserting the identity operator constructed from $  m  $-particle states, 
\begin{align}
\mathbbm{1}&=\sum_{m=0}^{\infty}\, \sum\limits_{\beta_1,\dots ,\beta_m}\int \dfrac{d^3\bfp_1}{(2\pi)^3}\dots  \dfrac{d^3\bfp_m}{(2\pi)^3}\left\rvert \{\bfp,\beta\}_{m}\right\rangle\, \left\langle \{\bfp,\beta\}_{m} \right\rvert \,,
\end{align}
between $\delta \cU$ and $\delta \cU^\dagger$ on the right-hand side of  \eqref{opt} we obtain the following version of the \textit{optical theorem} 
\begin{tcolorbox}[colback=brown!10!white, colframe=white] 
\begin{align}
 \label{UNIT}
& \bra{\{\bfk,\alpha\}_{n}} \delta \cU (\eta_0)\ket{ 0 } +\bra{\{\bfk,\alpha\}_{n}} \delta \cU  (\eta_0)^\dagger \ket{ 0 } = \\ \nonumber
&-\sum_{m=0}^{\infty}\,\sum\limits_{\beta_1,\dots ,\beta_m}\int \dfrac{d^3\bfp_1}{(2\pi)^3}\dots  \dfrac{d^3\bfp_m}{(2\pi)^3}\left\langle \{\bfk,\alpha\}_{n}\left\lvert \delta \cU  (\eta_0)\vphantom{)^\dagger}\right\rvert \{\bfp,\beta\}_{m}\right\rangle    \left\langle  \{\bfp,\beta\}_{m}\left\lvert \delta \cU  (\eta_0)^\dagger \right\lvert 0\right\rangle\,.
\end{align}
\end{tcolorbox}
This abstract equation is not of much use unless both sides can be related to the boundary observables. We will now show that, working perturbatively in the couplings, the matrix elements appearing in  \eqref{UNIT} can be directly written in terms of the coefficients of the wavefunction of the universe. In general, the time evolution operator may be gauge dependent. All of our result will be valid in any gauge as long as the same gauge is used throughout the calculation.


\subsection{Contact diagrams} 

To linear order in $  H_{\text{int}} $, i.e. for contact diagrams, the left-hand side of  \eqref{UNIT} should vanish on its own and the optical theorem simply gives
\begin{align}
\label{concop}
 \bra{\{\bfk,\alpha\}_n} \int_{-\infty}^{\eta_0} d\eta\, H_{\text{int}} (\eta) \ket{0}- \bra{\{\bfk,\alpha\}_n } \int_{-\infty}^{\eta_0} d\eta \, H_{\text{int}}^\dagger (\eta) \ket{0}=0\,.
\end{align}
This is just $H_{\text{int}}^\dagger=H_{\text{int}}$ sandwiched between the free vacuum and some $ n  $-particle state. Despite its trivial appearance, the Hermicity of the Hamiltonian will have interesting consequences for boundary correlators already at this order. 

Up to linear order in $g_n$, the left-hand side of  \eqref{concop} becomes\footnote{For interactions with at most one time derivative, $H_{\text{int}}$ is not necessarily $-{\cal L}_{\text{int}}$. Instead, it generically receives corrections of order $g_n^2$ due to the non-linear relation between the conjugate momentum and $\dot{\sigma}_\alpha$. This fact will not alter our contact term discussions, but we will come back to it when discussing exchange diagrams.}
\begin{align}
\nonumber
&\bra{\{\bfk,\alpha\}_{n}} \int_{{-\infty(1-i\epsilon)}}^{\eta_0} d\eta\, H_{\text{int}} (\eta) \ket { 0}= \\ \nonumber
&- g_n \int_{-\infty(1-i\epsilon)}^{\eta_0} d\eta\, a^{4-\sum s_a}\int_{\bfq_a}\, F_n\left[\frac{\bfq_a\cdot \bfq_b}{a^2(\eta)}\right](2\pi)^3 \delta^3\left(\sum \bfq_a\right)\langle 0| \prod_{a,b=1}^n a_{\bfk_a}( \alpha_a)  \partial_\eta^{s_b}\hat{\sigma}_{\alpha_b}(\bfq_b,\eta) |0\rangle \\ \label{cUexp}
&= - g_n \int_{-\infty(1-i\epsilon)}^{\eta_0} d\eta\, a^{4-\sum s_a} F_n\left[\dfrac{\bfk_a\cdot \bfk_b}{a^2(\eta)}\right] \prod_{a=1}^n \partial_{\eta}^{s_{a}} \sigma^+_{\alpha_a}(k_a,\eta)\,.\\ \nonumber
&\bra{\{\bfk,\alpha\}_{n}} \int_{-\infty(1-i\epsilon)}^{\eta_0} d\eta \, H_{\text{int}}^\dagger (\eta) \ket { 0}= \\ \nonumber
&- g^*_n \int_{{-\infty(1-i\epsilon)}}^{\eta_0} d\eta\, a^{4-\sum s_a} \int_{\bfq_a}\, F_n\left[\frac{\bfq_a\cdot \bfq_b}{a^2(\eta)}\right](2\pi)^3 \delta^3\left(\sum \bfq_a\right)\left \langle 0\right\lvert  \prod_{a,b=1}^n a_{\bfk_a}( \alpha_a) \partial_\eta^{s_b}\hat{\sigma}_{\alpha_b}(-\bfq_b,\eta) \left \rvert 0\right \rangle \\ 
&= - g^*_n \int_{-\infty(1-i\epsilon)}^{\eta_0} d\eta \,a^{4-\sum s_a} F_n\left[\dfrac{\bfk_a\cdot \bfk_b}{a^2(\eta)}\right] \prod_{a=1}^n \partial_{\eta}^{s_{a}}\sigma^+_{\alpha_a}(k_a,\eta)\,, \label{cUexp2}
\end{align}
where now we made explicit the $  i\epsilon $ prescription that projects onto the vacuum of the free theory at early times.
Since the annihilation operators sit to the left of $H_{\text{int}}$ in both matrix elements above, the negative frequencies (i.e. $\sigma^-_{\alpha_a}(k,\eta)$) did not make an appearance. Therefore, \eqref{cUexp} and \eqref{cUexp2} are intimately related to the wavefunction coefficients $\psi_n(\bfk_1,\dots ,\bfk_n)$ which similarly only incorporates positive frequencies (inside the bulk-to-boundary propagators) at the contact level. To see this, recall that the wavefunction coefficients are given by
\be
\psi'_n(\bfk_1,\dots ,\bfk_n)= -i r_n g_n\int_{-\infty(1-i\epsilon)}^{\eta_0} d\eta\, a(\eta)^{4-\sum s_a}\,F_n\left[\dfrac{\bfk_a\cdot \bfk_b}{a^2(\eta)}\right]\, \prod_{a=1}^n \partial_{\eta}^{s_{a}} K_{\alpha_a}(k_a,\eta)\,.
\ee
It will prove useful to have an alternative form for $\psi_n$ by absorbing the $i\epsilon$ part of the contour integral into the argument of the bulk-to-boundary propagators, and expressing the argument of $\psi_n$ in terms of the norm of momenta $k_a$ and the angles between every pair of them, i.e. 
\be
\label{analc}
\psi'_n(k_1,\dots ,k_n, \hat{k}_a\cdot\hat{k}_b)=-i r_n g_n \int_{-\infty}^{\eta_0} d\eta\, a(\eta)^{4-\sum s_a}\,F_n\left[\dfrac{\bfk_a\cdot \bfk_b}{a^2(\eta)}\right]\, \prod_{a=1}^n \partial_{\eta}^{s_{a}}K_{\alpha_a}(k_a-i\epsilon,\eta)\,,
\ee
where $  \hat{k}_{a}\equiv \bfk_{a}/k_{a} $. Therefore, the relationship between the matrix element in  \eqref{cUexp} and $\psi_n$ is 
\be
\bra{\{\bfk,\alpha\}_{n}} \int_{-\infty(1-i\epsilon)}^{\eta_0} d\eta\, H_{\text{int}} (\eta) \ket{ 0}=-\dfrac{i}{r_n}\psi'(k_1,\dots ,k_n,\hat{k}_a \cdot \hat{k}_b)\prod_{a=1}^n \sigma^+(k_a,\eta_0)\,.
\ee
As  \eqref{cUexp2} is not the complex conjugate of  \eqref{cUexp}, we need two additional results to relate  \eqref{cUexp2} to $\psi_n$:
\begin{itemize}
\item The bulk-to-boundary propagator in de Sitter obeys the following property:
\be
\label{BtBp}
K^*_\sigma(z,\eta)= K_\sigma(-z^*,\eta)\,,\qquad \text{Im}(z)< 0\,.
\ee 
\begin{proof}
For a scalar field with an arbitrary mass \eqref{quad}, using \eqref{due} and \eqref{uno} we find
\be
K_\sigma(z,\eta)=\dfrac{(-\eta)^{3/2} H^{(2)}_{\nu}(-c_\sigma z\eta)}{(-\eta_0)^{3/2} H^{(2)}_{\nu}(- c_\sigma z\eta_0)}\,.
\ee
By its definition, $\nu=\sqrt{\frac{9}{4}-\frac{m_\sigma^2}{H^2}}$ is either real or pure imaginary, i.e. $\nu^*=\pm\nu$. We also use the principal value for the Hankel function $H^{(2)}(x)$ which has a branch cut along the negative real axis, $x\in [-\infty, 0)$. 

Using the integral representation of the Hankel function of the second kind over the lower half complex plane \cite{Abraham}, one obtains, 
\be
H_\nu^{(2)}(x)=-\dfrac{\exp(\frac{i}{2}\nu \pi)}{i \pi} \int_{-\infty}^{+\infty} dt \exp(-i x \cosh t-\nu t)\,, \quad \text{Im}(x)<0\,,
\ee
and therefore
\begin{align}
& H^{(2) *}_\nu(x) = -\exp(-i\pi \nu) H_\nu^{(2)}(-x^*)\,,\qquad \nu \in \mathbb{R}\,,\\ \nonumber 
& H^{(2) *}_\nu(x)=- H_\nu^{(2)}(-x^*)\,,\qquad i \nu \in \mathbb{R}\,, 
\end{align}
for $\text{Im}(x)<0$, hence  \eqref{BtBp}. Notice that we don't expect this property to hold for generic modifications of the Bunch-Davies initial states, e.g. a general Bogoliubov transformation.
\end{proof}
\item For both massless and conformally-coupled fields with IR-finite contact interactions, $\psi_n(k_a,\hat{k}_a \cdot \hat{k}_b)$ is analytic in the whole complex plane of energies (up to poles), namely for $k_a\in \mathbb{C}$, and the analytical continuation to $k_a<0$ is straightforward in these cases. For other massive fields or when IR-divergent interactions are present, 
there is a natural analytical continuation of $\psi_n(k_a,\hat{k}_a \cdot \hat{k}_b)$ to the \textit{lower-half $ n  $-hyperplane of complex energies} (defined via $\mathbb{C}^{n -}\equiv \lbrace k_a, \text{Im}(k_a)<0\rbrace$), which is given by 
\begin{align}
\nonumber
&\psi'_n(k_a, \hat{k}_a \cdot \hat{k}_b) = \\ \label{analc2}
&  -i r_n g_n\int_{-\infty}^{\eta_0} d\eta\, a(\eta)^{4-\sum s_a}\,F_n\left[\dfrac{\bfk_a\cdot \bfk_b}{a^2(\eta)}\right]\, \prod_{a=1}^n \dfrac{d^{s_a}}{d\eta^{s_a}} K_{\alpha_a}(k_a,\eta)\,,\quad \text{Im}(k_a)<0\,.
\end{align}
This continuation is particularly useful when we approach the negative real energies from \textit{below}, i.e. $\psi_n(-|k_a|-i\epsilon, \hat{k}_a \cdot \hat{k}_b)$ with $  \epsilon >0 $. 
For $\text{Im}(k_a)<0$, the integral on the right-hand side of  \eqref{analc2} is convergent thanks to the asymptotic behaviour of $K(k,\eta)$, namely
\be
\lim_{\eta\to -\infty}\, K_a(k,\eta)\propto e^{+i \text{Re}(k)\eta } e^{ -\text{Im}(k)\eta }\,.
\ee
Therefore, $\psi_n(k_a)$ can diverge only when $k_a\in \mathbb{R}^-$ (notice that, by definition,  \eqref{analc} is regular at $k_a\in \mathbb{R}^+$). This can happen in two cases: (i) when IR divergences are present, leading to a branch cut at $k_T<0$, and/or  (ii) when at least one external field (e.g. with momentum $k_a$) is massive, for which the associated mode function has a branch cut at $k_a<0$. 
\end{itemize}
By virtue of these two properties, for $k_a\in \mathbb{R}^+$, we can express the second matrix element in  \eqref{concop} in terms of $\psi_n$  as 
\begin{align}
\bra{\{\bfk,\alpha\}_{n}} \int_{-\infty(1-i\epsilon)}^{\eta_0} d\eta \, H_{\text{int}}^\dagger (\eta) \ket{ 0}=\dfrac{i}{r_n}\left[\psi'_n(-k_a-i\epsilon,\hat{k}_a \cdot \hat{k}_b)\right]^*\prod_{a=1}^n \sigma^+(k_a,\eta_0)\,,
\end{align}
where $\psi_n(-k_a-i\epsilon,\hat{k}_a \cdot \hat{k}_b)$ is now defined by the analytical continuation in  \eqref{analc2}. Notice that for conformally coupled and massless scalars (with IR-finite interactions) the $-i\epsilon$ in  argument of $\psi_n(-k_a-i\epsilon)$ is unnecessary, as the limit is smooth. \\

So finally, we can write  \eqref{concop} for a contact diagram as 
\begin{tcolorbox}[colback=brown!10!white, colframe=white] 
\be
\label{unicont}
\qquad \psi'_n(k_a,\hat{k}_a \cdot \hat{k}_b)+\left[\psi'_n(-k_a-i\epsilon,\hat{k}_a \cdot \hat{k}_b)\right]^*=0\,,\qquad k_a\in \mathbb{R}^+\,.
\ee  
\end{tcolorbox}
Equipped with the analytical continuation to $k_a\in \mathbb{C}^{n -}$, we can easily generalize this optical theorem to 
\be\label{disc0}
\psi'_n(k_a,\hat{k}_a \cdot \hat{k}_b)+\left[\psi'_n(-k_a^*,\hat{k}_a \cdot \hat{k}_b)\right]^*=0\,,\qquad k_a\in \mathbb{C}^{n -}\,.
\ee
This relationship between wavefunction of the universe coefficients is shown diagrammatically from a bulk perspective in figure 1. Notice that it is only the energies $  k_{a} $ that are analytically continued, while the angles $  \hat{k}_a \cdot \hat{k}_b $ are left unchanged. This form of the optical theorem is the cosmological wavefunction equivalent of Hermitian analyticity of the amplitude in flat space \cite{Eden:1966dnq}. Indeed, one could simply rescale $  \psi_{n} $ in such a way that it obeys the same property as amplitudes, namely
\begin{align}
\Psi_{n}(k_{a})\equiv i  \psi'_n(i k_a,\hat{k}_a \cdot \hat{k}_b) \then \Psi^{\ast}_{n}(z_{i}^{\ast})=\Psi_{n}(z_{i})\,.
\end{align}
As for the discontinuity of amplitudes across the positive real axis, we can extract from \eqref{disc0} an expression for the discontinuity of $  \psi'_{n} $ across the negative imaginary axis. Evaluating the general expression
\begin{align}
2\Re \psi'_{n}(k_a,\hat{k}_a \cdot \hat{k}_b)&=\psi'_{n}(k_a,\hat{k}_a \cdot \hat{k}_b)+\left[ \psi'_{n}(k_a,\hat{k}_a \cdot \hat{k}_b) \right]^{\ast}\\
&=\psi'_{n}(k_a,\hat{k}_a \cdot \hat{k}_b)-  \psi'_{n}(-k_a^{\ast},\hat{k}_a \cdot \hat{k}_b)
\end{align}
at $  k_{a}=\e-i\lambda_{a} $ with $  \lambda_{a} $ and $  \e $ positive and real, and leaving the angle dependence implicit, we find in the limit $  \e\to 0 $
\begin{align}
2\lim_{\e\to0}\Re \psi'_{n}(\e -i\lambda_{a})&=\lim_{\e\to0} \left[ \psi'_{n}(\e-i\lambda_{a})-\psi'_{n}(-\e-i\lambda_{a}) \right]\\
&=\text{Disc}\left[ \psi'_{n}(-i\lambda_{a}) \right]\,.
\end{align}
At least for finite momenta $  k_{a}=-i\lambda_{a}\neq 0 $ in the complex lower-half plane, the wavefunction coefficients $  \psi_{n} $ should not have discontinuities because the integrals that defined them are convergent. This tells us that $  \psi_{n} $ must be purely imaginary on the negative imaginary axis. It would be interesting to understand what happens at the origin, $  \lambda_{a}=0 $, and in particular determine if a delta function can be hiding there. We leave this for future work.\\
 
\begin{figure}
\begin{center}
\includegraphics[scale=0.8]{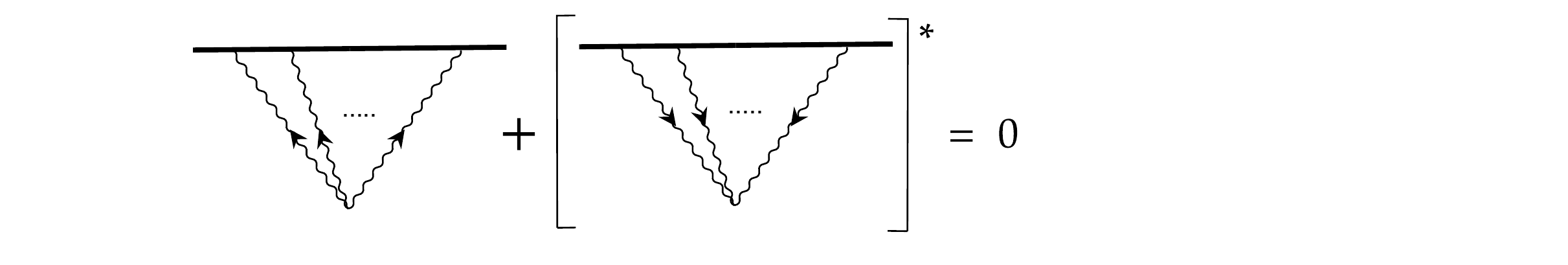}
\end{center}
\caption{Diagrammatic form of the bulk perspective on the optical theorem for contact terms. The direction of the arrow indicates the sign of the frequency inserted in the corresponding bulk-to-boundary propagator. It is positive when the arrow is towards the boundary and negative otherwise.\label{schemcon}}
\end{figure}

Let us now explore the consequences of the \textit{Cosmological Optical Theorem} (COT) for IR-finite interactions, leaving the non-trivial IR-divergent terms to Section \ref{NonTC}. By virtue of scale invariance, $\psi_n$ must be a homogeneous function of the momenta of degree three minus the scaling dimensions $ \Delta_\alpha $ of the fields under consideration\footnote{Notice that what we call $  \Delta $ is sometimes called $  \Delta^{+} $ in the literature, while $  \Delta^{-}=3-\Delta $.},
\be
\Delta_\alpha=\dfrac{3}{2}+\sqrt{\dfrac{9}{4}-\dfrac{m^2_\alpha}{H^2}}\in \mathbb{R}\,,
\ee 
where for simplicity we have assumed that all the scalars are light, $m_a<3H/2$.
Therefore, the wavefunction coefficient $\psi_n$ can always be written as
\be
\label{irfin}
\psi'^{\alpha_1\dots  \alpha_n}_n(k_a,\hat{k}_a \cdot \hat{k}_b)=k_T^{3(1-n)+\sum_a \Delta_{\alpha_a}}\,f_n\left( \dfrac{k_a}{k_T}, \hat{k}_a \cdot \hat{k}_b \right)\,.
\ee
When $\sum\limits_a \Delta_{\alpha_a}$ is irrational, following  \eqref{analc}, one should use the principal value of $\ln k_T$ (with a branch cut at $[-\infty,0]$), which is analytical over $\mathbb{C}^{n -}$. Using, 
\be
\label{logCOT}
\ln (-k_T-i\epsilon)=-i\pi+\ln (k_T)\,, 
\ee
and  \eqref{unicont} we arrive at, 
\be
\label{fnstar}
f_n^*=\exp(i \pi (3n-\sum\limits_a \Delta_{\alpha_a}))\, f_n\,.
\ee
In particular, for all massless fields, $\Delta_\alpha=3$, the COT implies
\be
\label{Realitymassless}
\text{Im}(\psi'^\phi_n)=0 \quad\quad \text{(massless field)}\,, 
\ee
whereas, for conformally-coupled fields, $\Delta_\alpha=2$, we have
\begin{align}
\label{vpicont}
& \text{Im}(\psi'^\vpi_n)=0\,,\qquad n=\text{even}\,,\\ \nonumber
& \text{Re}(\psi'^\vpi_n)=0\,,\qquad n=\text{odd}\,.
\end{align}
These examples already indicate that to express  perturbative unitarity at the contact level it is natural to work with the wavefunction coefficients, $\psi_n$, as opposed to the correlators. Up to linear order in the coupling, i.e. for contact interactions, the two quantities are related by
\be\label{bn}
B_n= -2 \left[ \prod_{a=1}^{n} \frac{1}{2\Re \psi_{2}'(k_{a})} \right] \Re \psi_n'\,,
\ee
where dilations impose the following scalings with momentum
\begin{align}\label{scaling}
B_{n}(\lambda \bfk_{a})= \frac{B(\bfk_{a})}{\lambda^{-3+\sum_{a}\Delta_{\alpha_{a}}}}\,.
\end{align}
For integer scaling dimensions $ \Delta$, such as for massless and conformally couples scalars, $\psi_n$ has only isolated poles but no branch cuts, i.e. it is ``analytical''. Then there is no ambiguity in evaluating the correlator at negative energies $k_a\to -k_a$. As a result, by taking the real part of \eqref{irfin} and using \eqref{bn} and \eqref{scaling} we find
\be
B_n(k_a, \hat{k}_a\cdot\hat{k}_b)\left[  1+(-1)^{3(1-n)+\sum_{a}\Delta_{\sigma_{a}}}\right]=0\,,\quad \quad  \text{(for $B_n$ analytic)}\,.
\ee
This equation tells us that any correlator $  B_{n} $ involving an odd number of conformally coupled fields ($  \Delta =2 $) and an arbitrary number of massless fields ($  \Delta =3 $) must vanish. We will check this fact explicitly in Section \ref{checkcontact}.


\subsection{Four-point exchange diagrams}
\label{fourpoint}
In this section, for simplicity, we focus on the quartic wavefunction for a single massless field $\phi$ mediated by the exchange of some other massless field $\sigma$. Assuming that the contribution is IR finite, this restriction enforces the final answer to be analytic for all complex energies. To begin with, let us assume that $\sigma$ interacts with $\phi$ via the following cubic vertices, 
\be
\label{cubic}
S_3=-\int d\eta d^3\bfx\, \left(g_A\,a^{4-N_A-M_A}\,\partial^{N_A}\phi\,\partial^{M_A}\phi\,\sigma+ g_B\,a^{4-N_B-M_B}\,\partial^{N_B}\phi\,\partial^{M_B}\phi\,\sigma \right)\,.
\ee
Where $N_{A,B}$ and $M_{A,B}$ are either 0 or 1, and derivatives on $\phi$ could be spatial or temporal. These interactions are special in that the $\sigma$ field does not carry a time derivative. The final result that we are about to derive holds even if the exchanging field held a time derivative or was identical to the external field $\phi$ (Appendix \ref{gencopt}). 
In Fourier space, the corresponding Hamiltonian can be expressed as, 
\begin{align}
H_{\text{int}} &=H^A+H^B+H_{4}\,,\\ \nonumber
H^A &= g_A \int d\eta\, a(\eta)^{4-n_A-m_A}\int \left(\prod_{a=1}^3 \dfrac{d^3\bfk_a}{(2\pi)^3}\right) (2\pi)^3 \delta^3\left(\sum \limits_{a=1}^3\bfk_a\right)\,\\ \nonumber
&\times F_A\left[\dfrac{\bfk_a\cdot \bfk_b}{a^2}\right]\,\partial_\eta^{n_A}\phi(\bfk_1,\eta)\partial_\eta^{m_A}\phi(\bfk_2,\eta)\sigma(\bfk_3,\eta)\,,\\ \nonumber
H^B &=g_B \int d\eta\, a(\eta)^{4-n_B-m_B}\int \left(\prod_{a=1}^3 \dfrac{d^3\bfk_a}{(2\pi)^3}\right) (2\pi)^3 \delta^3\left(\sum \limits_{a=1}^3\bfk_a\right)\,\\ \nonumber
&\times F_B \left[\dfrac{\bfk_a\cdot \bfk_b}{a^2}\right]\,\partial_\eta^{n_B}\phi(\bfk_1,\eta)\partial_\eta^{m_B}\phi(\bfk_2,\eta)\sigma(\bfk_3,\eta)\,,
\end{align}
where $n_A$ and $m_A$ denote the number of time derivatives and can be 0 or 1. Notice that although $H_A$ and $H_B$ are cubic in the fields, there is a possible additional term $H_4$, which starts at quartic order in the fields, that may arise in the presence of derivative interactions from the non-linear relation between $\dot{\phi}$ and the conjugate momentum of $\phi$. However, it is straightforwardly seen that $H_4$, if at all present, must be quadratic in $\sigma$ and therefore  cannot contribute to the matrix elements we compute below\footnote{When $\sigma$ is identical to $\phi$, the computed matrix elements do receive contributions from $H_4$, but even then, since $H_4$ contains only quartic \textit{contact} terms, such contributions cancel from the final optical theorem  \eqref{exchCOPT}. See Appendix \ref{gencopt} for further details.}.

Up to order ${\cal O}(g_A g_B)$,  \eqref{UNIT} implies
\begin{align}
\label{unitexch}
&\blangle \bfk_1,\phi;\dots ;\bfk_4,\phi\bl \delta \cU_{g_A g_B} \bl 0\brangle+\blangle \bfk_1,\phi;\dots ;\bfk_4,\phi \bl \delta \cU_{g_A g_B} ^\dagger \bl 0 \brangle=\\ \nonumber
&-\int \left(\prod_{a=1}^3 \dfrac{d^3\bfp_a}{(2\pi)^3}\right)\blangle \bfk_1,\phi;\dots ;\bfk_4,\phi\bl \delta \cU_{g_A}\bl \bfp_1,\phi;\bfp_2,\phi;\bfp_3,\sigma\rangle   \\ \nonumber
&\hspace{4cm}\times \blangle \bfp_1,\phi;\bfp_2,\phi;\bfp_3,\sigma \bl \delta \cU_{g_B}^\dagger \bl 0\brangle+ (A\leftrightarrow B)\,,
\end{align}
where on the left-hand side, $g_A g_B$ index denotes the ${\cal O}(g_A g_B)$ contribution to $\delta {\cal U}$, and similarly do $g_A$ and $g_B$ indices on the right-hand side, i.e. 
\begin{align}
\delta \cU_{g_A g_B} &= -\int_{-\infty}^{\eta_0} d\eta \int_{-\infty}^{\eta_0}d\eta' H^{A}(\eta)\,H^{B}(\eta')\,\theta(\eta-\eta')+(A\leftrightarrow B)\,,\\ \nonumber
\delta \cU_{g_A} &=-i \int_{-\infty }^{\eta_0} d\eta H^{A}(\eta)\,.
\end{align}
Here again we left implicit the projection $ -\infty    (1-i\epsilon) $ onto the vacuum at early times. Notice that in the identity operator inserted on the right-hand side of  \eqref{unitexch}, $n$-particle states other than the 3-particle states do not contribute at linear order in the couplings. Our task now is to relate the matrix elements appearing above to the wavefunction coefficients, namely $\psi_3^{\phi}$ and $\psi_4^{\phi}$. To that end, we need the following explicit expressions, 
\begin{align}
\label{delU}
\blangle \bfk_1,\phi;\dots ;\bfk_4,\phi\bl \delta \cU_{g_A g_B} \bl 0\brangle' &= i g_A g_B \int d\eta d\eta'\, G_{++}(s,\eta,\eta')a(\eta)^{4-n_A-m_A}a(\eta')^{4-n_B-m_B}\\ \nonumber
&\times \left(F_A \left[ \frac{\bfk_a\cdot \bfk_b}{a^2(\eta)} \right]\,\del^{n_A}_\eta \phi^+(k_1,\eta)\del^{m_A}_\eta \phi^+(k_2,\eta)+(k_1\leftrightarrow k_2)\right)\\ \nonumber
&\times \left(F_B\left[ \frac{\bfk_a\cdot \bfk_b}{a^2(\eta')} \right]\,\del^{n_B}_\eta \phi^+(k_3,\eta')\del^{m_B}_\eta \phi^+(k_4,\eta')+(k_3\leftrightarrow k_4)\right)\\ \nonumber
&+(A \leftrightarrow B)+(t\,,u \,\, \text{channels})\,,\\
\blangle \bfk_1,\phi;\dots ;\bfk_4,\phi\bl \delta \cU^\dagger_{g_A g_B} \bl 0\brangle' &= -i g_A g_B \int d\eta d\eta'\, G_{--}(s,\eta,\eta') a(\eta)^{4-n_A-m_A}a(\eta')^{4-n_B-m_B}\\ \nonumber
&\times \left(F_A\left[ \frac{\bfk_a\cdot \bfk_b}{a^2(\eta)} \right]\,\del^{n_A}_\eta \phi^+(k_1,\eta)\del^{m_A}_\eta \phi^+(k_2,\eta)+(k_1\leftrightarrow k_2)\right)\\ \nonumber
&\times \left(F_B\left[ \frac{\bfk_a\cdot \bfk_b}{a^2(\eta')} \right]\,\del^{n_B}_\eta \phi^+(k_3,\eta')\del^{m_B}_\eta \phi^+(k_4,\eta')+(k_3\leftrightarrow k_4)\right)\\ \nonumber
&+(A \leftrightarrow B)+(t\,,u \,\, \text{channels})\,.
\end{align}
Above and in the following, the $ t  $ ($ u  $) channel can be obtained by replacing $s\to t\,,\bfk_2\to \bfk_3, \bfk_3\to \bfk_2$ ($s\to u, \bfk_2\to \bfk_4, \bfk_4\to \bfk_2$).  
We have also defined, 
\begin{align}
\label{eq:curvedpropagators}
G_{++}(k,\eta,\eta') &= i\left[\sigma^-(k,\eta)\sigma^+(k,\eta')\theta(\eta-\eta')+\sigma^+(k,\eta)\sigma^-(k,\eta')\theta(\eta'-\eta)\right]\,,\\ \nonumber
G_{--}(k,\eta,\eta')&= G^*_{++}(k,\eta,\eta')\,.
\end{align}
Notice that these are two of the four propagators that appear in the in-in formalism \cite{Weinberg:2005vy}, but they can be easily related to the propagator of the wavefunction method via, 
\be
\label{props}
G(k,\eta,\eta')=G_{++}(k,\eta,\eta')-i\,\sigma^+(k,\eta)\sigma^+(k,\eta')\dfrac{\sigma^-(k,\eta_0)}{\sigma^+(k,\eta_0)}\,.
\ee
From  \eqref{delU}, we can directly see why  \eqref{unitexch} holds: the step functions add up to 1 (i.e. $\theta(\eta-\eta')+\theta(\eta'-\eta)=1$), which trades the sums of time-ordered and anti-time-ordered double-time integrals with a product of single-time integrals on the right-hand side. This observation might seem to suggest that  \eqref{unicont} is nothing but a trivial algebraic relation. However, as we saw for contact terms, the true value of  \eqref{unicont} is not in its bulk-integral manifestation, but in its interpretation in terms of boundary observables. 

Before relating the above matrix elements to $\psi_4^{\phi} $, let us first directly compute the right-hand side of  \eqref{unitexch}, after stripping off the momentum conserving delta function,
\begin{align}
\nonumber
& -\int \left(\prod_{a=1}^3 \dfrac{d^3\bfp_a}{(2\pi)^3}\right)\left\langle \bfk_1,\phi;...;\bfk_4,\phi\left\rvert\delta \cU_{g_A}\right\rvert \bfp_1,\phi;\bfp_2,\phi;\bfp_3,\sigma\right \rangle' \left\langle \bfp_1,\phi;\bfp_2,\phi;\bfp_3,\sigma\left\rvert \delta \cU_{g_B}^\dagger \right\lvert 0\right\rangle'+ (A\leftrightarrow B)\\ \nonumber
&= -g_A g_B \int d\eta\,a(\eta)^{4-n_A-m_A}\,\left( F_A\left [\dfrac{\bfk_a\cdot\bfk_b}{a^2(\eta)}\right]\del_\eta^{n_A}\phi^+(k_1,\eta)\del_\eta^{m_A}\phi^+(k_2,\eta)+(1\leftrightarrow 2) \right)\sigma^-(s,\eta)\\ \nonumber
&\times\int d\eta'\,a(\eta')^{4-n_B-m_B}\left(F_B\left[\dfrac{\bfk_a\cdot\bfk_b}{a^2(\eta')}\right]\del_\eta^{n_B}\phi^+(k_3,\eta')\del_\eta^{m_B}\phi^+(k_4,\eta')+(3\leftrightarrow 4) \right)\sigma^+(s,\eta')\\ \nonumber
&+(A\leftrightarrow B)+(t\,,u\,\,\,\,\text{channels})\\ \nonumber
&=\left(\prod_i \phi^+(k_i,\eta_0)\right)P_\sigma(s,\eta_0) \left[\psi'^{\phi\phi\sigma}_A(k_1,k_2,-s)\psi'^{\phi\phi\sigma}_B(k_3,k_4,s)+\psi'^{\phi\phi\sigma}_A(k_3,k_4,-s)\psi'^{\phi\phi\sigma}_B(k_1,k_2,s)\right]\\ 
&+(A\leftrightarrow B)+(t\,,u\,\,\,\,\text{channels})\,.
\end{align}
In the last equality, the time integral representation is written in terms of the cubic wavefunction coefficient $\psi'^{\phi\phi\sigma}_{A,B}$,. These are simply the contribution of each cubic interaction in  \eqref{cubic} to the cubic part of the wavefunction \eqref{WFu}. Due to the conservation of spatial momenta, $\psi'^{\phi\phi\sigma}_{A,B}$ is only a function of the norms, $k_a$, of the momenta.  Furthermore, above we had to employ the analytical continuations of $\psi_{A,B}$ to negative energies in the $\sigma$ field\footnote{We could have also used, for example, $-[\psi_A^{\phi\phi\sigma}(-k_1,-k_2,s)]^*$ instead of $\psi'^{\phi\phi\sigma}_A(k_1,k_2,-s)$ using the optical theorem for contact terms.}. 
\begin{figure}
\begin{center}
\includegraphics[scale=0.8]{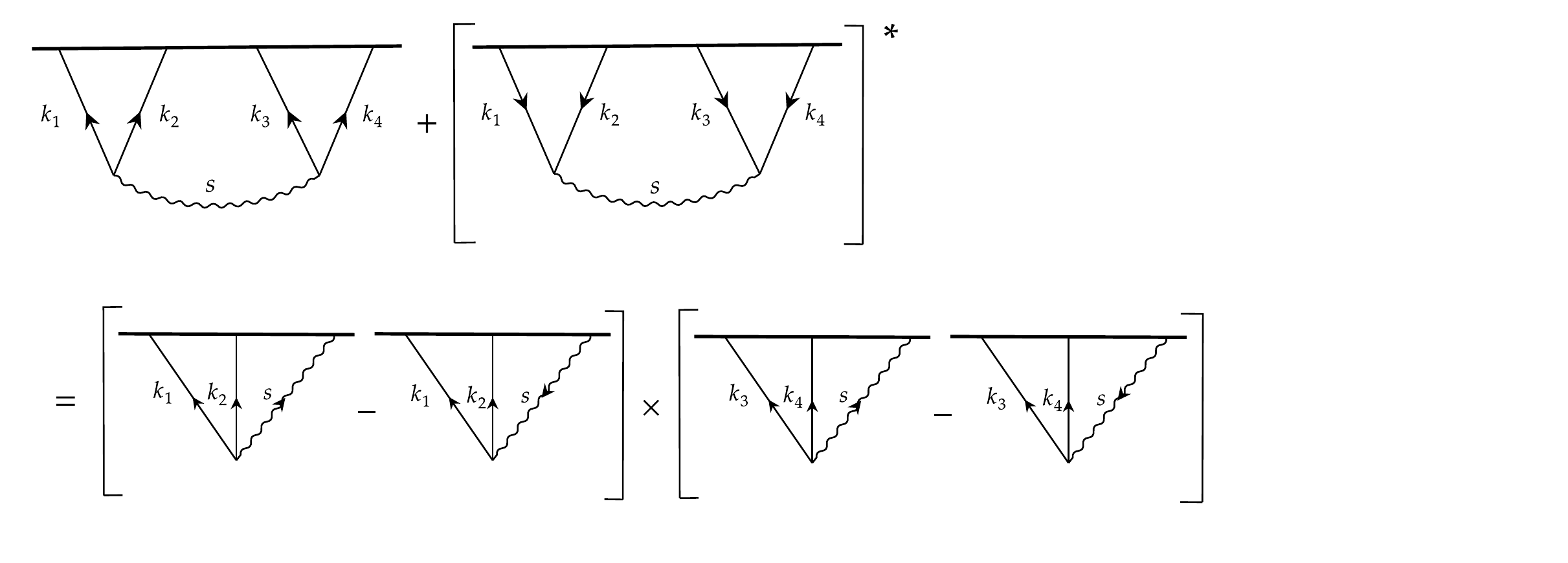}
\end{center}
\caption{A diagrammatic representation of the optical theorem for exchange diagrams from a \textit{bulk} perspective. The straight lines represent $  \phi $ propagators while the wiggly lines are $  \sigma $ propagators.}
\end{figure}
The final piece of our puzzle will be a relationship between the sum $\langle \bfk_1,...,|\delta \cU_{g_A g_B} | 0\rangle+\langle \bfk_1,...,|\delta \cU^\dagger_{g_A g_B} | 0\rangle$ and the quartic wavefunction coefficient $\psi_4^\phi$. Hereafter we use $\psi_4^s$ to refer to the $ s  $-channel contribution to $\psi_4^\phi$. Let us also denote the ${\cal O}( g_A g_B)$ contribution to $\psi_4^s$ with $\psi_4^{AB}$. 
Using  \eqref{props}, we find,
\begin{align}
\nonumber
& \blangle \bfk_1,\phi;\dots ;\bfk_4,\phi\bl \delta \cU_{g_A g_B} \bl 0\brangle'= -\left(\prod_{a=1}^4 \phi^+(k_a,\eta_0)\right)\\ \nonumber
&\times \,\left[\psi_4^{AB}(k_a,s,\hat{k}_a.\hat{k}_b)-\left(\psi_A'^{\phi\phi\sigma}(k_1,k_2,s)\psi_B'^{\phi\phi\sigma}(k_3,k_4,s)+(A\leftrightarrow B)\right)P_\sigma(s)\right]\\ 
& +(t,u\,\,\,\,\text{channels})\,,\\ \nonumber
& \blangle \bfk_1,\phi;\dots ;\bfk_4,\phi\bl \delta \cU^\dagger_{g_A g_B} \bl 0\brangle'=-\left(\prod_{a=1}^4 \phi^+(k_a,\eta_0)\right)\\ \nonumber
&\times \,\left[\left(\psi_4^{AB}(-k_a,s,\hat{k}_a \cdot \hat{k}_b)\right)^*-\left(\psi_A'^{\phi\phi\sigma}(k_1,k_2,-s)\psi_B'^{\phi\phi\sigma}(k_3,k_4,-s)+(A\leftrightarrow B)\right)P_\sigma(s)\right]\\ \nonumber
& +(t,u\,\,\,\,\text{channels})\,.
\end{align}
Notice also that for scalars, e.g. in the $ s  $-channel, the only possible inner products $\bfk_a\cdot\bfk_b$ appearing in $\psi_4^s$ are $\bfk_1\cdot\bfk_2$ and $\bfk_3\cdot \bfk_4$. These two can be written in terms of $s$ and $k_a$'s, therefore we can drop $\bfk_a\cdot\bfk_b$ in the argument of $\psi_4^s$.  The final Cosmological Optical Theorem takes the following form, 
\begin{align}
\label{exchCOPT0}
& \psi'^{AB}_4(k_1,k_2,k_3,k_4,s)+[\psi'^{AB}_4(-k_1,-k_2,-k_3,-k_4,s)]^*= P_\sigma(s)\\ \nonumber
& \times \left[\psi'^{\phi\phi\sigma}_A(k_1,k_2,s)- \psi'^{\phi\phi\sigma}_A(k_1,k_2,-s) \right]\,\left[\psi'^{\phi\phi\sigma}_B(k_3,k_4,s)- \psi'^{\phi\phi\sigma}_B(k_3,k_4,-s) \right]+(A\leftrightarrow B)\,.
\end{align}
A similar equation holds even if the two vertices $A$ and $B$ were identical, but then one should drop the last term, i.e. $A\leftrightarrow B$. Therefore, by adding the ${\cal O}(g_A^2)$ and ${\cal O}(g_B^2)$ contributions to $\psi_4^s$, and after defining the total three-point function as $\psi_3^{\phi\phi\sigma}=\psi'^{\phi\phi\sigma}_A+\psi'^{\phi\phi\sigma}_B$, we arrive at
\begin{tcolorbox}[colback=brown!10!white, colframe=white] 
\begin{align}
\label{exchCOPT}
& \psi'^s_4(k_1,k_2,k_3,k_4,s)+[\psi'^s_4(-k_1,-k_2,-k_3,-k_4,s)]^*= \\ \nonumber
& P_\sigma(s)\, \left[\psi'^{\phi\phi\sigma}_3(k_1,k_2,s)- \psi'^{\phi\phi\sigma}_3(k_1,k_2,-s) \right]\,\left[\psi'^{\phi\phi\sigma}_3(k_3,k_4,s)- \psi'^{\phi\phi\sigma}_3(k_3,k_4,-s) \right]\,.
\end{align}
\end{tcolorbox}
This case of identical vertices is shown diagrammatically from both a bulk, figure 2, and boundary, figure 3, perspective. A few remarks about this result:
\begin{itemize}
\item An analogous relation holds for the other two channels, $ t $ and $ u$, therefore, by summing over all channels we can write, 
\begin{align}\label{sum}
& \psi'_4(k_1,k_2,k_3,k_4,s,t,u)+[\psi'_4(-k_1,-k_2,-k_3,-k_4,s,t,u)]^*= \\ \nonumber
& P_\sigma(s)\left[\psi'^{\phi\phi\sigma}_3(k_1,k_2,s)- \psi'^{\phi\phi\sigma}_3(k_1,k_2,-s) \right]\,\left[\psi'^{\phi\phi\sigma}_3(k_3,k_4,s)- \psi'^{\phi\phi\sigma}_3(k_3,k_4,-s) \right]\,\\ \nonumber
& +P_\sigma(t)\, \left[\psi'^{\phi\phi\sigma}_3(k_1,k_3,t)- \psi'^{\phi\phi\sigma}_3(k_1,k_3,-t) \right]\,\left[\psi'^{\phi\phi\sigma}_3(k_2,k_4,t)- \psi'^{\phi\phi\sigma}_3(k_2,k_4,-t)\right]\,\\ \nonumber
& +P_\sigma(u)\, \left[\psi'^{\phi\phi\sigma}_3(k_1,k_4,u)- \psi'^{\phi\phi\sigma}_3(k_1,k_4,-u) \right]\,\left[\psi'^{\phi\phi\sigma}_3(k_2,k_3,u)- \psi'^{\phi\phi\sigma}_3(k_2,k_3,-u)\right]\,.
\end{align}

\item Although we have worked under the assumption that the intermediate field in the 4 point exchange diagram is distinct from the external fields, and that it carries no derivatives, one can indeed show that  \eqref{exchCOPT} remains the same,  regardless (see Appendix \ref{gencopt} for a discussion).
\item So far we have assumed that the intermediate field $\sigma$ is a scalar. However, for an intermediate spinning field (in particular a spin-2 field),  \eqref{exchCOPT} easily generalizes, as we illustrate in Subsection \ref{GExh} . We leave a generalization of our optical theorem to spinning external fields for future work.
\end{itemize}


\subsection{The Cosmological Optical Theorem for (analytic) correlators}

In  \eqref{exchCOPT}, $\psi_4$ is analytic (up to isolated poles), which in turn facilitates the derivation of a similar equation for correlators. We first need the explicit relation between $B_4^s$ (the $s $-channel contribution to the trispectrum of $  \phi $) and $\psi'^s_4$. To do this we invert the results from  \ref{eq:B4},
\begin{align}
\Re \psi'_3(k_1,k_2,k_3)&=-\frac{1}{2}\,B_3(k_1,k_2,k_3)\,\prod\limits_{a=1}^3 \frac{1}{P(k_a)}\\
\Re\psi'_4(k_1,k_2,k_3,k_4)&=\frac{1}{2}\left(\frac{B_3(k_1,k_2,s)B_3(k_3,k_4,s)}{P_\sigma(s)}+t+u-B_4(k_1,k_2,k_3,k_4)\right)\prod\limits_{a=1}^4 \frac{1}{P(k_a)}\,.\nonumber
\end{align}
For IR finite interactions, the COT for contact terms implies $\Im\left(\psi_3'\right)=0$. Moreover, it follows from dilatation symmetry that $B_3(k_a)=B_3(-k_a)$ and $P_\sigma(k)=-P_\sigma(-k)$. Putting everything together we find, 
\begin{tcolorbox}[colback=brown!10!white, colframe=white] 
\begin{multline}
\label{cotcor}
B_4^s(k_1,k_2,k_3,k_4,s)+B_4^s(-k_1,-k_2,-k_3,-k_4,s)=\\\frac{1}{2 P_\sigma(s)} \left[B_3(k_1,k_2,s)-B_3(k_1,k_2,-s)\right]\left[B_3(k_3,k_4,s)-B_3(k_3,k_4,-s)\right]\,.
\end{multline}
\end{tcolorbox}
Notice that, even when summed over all three channels, this relation is not directly observable because it involves the analytic continuation of the bispectrum to negative norms of the momenta.

 
\subsubsection{Four-point Exchange Diagrams: Graviton exchange}\label{sec:}
If we introduce into our field theory an additional massless spin-two particle, $\gamma_{ij}$, which we will call the graviton, then we expect the resulting wavefunction coefficients to be constrained in a similar manner to the scalar field case. We can express this spinning field in terms of a sum over polarisation tensors, 
\begin{equation}
\gamma_{ij}(\textbf{k},\eta)=\sum_{\lambda=+,\times}\epsilon_{ij}^\lambda (\textbf{k})\gamma^\lambda (k,\eta),
\end{equation}
which satisfy
\begin{equation}
\epsilon_{ij}^\lambda {\epsilon^{ij}}^{\lambda'}=2\delta^{\lambda\lambda'},\quad \epsilon^\lambda_{ii}=0,\quad \epsilon^\lambda_{ij}k^j=0.
\end{equation}
In terms of these two fields the quadratic action is given by
\begin{equation}
S_2=\int \frac{d^3\bfk}{(2\pi)^3} d\eta \frac{a^2}{2}\left({\phi'_{k}}^2-k^2\phi_{k}^2+\frac{1}{2}{{\gamma^+_{k}}'}^2-\frac{1}{2}k^2{\gamma_{k}^+}^2+\frac{1}{2}{{\gamma_{k}^{\times}}'}^2-\frac{1}{2}k^2{\gamma_{k}^{\times} }^2\right).
\end{equation}
This difference in the normalisation of the gravition fields slightly adjusts the propagators for the theory. Enforcing the canonical commutation relationship between $\gamma^\lambda$ and its conjugate momentum gives the two fundamental solutions,
\begin{equation}
{\gamma^\lambda}^\pm(k,\eta)=\frac{H}{\sqrt{k^3}}(1\mp ik\eta)\exp(\pm i k\eta)\,.
\end{equation}
Then the graviton propagators can be related to the scalar propagators by
\begin{align}
K^\lambda(k,\eta)&=\frac{{\gamma^\lambda}^+(k,\eta)}{{\gamma^\lambda}^+(k,\eta_0)}=K(k,\eta)\,,\\
G^\lambda(k,\eta,\eta)&=2G(k,\eta,\eta')\,.
\end{align}
From which we can find the classical solutions in terms of the value of the field at some boundary, $\bar{\gamma}^s$. In terms of these boundary fields we find multiple wavefunction coefficients defined by
\begin{multline}
iS=-\int \frac{d^3\bfk}{(2\pi)^3}\left(\frac{1}{2}\psi_2^{\phi\phi}(k)\bar{\phi}_{k}\bar{\phi}_{k}+\frac{1}{2}\psi_2^{++}(k)\bar{\gamma}^+_{k}\bar{\gamma}^+_k+\frac{1}{2}\psi_2^{\times\times}(k)\bar{\gamma}^{\times}_{k}\bar{\gamma}^{\times}_{k}\right)-\\ \int\frac{d^3\bfk_1d^3\bfk_2d^3\bfk_3}{(2\pi)^9}\left(\frac{1}{2}\psi_3^{\phi\phi+}(k_1,k_2,k_3)\bar{\gamma}_{k_1}^+\bar{\phi}_{k_2}\bar{\phi}_{k_3}+\frac{1}{2}\psi_3^{\phi\phi\times}(k_1,k_2,k_3)\bar{\gamma}_{k_1}^{\times}\bar{\phi}_{k_2}\bar{\phi}_{k_3}\right)-\\  \int\frac{d^3\bfk_1d^3\bfk_2d^3\bfk_3d^3\bfk_4}{(2\pi)^{12}}\left( \frac{1}{4!}\psi_4^{\phi\phi\phi\phi}(k_1,k_2,k_3,k_4) \bar{\phi}_{k_1}\bar{\phi}_{k_2}\bar{\phi}_{k_3}\bar{\phi}_{k_4}+\frac{1}{4}\psi_4^{\phi\phi++}(k_1,k_2,k_3,k_4) \bar{\phi}_{k_3}\bar{\phi}_{k_2}\bar{\gamma}^+_{k_1}\bar{\gamma}^+_{k_4}\right.\\\left.+\frac{1}{4}\psi_4^{\phi\phi\times\times}(k_1,k_2,k_3,k_4) \bar{\phi}_{k_3}\bar{\phi}_{k_2}\bar{\gamma}^{\times}_{k_1}\bar{\gamma}^{\times}_{k_4}+\frac{1}{2}\psi_4^{\phi\phi+\times}(k_1,k_2,k_3,k_4) \bar{\phi}_{k_3}\bar{\phi}_{k_2}\bar{\gamma}^+_{k_1}\bar{\gamma}^{\times}_{k_4}\right)
\end{multline}
Since the time dependence of the propagators is unchanged, the difference between the scalar and graviton case can be absorbed into the function $F$. As all the arguments about the analytic structure of the integrals were agnostic to the form of $F$ they will apply to interactions involving gravitons. So we expect the optical theorem to generalise to
\begin{align}\label{eq:internalg}
&\psi_4^{\phi\phi\phi\phi}(k_1,k_2,k_3,k_4,s)+\psi_4^{\phi\phi\phi\phi}(-k_1,-k_2,-k_3,-k_4,s)=\sum_{\lambda=+,\times}\frac{1}{2\Re\left\lbrace \psi_2^{\lambda\lambda}(s)\right\rbrace} \nonumber\\ &\times \left(\psi_3^{\phi\phi \lambda}(k_1,k_2,s)-\psi_3^{\phi\phi \lambda}(k_1,k_2,-s)\right)\left(\psi_3^{\phi\phi \lambda}(k_3,k_4,s)-\psi_3^{\phi\phi \lambda}(k_3,k_4,-s)\right)+t+u\,,\\\label{eq:externalg}
&\psi_4^{\phi\phi \lambda \lambda'} (k_1,k_2,k_3,k_4,s)+\psi_4^{\phi\phi \lambda \lambda'}(-k_1,-k_2,-k_3,-k_4,s)=\frac{1}{2\Re\left\lbrace \psi_2^{\phi\phi}(s)\right\rbrace} \nonumber\\ &\times\left(\psi_3^{\phi\phi \lambda}(k_1,k_2,s)-\psi_3^{\phi\phi \lambda}(k_1,k_2,-s)\right)\left(\psi_3^{\phi\phi \lambda'}(k_3,k_4,s)-\psi_3^{\phi\phi \lambda'}(k_3,k_4,-s)\right)+u\,.
\end{align}
The additional factor of 2 in the bulk-to-bulk propagator may seem to be cause for concern, however it is canceled by an additional factor of 2 from $\psi_2$. We confirm these expressions in Section \ref{GExh}, where we discuss the trispectrum from graviton exchange in slow-roll inflation.


\begin{figure}
\begin{center}
\includegraphics[scale=0.71]{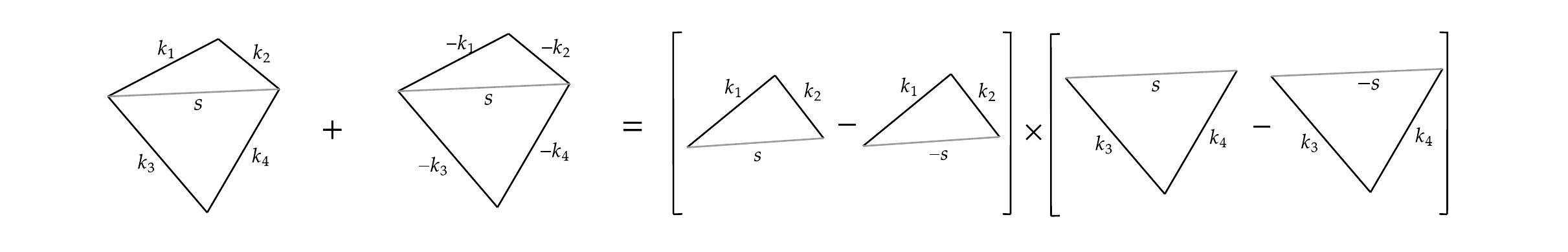}
\end{center}
\caption{A boundary depiction of the Cosmological Optical Theorem. The minus signs next to the $k_a$'s or $s$ indicate negative energies, which are accessible upon analytical continuation.}
\end{figure}

\section{Non-trivial checks}
\label{NonTC}

In this section we perform a series of non-trivial checks of the Cosmological Optical Theorem derived in the previous section. First we discuss separately contact interactions that are either IR finite or IR divergent. Then we move on to four-point diagrams featuring a particle exchange. In particular, we discuss the primordial trispectrum arising from the inflaton self-coupling in the Effective Field Theory of inflation \cite{Cheung:2007st} and from graviton exchange \cite{seery2009inflationary}.


\subsection{IR-finite contact diagrams}\label{checkcontact}

We start by checking one of the consequences of the Cosmological Optical Theorem discussed around \eqref{fnstar}, namely that for contact diagrams the wavefunction coefficients for an even (odd) number of conformally-coupled scalars must be real (imaginary). Consider a generic IR-finite self-interaction for a conformally-coupled field, schematically written as
\be
\label{contconf}
S_{\text{int}}=-g_n\,\int a^{4-m}d\eta\,d^3\bfx\, \, \partial^m\, \vpi^n\,.
\ee
In Fourier space, 
\be
S_{\text{int}}= -g_n\, \int d\eta\,a^{4-m}\,\left (\prod_{a=1}^n \dfrac{d^3\bfk_a}{(2\pi)^3}\right)\,\left(\prod \bfk_a\cdot \bfk_b\right)\,\left(\prod_{a=1}^l \vpi (\bfk_a,\eta)\,\right)\, \left(\prod_{a=l+1}^{n}\, \partial^{s_a}_\eta\, \vpi(\bfk_a, \eta)\right)\,,
\ee
where $  l $ is the number of fields without time derivatives. Above, $\lbrace s_a\neq 0 \rbrace$ are a set of integers that refer to the number of time derivatives acting on each field, the product $\prod \bfk_a\cdot\bfk_b\,$ accounts for the contracted spatial indices in \eqref{contconf}, and there are $m-\sum_{a}s_a$ spatial derivatives in total. For the integral to converge, one must have 
$m+l-4>0$. 
From the above action the wavefunction coefficients, $\psi_n^\vpi$, are
\begin{align}
\psi'^{\vpi}_n &=+i\,g_n\,\int\,d\eta\,a^{4-m}\,\left(\prod \bfk_a\cdot \bfk_b\,\right)\,\left(\prod_{a=1}^l\, K_\vpi(k_a,\eta)\right)\, \left(\prod_{a=l+1}^n\, \partial^{s_a}_\eta\, K_\vpi(k_a, \eta)\right) \,\\ \nonumber 
&+(n!-1)\,\,\text{permutations}\,\\ \nonumber
&=\dfrac{g_n\, (m+l-4)!}{\eta_0^n\,(-H)^{4-m}}\,\left(\prod \bfk_a\cdot \bfk_b\,\right)\left(\prod_{a=l+1}^n\, (-k_a)^{s_a-1}\right)\,(-1)^{n-l+1}\,i^{(m+n+\sum\limits_{a=l+1}s_a-2)}\\ \nonumber
& \left(\prod_{a=l+1}^n\,(s_a+k_a \frac{\partial}{\partial k_a})\right)\,k_T^{3-m-l}+(n!-1)\,\, \text{permutations}\,.
\end{align}
All the terms appearing above are real except for the factors of $i$. Since the number of spatial derivatives is even, $m+\sum s_a$ is even as well. Therefore,  in agreement with  \eqref{vpicont}, we find that $\psi'^\vpi_n$ is real when $ n  $ is even and it is purely imaginary if $ n  $ is odd.\\

%
%

A more interesting IR-finite example to put to the test is one that involves fields that are neither massless nor conformally coupled, $m^2\neq 0,2H^2$. For example, consider the three point function of two conformally-coupled fields and a generic light scalar (denoted by $\sigma$) generated via a simple cubic interaction $\vpi^2 \sigma$. The $\sigma$ field should be no lighter than $\sqrt{2}H$ for the time integral to converge in the IR. In this example, the wavefunction coefficient\footnote{Notice that due to spatial translation invariance, $\bfk_a\cdot\bfk_b$ can be written in terms of $k_1,k_2$ and $k_3$.} is
\be
\psi^{'\vpi\vpi\sigma}_3=k_T^{-2+\Delta}\, f_3 \left( \dfrac{k_a}{k_T} \right)\,,
\ee
where $f_3$ is fixed by de Sitter boosts, which in this example are taken to be a linearly-realized symmetry, in particular all speeds of sound are set to unity) \cite{Arkani-Hamed:2015bza}\footnote{In \cite{Arkani-Hamed:2015bza} the result is given up to a momentum independent factor. For our purpose, namely inspecting the contact term optical theorem, this prefactor is important, and we compute it in Appendix \ref{f3c}.}. The final answer is given by (see Appendix \ref{f3c} for the derivation),  
\begin{align}
\label{f3}
f_3 &= -\dfrac{\sqrt{\pi}\,2^{3-\Delta}}{H^4}\dfrac{\Gamma[2-\Delta]\Gamma[-1+\Delta]}{\Gamma[-\frac{3}{2}+\Delta]}\exp\left[  \dfrac{i \pi}{2}\left( 1-\Delta \right) \right] \\ \nonumber 
&\times (-\eta_0)^{-5+\Delta}\,\, \left(\frac{k_3}{k_T}\right)^{-2+\Delta}\,_2F_1 \left [\frac{1}{2}-\nu,\dfrac{1}{2}+\nu\,,1\,, \dfrac{1-\frac{k_1+k_2}{k_3}}{2}\right ]\,,
\end{align}
where $_2F_1[a,b,c,z]$ is the hypergeometric function. From this equation, one can directly see that  \eqref{fnstar} holds. In more detail, when $k_1,k_2,k_3$ and $\Delta$ are real, every term on the right-hand side is real except for the factor 
\begin{align}
\exp \left[  \dfrac{i \pi}{2}\left( \nu+\dfrac{1}{2} \right) \right]=\exp \left[  \dfrac{i \pi}{2}\left( 1-\Delta \right) \right]\,.
\end{align} 
This factor determines the complex phase of $f_3$ and its value is in agreement with  \eqref{fnstar}.


\subsection{IR-divergent contact diagrams}

Even at tree level, IR divergences are ubiquitous in de Sitter space, and they generically break dilatation symmetry of the correlators on the boundary (i.e. $\eta_0\to 0$) \cite{Weinberg:2005vy}. That is part of the reason for the theoretical efforts made to provide general proofs for the immunity of curvature perturbations to such IR infinities (see \cite{Senatore:2012ya, Senatore:2009cf, Senatore:2012nq, Pimentel:2012tw, Assassi:2012et, Cohen:2020php}). Here we show that the implication of the Cosmological Optical Theorem for these IR-divergent diagrams is more interesting than in the IR-finite cases. The reason lies in non-scale invariant singularities (e.g. branch cuts) induced by these interactions, which allow the resulting wavefunction coefficients to deviate from the behavior dictated by  \eqref{fnstar}.

As one illustration, consider the cubic self-interaction $g\,\phi^3$ (we also set $c_s=1$ by rescaling our spatial coordinate). The coefficient of the cubic term in the wavefunction (in the $\eta_0\to 0$ limit) is given by, 
\begin{align}
\label{phicube}
\psi'^{\phi}_3 &= \dfrac{g}{H^4} \Big[(-2+2 \gamma+i \pi ) e_1^3+(4-6 \gamma -3 i \pi ) e_2 e_1 \\ \nonumber
& +(2+6 \gamma +3 i \pi ) e_3+2 \left(e_1^3-3 e_2 e_1+3 e_3\right) \log \left(-e_1 \eta _0\right)+\dfrac{2i}{\eta_0}(e_1^2-2e_2)+\dfrac{2i}{ \eta_0^3}\Big]\,,
\end{align}
where $e_{1,2,3}$ are the elementary symmetric polynomials, 
\begin{align}
e_1 &=k_1+k_2+k_3\,, & e_2 &=k_1 k_2+k_1 k_3+k_2 k_3\,, & e_3 &=k_1 k_2 k_3\,.
\end{align}
The last two terms on the second line of \eqref{phicube} satisfy \eqref{unicont} by themselves. More interestingly, the optical theorem ties the logarithmic part $\log (-e_1 \eta_0)$, which has a branch cut at $e_1\in \mathbb{R}^-$, to the IR-finite imaginary part of $\psi_3$ via  \eqref{logCOT}. In more detail, 
\be
[ \psi'^{\phi}_3(-e_1-i\epsilon, e_2+i\epsilon, -e_3-i\epsilon)]^*\supset -2 \left(e_1^3-3 e_2 e_1+3 e_3\right) (\log \left(-e_1 \eta _0\right)+i \pi)\,.
\ee
The imaginary part of the right-hand side above is precisely twice the IR-finite imaginary part of $\psi_3$, in full agreement with  \eqref{unicont}. Notice that this delicate relationship between the branch cut and the finite imaginary parts of $\psi_3$ cannot be observed at the level of correlators, as $B_3$ only incorporates the real part of $\psi_3$.

\subsection{Exchange diagrams}

\subsubsection*{Flat Space}
The derivation of \eqref{unicont} and \eqref{exchCOPT} can be adapted to flat spacetime (i.e. $a(\eta)=1$), and the final results remain the same, primarily because \eqref{BtBp} applies to flat space as well\footnote{This might seem to suggest that the COT is background independent, but that is not true. On a generic FLRW background, we are not aware of a simple relation between $K^*(k,\eta)$ and $K(k,\eta)$.}. This motivates the study of a flat space example prior to the cosmological ones. 

Let's consider a massless scalar in flat space with the cubic coupling $g \phi^3$, for which we have
\begin{align}
&\psi'^{\phi}_3(k_1,k_2,k_3) =\dfrac{6 g}{k_T^{(3)}}\,, \\ \nonumber
&\psi'^s_4 =\dfrac{-36 g^2}{k_{T}^{(4)}(k_1+k_2+s)(k_3+k_4+s)}\,,\\ \nonumber
& P(k)=\dfrac{1}{2k}\,,
\end{align}
where the total energy $  k_{T}^{(n)} $ was defined in \eqref{kT}.
One interesting feature of the COT is that each term on the left-hand side of \eqref{exchCOPT} displays a total energy singularity ($k_T^{(4)}\to 0$), as was demonstrated in \cite{baumann2020cosmological}, however, the right-hand side only incorporates the singularities contained in $\psi_3$ (and its analytical continuation). Those singularities are located at $k_1+k_2\pm s=0$ and $k_3+k_4\pm s=0$. Therefore, the total energy singularities of the left-hand side of the COT should cancel out, i.e. 
\be
\psi'^s_4(k_1,\dots\,,k_4,s)+\psi'^s_4(-k_1,\dots\,,-k_4,s)\not\supset k_T^{(4)}\,\text{pole}\,.
\ee
From the example above, one can explicitly see that this is exactly the case and that  \eqref{exchCOPT} holds.


\subsubsection*{Trispectrum in the EFT of Inflation}

\begin{figure}
\begin{center}
\includegraphics[scale=0.8]{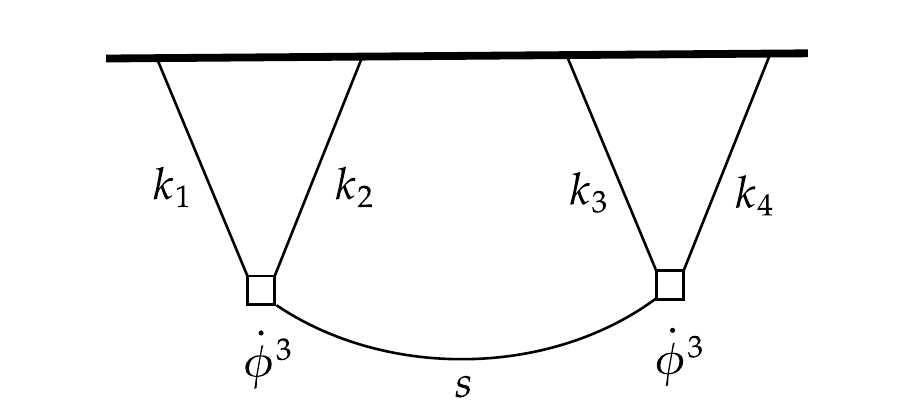}
\end{center}
\caption{The exchange diagram contribution to the trispectrum generated by inserting two copies of the cubic vertex $g\dot{\phi}^3$\label{ph3ph3}}
\end{figure}

The leading order interactions in the Effective Field Theory of Inflation are given by the cubic operators $\dot{\phi}^3$ and $\dot{\phi}(\partial_i \phi)^2$ \cite{Cheung:2007st}\footnote{Instead of the Goldstone boson of time translation $\pi$, we work with its canonically normalized version $\phi$.}. These cubic operators contribute to the four-point function via the exchange of the scalar field \cite{Chen:2009bc}. Let us consider the Feynman diagram depicted in Figure \ref{ph3ph3}. The trispectrum  associated with this diagram is given by\cite{Chen:2009bc, Arroja:2009pd}\footnote{Notice that from an in-in standpoint, the first term comes from the contact term $\dot{\phi}^4$, which is in turn induced by the non-linear relation between the conjugate moment and $\dot{\phi}$. In contrast, in a wavefunctional computation this term is already contained in the exchange diagram depicted in Figure \ref{ph3ph3}, without any reference to any contact term like $\dot{\phi}^4$. This subtlety is of no relevance to the COT, in that any contact term contribution to $B_4$ is enforced to cancel from the left hand side of  \eqref{cotcor} by dilation symmetry. See Appendix \ref{gencopt} for more discussion.},
\begin{align}
& B^s_4(k_a,s)=-\dfrac{108\, g^2\, H^8}{c_s^9}\,\dfrac{1}{(k_1 k_2 k_3 k_4)k_T^{5}}\\ \nonumber
&+\dfrac{9\, g^2 H^8}{c_s^9}\,\dfrac{s}{(k_1 k_2 k_3 k_4)}\, \left[ \dfrac{6}{k_T^5}\left(\dfrac{1}{E_R}+\dfrac{1}{E_L}\right)+\dfrac{3}{k_T^4}\left(\dfrac{1}{E_R^2}+\dfrac{1}{E_L^2}\right)+\dfrac{1}{k_T^3}\left(\dfrac{1}{E_R^3}+\dfrac{1}{E_L^3}\right)+\dfrac{1}{E_L^3\,E_R^3}\right]\,,
\end{align}
where 
\be
\nonumber
E_L\equiv k_1+k_2+s\,,\qquad E_R\equiv k_3+k_4+s\,.
\ee
The same cubic vertex $g\,\dot{\phi}^3$ induces the bispectrum below, 
\be
B_3(k_1,k_2,k_3)=\dfrac{3 g H^5}{c_s^6}\,\dfrac{1}{k_1 k_2 k_3\, (k_1+k_2+k_3)^3}\,.
\ee
By direct inspection, one can check that  \eqref{cotcor} is satisfied. Similar to the flat space example, here we see that all the three kinds of $k_T$ poles present in $B_4$ disappear from the left hand-side of the COT.

\subsubsection*{Graviton Exchange}
\label{GExh}

Consider now a theory with a single massless scalar and a massless graviton with the interaction term identified in \cite{maldacena2003non} as the only term relevant to the scalar trispectrum,
\begin{equation}
S=\int d^3x d\eta \frac{a^2}{2}\left(\phi'^2-\partial_i\phi\partial^i\phi+\frac{1}{4}{\gamma_{ij}}'{\gamma^{ij}}'-\frac{1}{4}\partial_k\gamma_{ij}\partial^k\gamma^{ij}+g\gamma^{ij}\partial_i\phi\partial_j\phi\right).
\end{equation}
Expressing this in terms of $\gamma^\lambda$ we find
\begin{multline}
S=\int \frac{d^3\bfk}{(2\pi)^3} d\eta \frac{a^2}{2}\left({\phi'_{k}}^2-k^2\phi_{k}^2+\frac{1}{2}{{\gamma^+_{k}}'}^2-\frac{1}{2}k^2{\gamma_{k}^+}^2+\frac{1}{2}{{\gamma_{k}^{\times}}'}^2-\frac{1}{2}k^2{\gamma_{k}^{\times} }^2\right)\\+\frac{g}{2}\sum_{\lambda=+,\times} \int\frac{d^3\bfk_1d^3\bfk_2d^3\bfk_3}{(2\pi)^9}d\eta a^2  \gamma^{\lambda}_{k_1}{\epsilon^\lambda_{ij}}(\bfk_1)k_2^ik_3^j\phi_{k_2}\phi_{k_3}\delta^3(\bfk_1+\bfk_2+\bfk_3)\,.
\end{multline}
Expanding in terms of our propagators and performing the time integrals gives the wavefunction coefficients:
\begin{align}
\psi_2^{\phi\phi}(k)&=-ia^2(\eta_0)K'(k,\eta_0)\,,\\
\psi_2^{\lambda\lambda}(k)&=-\frac{i}{2} a^2(\eta_0)K'(k,\eta_0)\,,\\
\psi_3^{\phi\phi \lambda}(k_1,k_2,k_3)&= g\,\epsilon^\lambda_{ij}(\bfk_1)k^i_2k^j_3 \left(\frac{i}{\eta_0} - \frac{e_1^3-e_1 e_2-e_3}{e_1^2}\right)\,, \\
\psi_4^{\phi\phi\phi\phi}(k_1,k_2,k_3,k_4)&=-2ig^2\sum_{\lambda=+,\times} \epsilon^\lambda_{ij}(\textbf{s})\epsilon^\lambda_{lm}(\textbf{s}) k_1^i k_2^j k_3^lk_4^m \,\mathcal{I}(s) +t+u\,,\\
\psi_4^{\phi\phi \lambda \lambda'}(k_1,k_2,k_3,k_4)&=-i g^2 \epsilon^\lambda_{ij}(\bfk_1) k^i_2 k_2^j \epsilon_{lm}^{\lambda'}(\bfk_4) k_3^lk_3^m\, \mathcal{I}(s) + u\,,
\end{align}
where the integral $\mathcal{I}(s)$ is defined in terms of the scalar propagators by
\begin{multline}
\mathcal{I}(s)=\int d\eta d\eta' a^2(\eta)a^2(\eta')K(k_1,\eta)K(k_2,\eta)K(k_3,\eta')K(k_4,\eta')G(s,\eta,\eta')= i\Bigg(\frac{1}{k_T}-\frac{s^2}{k_T E_LE_R} \\
+\frac{sk_1k_2}{k_TE_L^2E_R}+\frac{sk_3k_4}{k_TE_R^2E_L}+\frac{2sk_1k_2k_3k_4}{k_T^2E_L^2E_R^2}+\frac{k_1k_2}{k_T^2E_L}+\frac{k_3k_4}{k_T^2E_R}-\frac{s(k_1k_2+k_3k_4)}{k_T^2E_LE_R}+\frac{2k_1k_2k_3k_4}{k_T^3E_LE_R}\Bigg)\,.
\end{multline}
These wavefunction coefficients confirm the results in Eqs. \ref{eq:internalg} and \ref{eq:externalg} for this interaction. Furthermore, we can compare this to \cite{seery2009inflationary} to see that
\begin{multline}
\left\langle\phi^4\right\rangle = -\frac{2}{\prod\limits_a^4 2\Re{\psi_2^{\phi\phi}(k_a)}} \Bigg( \Re\left\lbrace \psi_4^{\phi\phi\phi\phi}(k_1,k_2,k_3,k_4)\right\rbrace \\
-\sum_{\lambda=+,\times} \frac{\Re\left\lbrace \psi_3^{\phi\phi \lambda}(k_1,k_2,s)\right\rbrace\Re\left\lbrace \psi_3^{\phi\phi \lambda}(k_3,k_4,s)\right\rbrace}{\Re\left\lbrace \psi_2^{\lambda\lambda}(s)\right\rbrace}+t+u\Bigg)\,,
\end{multline} 
which is the expected generalisation of  \ref{eq:B4} to interactions involving gravitons.

\section{Polology of cosmological correlators}\label{sec:polo}

In this section we provide a detailed and explicit derivation of the connection between the total-energy pole of cosmological correlators and flat-space amplitudes using perturbation theory and the (canonical) in-in formalism. Our final result for any tree-level diagram is \eqref{finalBtoA}, which simplifies to \eqref{eq:kTlim} for contact diagrams. The exceptional case of a logarithmic divergence is discussed around \eqref{eq:B3int}.


\subsection{Preliminaries: Amplitudes in flat spacetime}

In \cite{Maldacena:2011nz, Raju:2012zr}, it has been noted that the flat space correlator arising from contact interactions contains a simple pole at the total energy, the residue of which is equal to the real part of the amplitude of the same diagram (see \cite{Arkani-Hamed:2017fdk,Arkani-Hamed:2018kmz,Benincasa:2018ssx,baumann2020cosmological} for further discussions). In this section we will define a Feynmann esque methodology to calculate flat space amplitudes for a massless scalar field, defined to highlight similarities with the correlator, discussed in the following section. We define our theory with the second order action in  \ref{quad} except we have only a single species so we are free to define the speed of sound to be one, the scale factor is just a constant that we take to be $1$ and the mass is zero. The interaction terms similarly take the form of  \ref{contact}, 
\begin{equation}
S_n = g_n \int dt \int \left(\prod_{a=1}^n \dfrac{d^3\bfk_a}{(2\pi)^3}\right) (2\pi)^3 \delta^3\left(\sum \limits_{a=1}^n\bfk_a\right)F_n\left[\bfk_a\cdot \bfk_b\right] \left[ \prod_{a}^{n}\partial_t^{s_a}\phi(\bfk_a,t) \right].
\end{equation}
For this interaction the $n$-particle amplitude is given by
\begin{equation}
A_n=-i\lim_{t_0,T\rightarrow \infty}\left\langle 0 \right\rvert \prod_i^N \sqrt{2k_i}a_{\textbf{k}}\mathcal{U}(t_0,-T)\left\lvert 0\right\rangle,
\end{equation}
where $\mathcal{U}$ is the interaction picture time evolution operator, defined in  \ref{eq:timeevo}. Under the assumption that the interaction is weak we could, for example, expand this operator in powers of the interaction Hamiltonian and perform the resulting integrals using the Feynman rules. The free field operators are found to be
\begin{equation}
\hat{\phi}(t,\textbf{k})=\phi(t,k)a_{\textbf{k}}+\phi^*(t,k)a_{-\textbf{k}}^\dagger \quad \textrm{with}\quad \phi(t,k)=\frac{1}{\sqrt{2k}}e^{ik t}.
\end{equation}
It is therefore possible to express time derivatives as
\begin{equation}\label{eq:flatderivative}
\partial_t^{s}\phi(k,t)=f_s(k)\phi(k,t),
\end{equation}
this momentum dependence is typically absorbed into the function $F_n$ and included in the Feynman rules as a vertex contribution, however in de Sitter this will not be possible, so here we will instead include this term in our definition of the propagators which are therefore given by
\begin{align}
G_{\text{flat}}(k,t,t')&=\int \frac{d^3\bfq}{(2\pi)^3}\langle 0\rvert \mathcal{T} \left[ \partial_{t'}^{s'}\hat{\phi}(t',\textbf{k})\partial_t^s\hat{\phi}(t,\textbf{q}) \right]\lvert 0\rangle\\
&=\theta(t'-t)\frac{f_{s}^*(k)f_{s'}(k)}{2k}e^{-ik(t-t')}+\theta(t-t')\frac{f_s(k)f_{s'}^*(k)}{2k}e^{ik(t-t')}.
\end{align}
Furthermore, as we will see in the next section, in de Sitter we need to define bulk-to-boundary propagators to account for the way in which our external fields enter the calcultation. These are not present in typical Feynman rules so we will choose to represent the external lines in our amplitude diagrams as bulk-to-boundary propagators. They therefore come from the contraction of external $a_{\textbf{k}}$ with the $a_{\textbf{q}}^\dagger$ from the vertex operators. We need not worry about any annihilation operators from the time evolution operator at the boundary as those that do not appear as contractions will annihilate the vacuum
\begin{equation}
G_{B,\text{flat}}(k,t)=\int \frac{d^3 \bfq}{(2\pi)^3} \langle 0 \rvert \sqrt{2 k} a_{\textbf{k}}\partial^s_t\phi^*(t,-\textbf{q})a^\dagger_{-\textbf{q}}\lvert 0\rangle=f^*_s(k) e^{-ikt}.
\end{equation}
Using this notation we can construct amplitudes from Feynman diagrams using the following rules:
\begin{itemize}
\item Add a $G_{\text{flat}}$ propagator for each internal line in the diagram and a $G_{B,\text{flat}}$ propagator for each external line,
\item Conserve momentum at each vertex and overall.
\item Integrate over any undetermined momenta.
\item Integrate over time for each vertex from $-\infty$ to $\infty$.
\item Introduce a factor of $-igF(\textbf{k})$ for each vertex.
\end{itemize}
These rules can be expressed as
\begin{equation}
A_n=(2\pi)^3\delta^3\left(\sum_a^n\textbf{k}_a\right) i^{-V-1}\int\prod_\alpha^V g_\alpha F_\alpha(\textbf{k}) dt_\alpha \prod_\beta^L\frac{d^3 \bfk_\beta}{(2\pi)^3}\prod_\gamma^I G_{\text{flat}}(K_\gamma;t,t') \prod_a^nG_{B,\text{flat}}(k_a,t),
\end{equation}
where $V$ is the number of vertices, $L$ the number of loops and $I$ the number of internal lines in the diagram. If we restrict ourselves to diagrams without loops then performing these time integrals (details in Appendix \ref{amp}) gives
\begin{tcolorbox}[colback=brown!10!white, colframe=white] 
\begin{equation}
A_n'=-\prod_\alpha^V g_\alpha F_\alpha (\textbf{k})\prod_\beta^I \mathcal{G}_\beta(K_\beta)\prod_a^n f_{s_a}^*(k_a).
\end{equation}
\end{tcolorbox}
Where the prime indicates that we have stripped off a factor of $(2\pi)^4 \delta^4\left(\sum_a^n k_a^\mu\right)$ and we have defined
\begin{equation}\label{eq:P(K)}
\mathcal{G}(K)=\frac{f_{s'}^*(K)f_{s}(K)}{2K}\frac{1}{K-\sum_b^m k_b}+ \frac{f_{s'}(K)f_{s}^*(K)}{2K}\frac{1}{K+\sum_b^m k_b},
\end{equation}
where $\textbf{K}=\sum\limits_b^m \textbf{k}_b$ is the momentum of the internal line and the sum is over all other lines coming into the diagram. This result is simply a restructuring of the results found in any textbook on scattering amplitudes, for example \cite{WeinbergBook1}, which can be seen if we take $s=s'=0$ so that $\mathcal{G}$ becomes
\begin{equation}
\mathcal{G}(K)=\frac{1}{K^2-\left(\sum_b^mk_b\right)^2}=\frac{1}{\sum_b^m{k_b}^\mu\sum_b^m{k_b}_{\mu}}.
\end{equation}
Therefore we can rewrite our expression for the amplitude as 
\begin{equation}
A_n'=-i\prod_\alpha^V- i g_\alpha F_\alpha (\textbf{k})\prod_{a}^{n_{\alpha}}f_{s_a}\prod_\beta^I  \frac{i}{K_I^\mu {K_I}_\mu}\,,
\end{equation}
which is a more familiar form of the amplitude. 

\subsection{Curved spacetime: the total-energy pole}
In this section, we will make clear why we went to such lengths to rewrite the Feynmann rules by constructing equivalent rules for the calculation of de Sitter correlators, we will then demonstrate that the amplitude is recovered as the residue of a pole at the total energy with an order that is fixed by the scaling dimension of the interactions. In the in-in formalism\footnote{We use the in-in formalism rather than the wavefunction of the universe because when we allow derivative interactions the alterations to the conjugate momentum affect the in-in calculation in the same way as the amplitude, unlike for the wavefunction of the universe which has different diagrams, complicating the comparison between the two.}, vacuum correlation functions at a given time $\eta_0$ are found by evolving the vacuum state backwards in time to the infinite past where we consider the universe to be in the Bunch-Davies vacuum. We define the $ n  $-particle correlator to be the vacuum expectation value of the product $  n $ $\phi$-fields,
\begin{equation}
B_n=\left\langle\prod_a^n\hat{\phi}(\eta_0,\textbf{k}_a)\right\rangle=\lim_{\eta\rightarrow -\infty}\left\langle0\right\rvert \mathcal{U}^\dagger(\eta_0,\eta)\prod_a^n\hat{\phi}(\eta_0,\textbf{k}_a)\mathcal{U}(\eta_0,\eta)\left\lvert0\right\rangle,
\end{equation}
where $\mathcal{U}$ is defined as before. The interaction terms in the action are now exactly those in   \ref{contact}
\begin{equation}
S_n = g_n \int d\eta\, a(\eta)^{4-\sum\limits_a^n s_a}\,\int \left(\prod_{a=1}^n \dfrac{d^3\bfk_a}{(2\pi)^3}\right) (2\pi)^3 \delta^3\left(\sum \limits_{a=1}^n\bfk_a\right) F_n\left[\frac{\bfk_a\cdot\bfk_b}{a^2}\right]\,\prod_{a=1}^{n}\partial_\eta^{s_a}\phi(k_a,\eta),
\end{equation}
Typically we want to explore these correlators at $\eta_0=0$ but this introduces the potential additional complication of a IR-divergence. We will therefore keep $\eta_0$ finite during the following derivation. Because the mode functions and the scale factor are finite everywhere except at $\eta=0$ and $\eta=-\infty$ this restriction means that all singularities must come from the limit $\eta\rightarrow-\infty$\footnote{Here we left the Wick rotation implicit.}. Therefore, the pole structure will be dominated by the behaviour in the infinite past, where all fields can be approximated to be massless, thus we will focus the discussion here on massless correlators for which the mode function is 
\begin{equation}
\phi(\eta,k)=\frac{H}{\sqrt{2k^3}}(1-ik\eta)e^{ik\eta},
\end{equation}
which has time derivatives that are given by
\begin{equation}
\frac{d^s}{d\eta^s}\phi(\eta,k)=\frac{f_s(k)H}{\sqrt{2k^3}}(1-s-ik\eta)e^{ik\eta}\,,
\end{equation}
where $f_s(k)$ is the same function as defined in  \ref{eq:flatderivative}. There remain two significant differences between this calculation and the one for amplitudes. First, the upper limit of the integrals is $\eta_0$ rather than infinity. Second, when we expand the time evolution operator to some order in the coupling we receive contributions from both $\mathcal{U}$ and $\mathcal{U}^\dagger$. In the construction of Feynman diagrams, this manifests itself in the fact that we must colour each vertex depending on whether it comes from $\mathcal{U}$ or $\mathcal{U}^\dagger$. These vertices will be referred to as time ($+$) and anti-time ($-$) ordered vertices respectively and come with different propagators, a fact that we must take into account when defining our Feynman rules. For two time ordered vertices we will recover the analogue of the Feynman propagator, the time ordered propagator,
\begin{multline}
G_{++}(k,\eta,\eta')=\theta(\eta'-\eta)\frac{H^2 f^*_s(k)f_{s'}(k)}{2k^3}\left(-H\eta\right)^{s}\left(-H\eta'\right)^{s'}(1-s+ik\eta)(1-s'-ik\eta')e^{-ik(\eta-\eta')}\\+\theta(\eta-\eta')\left(-H\eta\right)^s\left(-H\eta'\right)^{s'}\frac{H^2 f_s(k)f_{s'}^*(k)}{2k^3}(1-s-ik\eta)(1-s'+ik\eta')e^{ik(\eta-\eta')}\,.	
\end{multline}
This is related to the previously defined propagators,  \ref{eq:curvedpropagators}, but includes derivative terms. We have not introduced a new notation to distinguish these terms as this is the definition that will be used from here and any additional marks will just cause notational clutter. We also have boundary propagators that now include the mode functions at $\eta_0$ and are given by
\begin{equation}
G_{B+}(k,\eta)=\frac{H^2 f_s^*(k)}{2k^3}\left(-H\eta\right)^s(1-s+ik\eta)(1-ik\eta_0)e^{-ik(\eta-\eta_0)}\,.
\end{equation}
The anti-time-ordered expressions are simply related to these by complex conjugation as they come from the Hermitian conjugate operator. Finally, we must consider propagators from a time-ordered to an anti-time-ordered vertex. As the time-ordered piece will always contribute a creation operator and the anti-time-ordered piece an annihilation operator this gives
\begin{equation}
G_{+-}(k,\eta,\eta')=\frac{H^2 f_s^*(k)f_{s'}(k)}{2k^3}\left(-H\eta\right)^{s}\left(-H\eta'\right)^{s'}(1-s+ik\eta)(1-s'-ik\eta')e^{-ik(\eta-\eta')}\,.
\end{equation}
Where $\eta$ comes from the time-ordered vertex and $\eta'$ from the anti-time-ordered vertex. The Feynman rules are then:
\begin{itemize}
\item Add a propagator for each line in the diagram.
\item Conserve momentum at each vertex and overall.
\item Integrate over any undetermined momentum.
\item Integrate over time for each vertex from $-\infty$ to $\eta_0$.
\item Introduce a factor of $-ig\,F\left[\frac{\bfk_a\cdot\bfk_b}{a^2}\right]$ for each $+$ vertex and $ig\,F\left[\frac{\bfk_a\cdot\bfk_b}{a^2}\right]$ for each $-$ vertex.
\end{itemize} 
As for the amplitude this can be expressed as an integral,
\begin{equation}
B_n= i^{V_--V_+}\int\prod_\alpha^{V_++V_-}\frac{ g_\alpha\, \tilde{F}_\alpha (\textbf{k})}{a^{R_\alpha-4 }(\eta_\alpha)} d\eta_\alpha  \prod_\beta^L\frac{d^3 \bfk_\beta}{(2\pi)^3}\prod_\gamma^I G_{++}(K_\gamma;\eta,\eta') \prod_a^nG_{B+}(k_a,\eta).
\end{equation}
Here we have removed the factors of the scale factor from $F_\alpha$, redefining it as 
\begin{equation}
\tilde{F}_\alpha(\textbf{k}) \equiv F_\alpha\left[\bfk_a\cdot\bfk_b\right]=a^{R_\alpha}F_\alpha\left[\frac{\bfk_a\cdot\bfk_b}{a^2}\right] \,,
\end{equation}
where $R_\alpha$ is the total number of spatial derivatives at the vertex $\alpha$. This is exactly the term that appears in the amplitude. We will, as for the amplitude, begin by ignoring loops, see Appendix \ref{loop} for a discussion of how our results generalise to loop diagrams. For diagrams that contain both time-ordered and anti-time-ordered vertices there must be at least one $G_{+-}$ propagator. The presence of this propagator means that the integral is separable into a left and a right side connected by some propagator with a momentum $\textbf{K}$,
\begin{equation}\label{eq:leftright}
B_n^{+-}(k_a)=\frac{2K^3}{H^2(1+K^2\eta_0^2)} B_L^+(k_{a_L},K) B_R^-(k_{a_R},K).
\end{equation}
The factor of $\frac{2K^3}{H^2(1+K^2\eta_0^2)}$ comes from the difference between the product of two bulk-to-boundary propagators and the bulk-to-bulk propagator that separates the two vertices. We are interested in exploring the total energy pole and, as neither $B_L$ or $B_R$ know anything about each other, neither can have a dependence on the total energy. Hence, we need only consider diagrams with either all time-ordered vertices or all anti-time-ordered vertices. As the two are related by complex conjugation it is enough to study just one of these, which we will choose to be the time-ordered diagrams.


\paragraph{Contact diagrams} We can begin by looking at contact diagrams, for which the correlator is
 \begin{equation}\label{eq:contactBN}
B_n^+= -i\int_{-\infty}^{\eta_0} (-\eta H)^{R-4}g\,\tilde{F}(\textbf{k}) d\eta \prod_a^n\frac{H^2f^*_{s_a}(k_a)}{2k_a^3}\left(-H\eta\right)^{s_a}(1-s_a+ik_a\eta)(1-ik_a\eta_0)e^{-ik_a(\eta-\eta_0)}.
\end{equation}
By considering this product as a sum over powers of $\eta$ we can perform this integral (additional details in Appendix \ref{corr}) to find that, provided $n+N>3$, the worst singularity in the total energy is
\begin{align}\label{eq:BN+}
\lim_{k_T\rightarrow 0} B_n^+ &=(-1)^{n+1}iH^{2n+N-4}g\tilde{F}(\textbf{k})\prod_a^n\frac{(i+k_a\eta_0) f_{s_a}^*(k_a)}{2k_a^2} \frac{(N+n-4)!}{(-ik_T)^{N+n-3}}\\
&=(-1)^n iH^{2n+N-4} A_n'\frac{(N+n-4)!}{(-ik_T)^{N+n-3}}\prod_a^n\frac{i+k_a\eta_0}{2k_a^2},
\end{align}
where $  N $ is the total number of derivatives including spatial and temporal alike,
\begin{align}
N\equiv R+\sum_a s_a =R+S\,.
\end{align} 
Therefore, for a contact term, the total correlator is the sum of the time and anti-time ordered contributions, which are complex conjugates of each other, so we have that
\begin{equation}\label{eq:kTlim}
\lim_{k_T\rightarrow 0} B_n=2(-1)^{n}H^{2n+N-4}(n+N-4)!\Re\left\lbrace \frac{iA_n'}{(-ik_T)^{N+n-3}}\prod_a^n\frac{i+k_a\eta_0}{2k_a^2}\right\rbrace.
\end{equation}


\paragraph{The special case of $\phi^3$}
For local interactions $N\geq 0$ whilst $n>2$ as any second order terms will be included in the background equation of motion so will not enter $\mathcal{U}$. Therefore the only interaction for which $n+N\leq 3$ is $\phi^3$ which we must consider separately. The contact diagram in this case has a correlator given by
\begin{equation}\label{eq:B3int}
B_3^+=-i\int_{-\infty}^{\eta_0}\frac{gd\eta}{(H\eta)^4} \frac{H^6}{8k_1^3k_2^3k_3^3}(1+ik_T\eta-e_2\eta^2-ie_3\eta^3)(1-ik_T\eta_0-e_2\eta_0^2+ie_3\eta_0^3)e^{-ik_T(\eta-\eta_0)}.
\end{equation}
So in the limit $k_T\rightarrow 0$, we can use the fact that for such an interaction $A_3'=-g$ so that,
\begin{equation}
\lim_{k_T\rightarrow0} B_3=  2 (-1)^3 H^{2\times3-4}\log(-\eta_0 k_T) \Re\left\lbrace iA_3' \prod_a^3\frac{i+k_a\eta_0}{2k_a^2}\right\rbrace\,.
\end{equation}
Therefore we have replaced $$\frac{(n+N-4)!}{(-ik_T)^{n+N-3}} \rightarrow \log(-\eta_0 k_T)$$ in  \ref{eq:kTlim}. This logarithmic divergence in the place of an expected zeroth order pole is fairly typical behaviour when studying poles.  The non-singular terms in $k_T$ are also related to the amplitude but in a much less trivial way as there is a contribution from multiple terms in the integral in  \ref{eq:B3int}. This relationship also won't be simple to extend to other diagrams. 


\paragraph{General diagrams} To understand the behaviour of non-contact terms we will look to integrate out a single vertex from the totally time-ordered diagram. Using the fact that there must be at least one vertex, labeled by $V$, that is connected to only one other, labeled $V-1$, we perform this integral, keeping only the highest power of $\eta_{V}$ and $\eta_{V-1}$, we recover the same function, $\mathcal{G}_I(K_I)$, that appeared in the amplitude multiplying a correlator with one fewer vertex,
\begin{multline}
\lim_{k_T\rightarrow 0} B_n^+=\lim_{k_T\rightarrow 0}  i^{-(V-1)}g_V F_V(\textbf{k})\mathcal{G}_I(K_I)H^{2I+2n+\sum\limits_\alpha^V(N_\alpha-4)}\\ \int\prod_a^{V-1}\frac{ g_\alpha \tilde{F}_\alpha (\textbf{k})}{(-\eta_a)^{4-R_a}}d\eta_a \prod_\beta^{E-1} G_{++}(K_\beta;\eta,\eta') \prod_a^{n-m}G_{B+}(k_a,\eta)\prod_b^mG_{B+}(k_b,\eta_{V-1})(-\eta)_{V-1}^{N_V+s_{V-1}-2}.
\end{multline}
Where we note that, as we are only interested in the most singular terms, we have recovered the bulk-to-boundary propagators that were associated with the vertex at $\eta_V$ but they have moved to the vertex to which it was connected whilst introducing an extra factor of $(-\eta)_{V-1}^{N_V+s_{V-1}-2}$. If we repeat this to remove all of the vertices except one we are left with
\begin{multline}
\lim_{k_T\rightarrow 0} B_n^+= - iH^{2I+2n+\sum\limits_\alpha^V(N_\alpha-4)} \prod_\alpha^{V}g_\alpha \tilde{F}_\alpha(\textbf{k})\prod_\beta^{I}\mathcal{G}_\beta(K_\beta)\\  \lim_{k_T\rightarrow 0}  \int_{-\infty}^{\eta_0} (-\eta)^{\sum_\alpha^V R_\alpha+\sum_{\beta}^I\left(s_{\beta}+s'_{\beta}-2\right)-4} d\eta  \prod_d^{N}G_{B+}(k_d,\eta).
\end{multline}
This is exactly the form of a contact term but we have replaced $R$, the number of spatial derivatives with $\sum\limits_\alpha^V R_\alpha +\sum\limits_{\beta}^I\left(s_{\beta}+s'_{\beta}-2\right)$, which is similarly a counting of derivatives, therefore the $N$ in  \ref{eq:BN+} becomes 
\begin{align}
\sum\limits_\alpha^V R_\alpha +\sum\limits_{\beta}^I\left(s_{\beta}+s'_{\beta}-2\right)+\sum\limits_a^n s_a=\sum\limits_\alpha^V(N_\alpha-2)\,.
\end{align}
In $(-1)^N$ we can drop the two whereas 
\begin{align}
N+n\rightarrow\sum\limits_\alpha^V(N_\alpha-2)+n=\sum\limits_\alpha^V \left(N_\alpha+n_\alpha-4\right)+4\,,
\end{align}
where $n_\alpha$ is valence of each vertex, i.e. the total number of internal and external lines connected to it. This sum of fields plus derivatives at a vertex is just the dimension $  D_{\alpha} $ of the operator
\begin{align}
D_{\alpha}\equiv n_{\alpha} + N_{\alpha}\,.
\end{align}
Therefore the $k_T$ pole of the entire correlator is given by
\begin{tcolorbox}[colback=brown!10!white, colframe=white] 
\begin{equation}\label{finalBtoA}
\lim_{k_T\rightarrow 0} B_n=2\,(-1)^{n} H^{2I+2n+\sum\limits_\alpha^V(N_\alpha-4)} \left(\sum\limits_\alpha^VD_\alpha-4\right)! \times \Re\left\lbrace  \frac{iA_n'}{(-ik_T)^{p} }\prod_a^n\frac{i+k_a\eta_0}{2k_a^2}\right\rbrace,
\end{equation}
\end{tcolorbox}
where the order $  p $ of the total energy pole of any tree-level diagram is
\begin{align}
p\equiv 1+\sum\limits_\alpha^V \left( D_\alpha -4\right)\,.
\end{align}
This agrees with the result first derived in \cite{Enrico} using dimensional analysis and scale invariance. It is also possible to extend this result to include loop diagrams, see Appendix \ref{loop} for an outline of this argument, in this case however it is only at the level of the integrand that the amplitude is recovered i.e. if we define the loop integrands, $\mathcal{A}_n$ and $\mathcal{B}_n$,
\begin{equation}
A_n'=\int \prod_\beta^L\frac{d^3 \bfk_\beta}{(2\pi)^3}\mathcal{A}_n,\ B_n=\int \prod_\beta^L\frac{d^3 \bfk_\beta}{(2\pi)^3}\mathcal{B}_n,
\end{equation}
then the generalisation of our result to diagrams that include loops is
\begin{equation}
\lim_{k_T\rightarrow 0}\mathcal{B}_n=2\, (-1)^{n}H^{2I+2n+\sum\limits_\alpha^V(N_\alpha-4)}  \left(\sum\limits_\alpha^VD_\alpha-4\right)!\times \Re\left\lbrace\frac{i\mathcal{A}_n }{(-ik_T)^{p}}\prod_a^n\frac{i+k_a\eta_0}{2k_a^2}\right\rbrace.
\end{equation}


\subsection{Other poles}\label{sec:op}

We have seen that when the total energy of a diagram is taken to zero we recover the amplitude at that pole, however we also saw in  \ref{eq:leftright} that there is a term that enters the correlator that can be separated into two correlators. Therefore, when the total energy of either of these correlators is taken to zero we will recover the sub diagram amplitudes,
\begin{equation}\label{eq:BN+-}
\lim_{E_L\rightarrow0}B_n^{+-}=\frac{KH^{2I_L+2n_L+\sum\limits_\alpha^{V_L}(N_\alpha-4)}}{1+iK\eta_0} (-1)^{n_L}A_L'(k_b,K) \frac{\left(\sum\limits_\alpha^{V_L}D_\alpha-4\right)!}{\left(-iE_L\right)^{1+\sum\limits_\alpha^{V_L}(D_\alpha-4)  }}\prod_b^{n_L}\frac{i+k_b\eta_0}{2k_b^2}  B_R^-(k_{a},K).
\end{equation}
 It is important to note that it is not the energy at a vertex that is taken to zero but rather the sum of the energies of the external lines coming into the vertex and the energy of the exchanged momentum. However, unlike for the total energy pole more than one diagram will now contribute. To see the remaining poles consider the terms in  \ref{eq:BNtotal}. There appear to be two poles that we have discovered on integrating out one of the vertices which occur when we take
\begin{equation}
\tilde{E}_L=\sum_a^mk_a-K\rightarrow 0, \ E_L=\sum_a^m k_a + K\rightarrow 0.
\end{equation}
We begin be exploring the limit $\tilde{E}_L\rightarrow 0$, the apparent singularity comes from the first integral from $\eta_{V-1}$ to $\eta_0$ but if we evaluate this integral in this limit we find that it is
\begin{equation}
\lim_{\tilde{E}_L\rightarrow0}\int_{\eta_{V-1}}^{\eta_0} d\eta_V\prod_b^m \eta_V^{s_b}(1-s_b+ik_b\eta_V) \eta_V^{R_V+s_V-4}(1-s_V-iK\eta_V)e^{-i\tilde{E}_L\eta_V}
\end{equation}
which is, in fact, finite. This is exactly as expected because this singularity is at a physically realisable value of the momenta. However, the other limit is a true singularity the worst of which will, as for the total energy pole, come from the highest power of $\eta_{V}$, but this time we take the term with $k=0$ as it is most singular in energy configuration. It is possible to express the $\eta_{V-1}$ term from the time ordered propagator as a boundary propagator using the fact that $G_{B+}^*(K,\eta)=G_{B+}(-K,\eta)$, this turns the integral into a new correlator, whilst the other terms can be expressed as the amplitude of the contact diagram arising from the vertex that was removed,
\begin{multline}\label{eq:BN++}
\lim_{E_L\rightarrow 0} B_n^+=\\-\frac{({N_V+m+1-4})! }{ \left(-iE_L\right)^{N_V+m+1-3}}(-1)^{m} H^{N_V+2m-4}  A_L'(k_b,K)\prod_b^m\frac{i+k_b\eta_0}{2k_b^2} B_R^+(k_a,-K) \frac{K}{1+iK\eta_0}.
\end{multline}
The additional $+1$ that appears in these terms is due to the inclusion of the exchange line in the counting of boundary fields. Had we wanted to explore this same limit for an internal vertex it would have been possible to remove the vertices one at a time just as for the total energy pole, until the vertex of interest had only one connecting line then it will have all the external lines from these vertices attached to it and additional factors of $\eta$ for each of the removed lines and derivatives. Therefore, we will still recover the amplitude of this reduced diagram, due to this procedure the pole will be where the external energy coming into the vertex sums to zero, rather than when the energy at the vertex sums to zero, just we saw in  \ref{eq:BN+-}. At this pole the correlator is
\begin{multline}\label{eq:BLBn+}
\lim_{E_L\rightarrow 0} B_n^+=\\-\frac{KH^{2I_L+2n_L+\sum\limits_\alpha^{V_L}(N_\alpha-4)} }{1+iK\eta_0}\frac{\left({\sum\limits_\alpha^{V_L}D_\alpha-4}\right)! }{ \left(-iE_L\right)^{1+\sum\limits_\alpha^{V_L}(D_\alpha-4)}}(-1)^{n_L}  A_L'(k_b,K)\prod_b^{n_L}\frac{i+k_b\eta_0}{2k_b^2} B_R^+(k_a,-K) .
\end{multline}
 Although the left half of the diagram must be entirely either time or anti-time ordered to maintain the pole this restriction does not apply to the right hand of the diagram so we must account for all possible time orderings of the right hand side, except for the vertex connecting it to the left hand side for which the time ordering fixes whether it contributes according to  \ref{eq:BN++} or  \ref{eq:BN+-}, so the total correlator is given, in this limit, by
 \begin{tcolorbox}[colback=brown!10!white, colframe=white] 
\begin{align}
&\lim_{E_L\rightarrow0} B_n=2H^{2I_L+2n_L+\sum\limits_\alpha^{V_L}(N_\alpha-4)}\\ 
&\Re\left\lbrace \frac{(-1)^{n_L}  K}{1+iK\eta_0}\frac{ A_L'(k_b,K) \left({\sum\limits_\alpha^{V_L}D_\alpha-4}\right)! }{ \left(-iE_L\right)^{1+\sum\limits_\alpha^{V_L}(D_\alpha-4)}}\prod_b^{n_L}\frac{i+k_b\eta_0}{2k_b^2} \left(B_R^-(k_a,K)- B_R^+(k_a,-K)\right)\right\rbrace \nonumber
\end{align}
\end{tcolorbox}
Where the superscripts on the correlators on the final line refer to the time ordering of the vertex connecting them to the left hand side. This is a generalisation of the results in Section 6 from \cite{baumann2020cosmological} to diagrams with multiple vertices. 


\subsection{Relation to the Cosmological Optical Theorem}

The subdiagram poles in the 4-point correlator can be seen directly from the optical theorem,
\begin{multline}
\lim_{E_L\rightarrow 0} B_4^s(k_1,k_2,k_3,k_4,s)+B_4^s(-k_1,-k_2,-k_3,-k_4,s)=\lim_{E_L\rightarrow 0} B_4^s(k_1,k_2,k_3,k_4,s) \\=\frac{1}{2 P(s)} \lim_{E_L\rightarrow 0} B_3(k_1,k_2,s)\left(B_3(k_3,k_4,s)-B_3(k_3,k_4,-s)\right)\,.
\end{multline}
Where the first equality follows due to the fact that there is no singularity at $\tilde{E}_L$ and hence the second term is finite in this limit. Whilst the second equality follows from the total energy pole in $B_3$. None of the other terms on right hand side have singularities here. In the case $n=4$ we can rewrite  \ref{eq:BLBn+} as
\begin{equation}
\lim_{E_L\rightarrow 0} B_4^+=\lim_{E_L\rightarrow 0}\frac{1}{P(s)}  B_3^+(k_1,k_2,s) B_3^+(k_3,k_4,-s) .
\end{equation}
Whilst we also have
\begin{equation}
\lim_{E_L\rightarrow 0} B_4^{+-}=\lim_{E_L\rightarrow0} \frac{1}{P(s)}   B_3^+(k_1,k_2,s) B_3^-(k_3,k_4,s) .
\end{equation}
This means that
\begin{equation}
\lim_{E_L\rightarrow 0} B_4=\frac{2}{P(s)} \lim_{E_L\rightarrow 0} \Re\left\lbrace B_3^+(k_1,k_2,s)( B_3^-(k_3,k_4,s)-B_3^+(k_3,k_4,-s) ) \right\rbrace .
\end{equation}
However, we saw that for IR finite contact interactions
\begin{equation}
\Im(\psi_n')=0\rightarrow \Re(\psi_n')=\psi_n'=-\frac{1}{2}\prod_{a=1}^3\frac{1}{P(k_a)}B_3\rightarrow \Im(B_3)=0\rightarrow B_3^-=B_3^+=\frac{1}{2}B_3
\end{equation}
Therefore we can rewrite this expression as
\begin{equation}
\lim_{E_L\rightarrow 0} B_4=\frac{1}{2P(s)} \lim_{E_L\rightarrow 0} B_3(k_1,k_2,s)( B_3(k_3,k_4,s)-B_3(k_3,k_4,-s) ) ,
\end{equation}
which is exactly what we saw from taking this limit from the optical theorem. For higher point correlators we have not derived the form of the optical theorem, however, we do know the position of all the poles of the diagram, it may therefore be possible to apply this relationship in reverse, by writing out the poles of the correlator and using this expression to extend the optical theorem to these more complicated correlators.

\section{Conclusion}\label{sec:conc}

In this work we have derived a Cosmological Optical Theorem (COT) for the coefficients of the wavefunction of the universe as a consequence of unitary time evolution. Specifying the Cosmological Optical Theorem to contact diagrams and exchange diagrams lead to the powerful unitarity constraints in \eqref{unicont} and \eqref{exchCOPT}, respectively. Among the many interesting lemmas, our results imply a simple relation between the trispectrum and the bispectrum, \eqref{cotcor}. As the name suggests, our results can be thought of as the cosmological analog of the well-known (generalized) optical theorem for amplitudes.

In the following we mention some avenues for future investigation.
\begin{itemize}
\item Exchange diagrams with massless mediating particles are extremely constrained by the Cosmological Optical Theorem in \eqref{exchCOPT}. This result is independent from, and does not rely on invariance under de Sitter boosts. Indeed we have checked it for operators that break de Sitter boosts, such as the $  \dot\phi^{3} $ interaction arising form coupling with the inflaton foliation in non-canonical models. It is natural to use the COT, in conjunction with a number of other assumptions about the analytic structure and limits of correlators summarized in \cite{Enrico}, to bootstrap the four-point function of scalars and/or gravitons (e.g. in the EFT of inflation) from the three-point function, in a way that is analogous to on-shell methods for amplitudes. This would generalize the very interesting results recently derived in \cite{Arkani-Hamed:2018kmz,Baumann:2019oyu,baumann2020cosmological} using de Sitter isometries to more phenomenologically interesting models. 
\item Locality and unitarity are two of the main pillars of the scattering amplitude program. They are manifest in the type of singularities that amplitudes can have and the way they must factorize on these poles \cite{WeinbergBook1}. In hindsight, similar properties such as factorization on the subdiagram poles pertain to cosmological correlators as well. Nevertheless, a first principle study of the combined implications of locality and unitarity for Cosmological boundary observables remains elusive. It would be desirable to derive some non-perturbative results, as it is possible to do in flat spacetime \cite{Schwartz:2013pla}.
\item On-shell methods for amplitudes are most useful when applied to spinning particles. In fact, one of the greatest achievements of the amplitude program is the on-shell construction of all tree level amplitudes of YM and GR \cite{CheungReview, Benincasa:2007xk, Britto:2005fq, ArkaniHamed:2008yf, Benincasa:2007qj}.
Hence, it will be of utmost interest to understand the implications of unitarity for the correlators of spinning fields, in the hope for achieving the similar level of sophistication for correlators. This might involve a derivation of the COT for higher-point correlators of the spinning fields. 
\item The four point function induced by the exchange of a massive particle generically has branch cut singularities\cite{Arkani-Hamed:2015bza, Arkani-Hamed:2018kmz}. That is the reason we did not discuss those cases in this paper. Nonetheless, we expect that the relation \eqref{exchCOPT} will remain true, albeit after a careful definition of the meaning of the analytical continuation of the four point function to negative real energies. We leave the rigorous proof of such relation to future work. 
\item Our paper was exclusively focused on scale-invariant correlators. However, we suspect there might be hidden constraints from unitarity even in cases where dilatation is explicitly broken by the time dependent background. If true, this will at least give some leverage on the space of possibilities, for example, in the consistent non-gaussianity shapes that can be generated by oscillations or features in the inflationary potential \cite{Flauger:2009ab, Behbahani:2011it}.
\item The consequences of unitarity in AdS have been recently studied in \cite{Meltzer:2020qbr}, where ``AdS cutting rules'' were derived. Those results display some remarkable similarity with our findings here, in particular for example for four-particle exchange diagrams at tree level. It would be interesting to better understand this relationship.
\end{itemize}


\section*{Acknowledgements}
We would like to thank Aliakbar Abolhasani, Daniel Baumann, Tanguy Grall, Sebastien Renaux-Petel and David Stefanyszyn for stimulating discussions. We are also particularly thankful to Garrett Goon, Austin Joyce and Guilherme L. Pimentel for commenting on a draft of this manuscript. E.P. and S.J. have been supported in part by the research program VIDI with Project No. 680-47-535, which is (partly) financed by the Netherlands Organisation for Scientific Research (NWO).


\appendix


\section{Wave Function Method} 
\label{wfu}
In this appendix we provide a pedagogical review of the wavefunction approach to the computation of cosmological correlators\cite{Maldacena:2011nz}. We closely follow Weinberg's discussion of Path-Integral methods in \cite{WeinbergBook1} and an unpublished manuscript from Garrett Goon. 
In Quantum Mechanics we define an orthonormal  basis of eigenstates of the position operator $\hat{x}$,
\begin{equation}
\hat{x}\lvert \textbf{x}\rangle=x\lvert\textbf{x}\rangle.
\end{equation}
We can expand any state in terms of this basis,
\begin{equation}
\lvert \Psi(t)\rangle =\int \Psi(t,\textbf{x})\lvert \textbf{x}\rangle d^3 x,
\end{equation}
where the function $\Psi(t,\textbf{x})$ is called the position space wavefunction (often just the wavefunction). We can then use the orthogonality between our basis states to express this function as an inner product 
\begin{equation}
\langle \textbf{y}\lvert \Psi(t)\rangle =\int \Psi(t,\textbf{x})\langle \textbf{y}\lvert \textbf{x}\rangle d^3 x=\Psi(t,\textbf{y}).
\end{equation}

Due to its relevance to inflation we will explore a quantum field theory with a single scalar field, $\phi$, but this discussion will simply extend to systems with more than one field. Just as in Quantum Mechanics we choose our basis to be adapted to the eigenstates of this field, by defining the states $\lvert \phi;\eta\rangle$ so that they are the eigenstates of the Heisenberg picture operator $\hat{\phi}_{\text{H}}$ with the same eigenvalues as the eigenstates, $\langle\phi(\eta)\rvert_{\text{S}}$ of the Schroedinger picture operator, $\hat{\phi}_{\text{S}}$, at some fixed initial time $\eta_{\text{I}} $. We similarly define the states $\lvert \pi;\eta\rangle$ from the eigenstates of the conjugate momentum operator $\hat{\pi}_{\text{S}}$,
\begin{align}
\hat{\phi}_{\text{S}}\lvert \phi (\eta_{\text{I}}) \rangle_{\text{S}}&=\phi \lvert \phi(\eta_{\text{I}})\rangle_{\text{S}}&&\rightarrow& \hat{\phi}_{\text{H}}(\eta) \lvert\phi;\eta\rangle&=\phi\lvert\phi;\eta\rangle &&\rightarrow &  \lvert \phi ;\eta \rangle&=U^\dagger(\eta_{\text{I}},\eta) \lvert \phi(\eta_{\text{I}}) \rangle_{\text{S}}  \\ \hat{\pi}_{\text{S}}\lvert \pi (\eta_{\text{I}}) \rangle_{\text{S}}&=\pi \lvert \pi(\eta_{\text{I}}) \rangle_{\text{S}}&&\rightarrow& \hat{\pi}_{\text{H}}(\eta)\lvert\pi,\eta\rangle&=\pi\lvert\pi;\eta\rangle& &\rightarrow&   \lvert \pi ;\eta \rangle&=U^\dagger(\eta_{\text{I}},\eta)\lvert \pi(\eta_{\text{I}}) \rangle_{\text{S}}.
\end{align}
For some time independent $\phi,\ \pi$  and where we have defined the time evolution operator
\begin{equation}
U(\eta_1,\eta_2)=T \exp(-i\int_{\eta_1}^{\eta_2} \hat{H}_{\text{S}} d\eta' ).
\end{equation}
These states obey equal-time orthogonality and completeness relations,
\begin{equation}
\langle \phi';\eta\rvert \phi;\eta\rangle=\delta(\phi'-\phi),\ \ \langle \pi';\eta\rvert \pi;\eta\rangle=\delta(\pi'-\pi),\ \  \int  d\phi \lvert \phi ;\eta\rangle\langle \phi;\eta\rvert=1,\ \ \int d\pi\lvert \pi;\eta\rangle\langle \pi;\eta\rvert=1,
\end{equation}
and their inner product with each other is given by
\begin{equation}\label{eq:qt-pt}
\langle \phi;\eta\rvert \pi;\eta\rangle = \frac{1}{\sqrt{2\pi}}\exp(i\phi\pi).
\end{equation}
By analogy with quantum mechanics we define the wavefunctional, $\Psi[\phi;\eta]$, (in accordance with the typical nomenclature this will be referred to as the wavefunction from here)  by
\begin{equation}
\lvert \Psi(\eta)\rangle=\int d\phi \Psi[\phi ;\eta] \lvert \phi;\eta\rangle,
\end{equation}
which can be computed using the orthogonality of our basis,
\begin{equation}
\langle \bar{\phi} ;\eta_0\lvert \Psi(\eta)\rangle=\int  d\phi \Psi[\phi;\eta]\langle \bar{\phi};\eta_0 \lvert \phi;\eta\rangle= \Psi[\bar{\phi};\eta_0].
\end{equation}
In inflation the wavefunction of interest is the one for the vacuum so $\lvert\Psi(\eta)\rangle=\lvert \Omega\rangle$ (we are considering the Heisenburg picture vaccum so the state is constant). As we do not know the behaviour of this inner product at the arbitrary time $\eta_0$ we must relate it to the early time behaviour, where we can impose some initial condition, to do this we insert the identity at this early time,
\begin{equation}\label{eq:wave}
\langle \bar{\phi};\eta_0\lvert \Omega\rangle=\int  d\phi \langle \bar{\phi};\eta_0\rvert \phi;\eta\rangle\langle \phi;\eta\lvert \Omega\rangle.
\end{equation}
In order to calculate $\langle \bar{\phi};\eta_0\rvert \phi;\eta\rangle$ we start by considering an infinitesimal change in time, $\eta=\eta_0-d\eta$,
\begin{align}
\left\langle \left. \bar{\phi};\eta_0\right\rvert \phi;\eta_0-d\eta\right\rangle&=\left\langle\bar{\phi}(\eta_{\text{I}})\left\rvert_{\text{S}} U(\eta_{\text{I}},\eta_0)U^{\dagger}(\eta_{\text{I}},\eta_0-d\eta)\right\lvert\phi(\eta_{\text{I}})\right\rangle_{\text{S}}\\&=\left\langle\bar{\phi}(\eta_{\text{I}})\left\rvert_{\text{S}}U(\eta_0-d\eta,\eta_0)\right\lvert\phi(\eta_{\text{I}})\right\rangle_{\text{S}}\\&=\left\langle\bar{\phi};\eta_0 \left\rvert U^\dagger(\eta_{\text{I}},\eta_0) U(\eta_0-d\eta,\eta_0)U(\eta_{\text{I}},\eta_0) \right\lvert\phi;\eta_0 \right\rangle\\&=\left\langle\bar{\phi};\eta_0 \left\rvert T \exp\left(-i\int_{\eta_0-d\eta}^{\eta_0} \hat{H}_{\text{H}} d\eta' \right)\right\lvert\phi;\eta_0 \right\rangle\\&=\left\langle\bar{\phi};\eta_0 \left\rvert 1-i \hat{H} \left(\hat{\phi}_{\text{H}}(\eta_0),\hat{\pi}_{\text{H}}(\eta_0)\right) d\eta \right\lvert\phi;\eta_0 \right\rangle
\end{align}
We now rewrite $\hat{H}$ so that all $\hat{\pi}_{\text{H}}$'s appear on the right and all $\hat{\phi}_{\text{H}}$'s on the left, using the commutation relations between $\hat{\pi}_{\text{H}}$ and $\hat{\phi}_{\text{H}}$, $\left[\hat{\phi}_{\text{H}}(\textbf{x}),\hat{\pi}_{\text{H}}(\textbf{y})\right]=i\delta(\textbf{x}-\textbf{y})$. The $\hat{\phi}_{\text{H}}$'s act on the bra state to give $\phi$ but to deal with the $\hat{\pi}_{\text{H}}$'s we must insert the identity operator,
\begin{equation}
\langle \bar{\phi};\eta_0\rvert \phi;\eta_0-d\eta\rangle=\int d\pi\langle \bar{\phi};\eta_0\rvert 1-iH(\bar{\phi},\hat{\pi}_{\text{H}}) d\eta\lvert \pi;\eta_0 \rangle\langle \pi;\eta_0\rvert \phi;\eta_0\rangle
\end{equation}
and the $\hat{\pi}_{\text{H}}$'s act on the $\pi$ ket states. This leaves the exponetial of some complex numbers inbetween two states so we can just move this outside the inner product and use  \ref{eq:qt-pt} to give
\begin{equation}
\langle \bar{\phi};\eta_0\rvert \phi;\eta_0-d\eta\rangle=\int \frac{d\pi}{2\pi} \exp(i\bar{\phi}\pi-i\phi\pi) \exp\left(-iH(\bar{\phi},\pi) d\eta \right)
\end{equation}
We then extend this infinitesimal inner product to a finite difference by inserting $N$ copies of the identity,
\begin{equation}
\langle \bar{\phi};\eta_0\rvert \phi;\eta\rangle=\int \prod_{i=0}^N d\phi_{i} \langle \phi_{i};\eta_{i}\rvert \phi_{i+1};\eta_{i+1}\rangle 
\end{equation}
Where we have defined, $\eta=\eta_{N+1}$, $\eta_{i+1}=\eta_{i}-d\eta$ and $d\eta=(\eta_0-\eta)/(N+1)$ then we find
\begin{equation}
\langle \bar{\phi};\eta_0\rvert \phi;\eta\rangle=\int \prod_{i=0}^N d\phi_{i} \int \prod_{i=0}^N \frac{d\pi_{i}}{2\pi} \exp\left[i\sum_{i=0}^N\left\lbrace (\phi_{i}-\phi_{i+1})\pi_{i}-H(\phi_{i},\pi_{i}), d\eta \right\rbrace\right]
\end{equation}
taking $N\rightarrow \infty$ converts the sum to an integral and the integrals become path integrals,
\begin{equation}
\langle \bar{\phi};\eta_0\rvert \phi;\eta\rangle=\int\limits_{\substack{\phi(\eta)=\phi\\\phi(\eta_0)=\bar{\phi}}}\mathcal{D}\phi \int\frac{\mathcal{D}\pi}{2\pi} \exp\left[i\int_{\eta_{\text{I}}}^{\eta_0} d\eta\left\lbrace {\phi'}(\eta)\pi(\eta)-H(\phi(\eta),\pi(\eta)) \right\rbrace\right].
\end{equation}
The wavefunction is related to this by an integral over all possible values of the initial value, $\phi$, which we can express by dropping the lower integration bound,
\begin{equation}
\Psi[\bar{\phi};\eta_0]=\langle \bar{\phi};\eta_0\lvert \Omega\rangle=\int\limits_{\phi(\eta_0)=\bar{\phi}}\!\!\!\!\mathcal{D}\phi \int \frac{\mathcal{D}\pi}{2\pi} \exp\left[i\int_\eta^{\eta_0} d\eta \left\lbrace {\phi'}(\eta)\pi(\eta)-H(\phi(\eta),\pi(\eta)) \right\rbrace\right]\langle \phi;\eta\lvert \Omega\rangle
\end{equation}
So, by this process we have related the wavefunction at one time to that at another. Now we take $\eta_{\text{I}}$ to correspond to $\eta$ and $\eta \rightarrow -\infty$ as this is a convenient place to specify our initial conditions. We further specify that in this limit we are in a de Sitter background and that in this limit the scalar field is both free and massless, so the Lagrangian is given by,
\begin{equation}
\mathcal{L}=\frac{a^2}{2}\left({\phi'} ^2-c_s^2(\nabla\phi)^2\right)\,,
\end{equation}
so that
\begin{equation}
\pi=\frac{\delta \mathcal{L}}{\delta \dot{\phi}}=a^2{\phi}'.
\end{equation}
In the interaction picture the corresponding operators are given by
\begin{align}
\hat{\phi}(\textbf{x},\eta )&=\int \frac{d^3p}{(2\pi)^3} \phi(p,\eta )e^{i\textbf{p}\cdot \textbf{x}} a_{\textbf{p}}+\phi^*(p,\eta )e^{-i\textbf{p}\cdot \textbf{x}} a_{\textbf{p}}^\dagger\,,\\
\hat{\pi}(\textbf{x},\eta)&= \int  \frac{d^3p}{(2\pi)^3} a^2 {\phi'}(p,\eta)e^{i\textbf{p}\cdot \textbf{x}} a_{\textbf{p}}+a^2{\phi'}^*(p,\eta )e^{-i\textbf{p}\cdot \textbf{x}} a_{\textbf{p}}^\dagger.
\end{align}
Where in the infinite past the mode functions and their derivatives are 
\begin{equation}
\lim_{\eta \rightarrow-\infty}\phi(\eta,k)=i\frac{H\eta e^{-ic_sk\eta}}{\sqrt{2c_sk}},\quad \lim_{\eta \rightarrow-\infty}{\phi'}(\eta,k)=\frac{H\eta c_sk e^{-ic_sk\eta}}{\sqrt{2c_sk}}.
\end{equation}
Therefore,
\begin{multline}
\lim_{\eta \rightarrow-\infty(1-i\epsilon)}\langle \phi,\eta\rvert \int d^3 x\frac{-i\sqrt{2c_sk}}{2H\eta}\hat{\phi} e^{-i\textbf{k}\cdot\textbf{x}}+\frac{\sqrt{2c_s k}}{2H\eta a^2}\hat{\pi}e^{-i\textbf{k}\cdot\textbf{x}}\lvert \Omega\rangle=\lim_{\eta \rightarrow-\infty(1-i\epsilon)}\frac{\sqrt{2c_s k}}{2H\eta}\\ \left[\left(\frac{{\phi'}(\eta,\textbf{k})}{c_s k}-i\phi(\eta,\textbf{k})\right)\langle \phi,\eta\rvert  a_{\textbf{k}}\lvert\Omega\rangle+\left(\frac{{\phi'}^*(\eta,-\textbf{k})}{c_s k}-i\phi^*(\eta,-\textbf{k})\right)\langle \phi,\eta\rvert a^\dagger_{-\textbf{k}}\lvert\Omega\rangle\right].
\end{multline}
in this limit the $a^\dagger_{-\bfk}$ terms vanish and $a_{\textbf{k}}$ is defined to annihilate the vacuum so this term must also be zero. We also use the fact that in the infinite past the states and operators in all pictures are the same,
\begin{align}
\lim_{\eta\rightarrow-\infty} \hat{\phi}\lvert\phi,\eta\rangle &= \phi\lvert\phi,\eta\rangle\,, \\
\lim_{\eta\rightarrow-\infty} \hat{\pi}\lvert\phi,\eta\rangle &=-i\frac{\delta}{\delta \phi}\lvert\phi,\eta\rangle, 
\end{align}
so, on making the Wick rotation explicit one again, we have 
\begin{equation}
\lim_{\eta \rightarrow-\infty(1-i\epsilon)} \frac{i\sqrt{2c_sk}}{2}\int d^3 xe^{-i\textbf{k}\cdot\textbf{x}}\left(\frac{-1}{H\eta}\phi +H\eta\frac{\delta}{\delta\phi}\right) \langle \phi,\eta\rvert \Omega\rangle=0\,.
\end{equation}
The first term will vanish in this limit but in order for the second term to be zero we need
\begin{equation}
\lim_{\eta \rightarrow-\infty(1-i\epsilon)} \frac{\delta}{\delta\phi} \langle \phi,\eta\rvert \Omega\rangle=0\quad \Rightarrow \quad \lim_{\eta \rightarrow-\infty(1-i\epsilon)} \langle \phi,\eta\rvert \Omega\rangle=const.
\end{equation}
This constant will be fixed by the normalisation condition on the wavefunction of the universe so that
\begin{equation}
\Psi[\bar{\phi};\eta_0]=\mathcal{N}\lim_{\eta\rightarrow -\infty}\!\!\!\!\int\limits_{\phi(\eta_0)=\bar{\phi}}\!\!\!\!\mathcal{D}\phi\int\mathcal{D}\pi \exp\left[i\int_\eta^{\eta_0} d\eta\left\lbrace {\phi}'(\eta)\pi(\eta)-H(\phi(\eta),\pi(\eta)) \right\rbrace\right].
\end{equation}
The importance of the wavefunction comes from its use in calculating correlation functions,
\begin{equation}
\langle \Psi(\eta_0)\rvert \mathcal{O}(\hat{\phi}(\eta_0),\hat{\pi}(\eta_0)) \lvert \Psi(\eta_0)\rangle=\int d\bar{\phi} \int d\phi \Psi^*[\phi,\eta_0] \Psi[\bar{\phi},\eta_0]\langle \phi,\eta_0  \rvert \mathcal{O}(\hat{\phi}(\eta_0),\hat{\pi}(\eta_0)) \lvert \bar{\phi},\eta_0\rangle.
\end{equation}
To address the possible dependence of $\mathcal{O}$ on $\hat{\pi}$ we introduce the identity, as an integral over $\pi$,
\begin{equation}
\langle \Psi(\eta_0)\rvert \mathcal{O}(\hat{\phi},\hat{\pi}) \lvert \Psi(\eta_0)\rangle=\int d\bar{\phi} \int d\phi \int d\pi \Psi^*[\phi,\eta_0] \Psi[\bar{\phi},\eta_0]\langle \phi,\eta_0  \rvert  \mathcal{O}(\hat{\phi},\hat{\pi})\lvert \pi,\eta_0\rangle\langle \pi,\eta_0 \lvert \bar{\phi},\eta_0\rangle .
\end{equation}
If we choose to define our operator $\mathcal{O}$ with all $\hat{\phi}$ on the left and all $\hat{\pi}$ on the right (this is not strictly necessary but simplifies the calculation) then we find
\begin{equation}
\langle \Psi(\eta_0)\rvert \mathcal{O}(\hat{\phi},\hat{\pi}) \lvert \Psi(\eta_0)\rangle=\int d\bar{\phi} \int d\phi \int d\pi \Psi^*[\phi,\eta_0] \Psi[\bar{\phi},\eta_0] \frac{1}{2\pi} \exp(i\phi \pi) \mathcal{O}(\phi,\pi) \exp(-i\bar{\phi}\pi) .
\end{equation}
Due to the exponential behaviour it is possible to rewrite the $\pi$ dependence in $\mathcal{O}(\phi,\pi)$ in terms of $\phi$ derivatives, integration by parts then moves these derivatives to the wavefunction so that  performing the $ \pi $ integral gives a delta function that makes the $\phi$ integral trivial and we have
\begin{equation}
\langle \Psi(\eta_0)\rvert \mathcal{O}(\hat{\phi}(\eta_0),\hat{\pi}(\eta_0)) \lvert \Psi(\eta_0)\rangle=\int d\bar{\phi}  \Psi^*[\bar{\phi},\eta_0] \mathcal{O}\left(\bar{\phi},-i\frac{\delta}{\delta \bar{\phi}}\right)  \Psi[\bar{\phi},\eta_0]   .
\end{equation}
The expectation values of most interest are those involving products of the field $\hat{\phi}$ at $\eta_0$ for which we have
\begin{equation}
\langle \Omega\rvert \prod_i^n\hat{\phi}_{\textbf{p}_i}(\eta_0) \lvert \Omega \rangle=\int d\bar{\phi} \prod_i^n\bar{\phi}_{\textbf{p}_i} \left\lvert  \Psi[\bar{\phi},\eta_0]  \right\rvert^2.
\end{equation}
To expedite the calculation of these correlation functions it is convenient to express the wavefunction in the form
\begin{equation}
\Psi[\bar{\phi},\eta_0]\propto\exp\left\{-\sum_{n=2}^\infty \frac{1}{n!}\int \left[ \prod_i^n \bar{\phi}_{\textbf{k}_i} \frac{d^3\bfk_i}{(2\pi)^3}  \right]\,\psi_n'(\textbf{k}_i) \,\delta_{D}^{3}\left(\sum_i^n\textbf{k}_i\right)\right\}
\end{equation}
where the $\psi_n'$ have been defined to be symmetric in their arguments. From this we have, to leading order,
\begin{align}\label{eq:B4}
\langle\phi_{\textbf{p}}(\eta_0)\phi_{-\textbf{p}}(\eta_0)\rangle'&=\frac{1}{2\Re \psi_2'(p)} , \\
\label{phi3}
\left\langle \prod_{a=1}^{3}\phi_{\textbf{p}_a}(\eta_0) \right \rangle'&=-2\prod\limits_{a=1}^3\frac{1}{2\Re \psi_2'(p_{a})  }\Re\left\lbrace\psi_3'(\textbf{p}_1,\textbf{p}_2,\textbf{p}_3)\right\rbrace,\\ 
\left \langle\prod_{a=1}^{4}\phi_{\textbf{p}_a}(\eta_0) \right \rangle'&=-2\prod\limits_{a=1}^4\frac{1}{2\Re \psi_2'(p_{a})} \left[\vphantom{\frac{\Re}{\Re}}\Re\left\lbrace\psi_4'(\textbf{p}_1,\textbf{p}_2,\textbf{p}_3,\textbf{p}_4)\right\rbrace  \nonumber\right. \\ 
\label{phi4}
& \left.  -\frac{\Re\left\lbrace \psi_3'(\textbf{p}_1,\textbf{p}_2,-\textbf{s})\right\rbrace\Re\left\lbrace \psi_3'(\textbf{p}_3,\textbf{p}_4,\textbf{s})\right\rbrace}{\Re \psi_2'(s)}-t-u\right].
\end{align}
In order to expand the wavefunction in this way it is first necessary to perform the $\pi$ integral. For non-derivative interactions (it is suggested in \cite{chen2017schwinger} that this holds for all interactions), the integral gives
\begin{equation}
\Psi[\bar{\phi};\eta_0]=\mathcal{N}\lim_{\eta\rightarrow -\infty}\!\!\!\!\int\limits_{\phi(\eta_0)=\bar{\phi}}\!\!\!\!\mathcal{D}\phi e^{iS[\phi]}.
\end{equation}
This is dominated by the contribution from the extremum of the classical action
\begin{equation}
\Psi[\bar{\phi};\eta_0]\approx\mathcal{N}e^{iS[\phi_{\text{cl}}(\bar{\phi})]},
\end{equation}
which means that these wavefunction coefficients are given by
\begin{equation}
iS[\phi_{\text{cl}}(\bar{\phi})]=-\sum_{n=2}^\infty \frac{1}{n!}\int \left[ \prod_i^n \bar{\phi}_{\textbf{k}_i} \frac{d^3\bfk_i}{(2\pi)^3}\right]  \psi_n'(\textbf{k}_i) \,\delta^{3}_{D}\left(\sum_i^n\textbf{k}_i\right).
\end{equation}

To determine the coefficients of the wavefunction of the universe it is necessary to find the classical solution that extremises this action, i.e. that solves the Euler-Lagrange equations,
\begin{equation}\label{eq:EL}
\mathcal{O}(\textbf{x},\eta)\phi \equiv\frac{\partial }{\partial \eta}\frac{\delta \mathcal{L}_2}{\delta{\phi}'}-\frac{\delta\mathcal{L}_2}{\delta\phi}=-\frac{\partial }{\partial \eta}\frac{\delta \mathcal{L}_{\text{int}}}{\delta{\phi}'}+\frac{\delta\mathcal{L}_{\text{int}}}{\delta\phi}.
\end{equation}
Where we have split the Lagrangian into a quadratic term, $\mathcal{L}_2$ and a, small, interaction term, $\mathcal{L}_{\text{int}}$. Whilst we also require it to satisfy the boundary conditions that 
\begin{equation}
\phi_{\text{cl}}(\eta_0)=\bar{\phi}, \quad \lim\limits_{\eta\rightarrow-\infty}\phi_{\text{cl}}(\eta)=0.
\end{equation}
We solve this using the method of Greens functions and so we seek the solutions to
\begin{equation}
\mathcal{O}(\textbf{x},\eta)K(\textbf{x},\textbf{y},\eta)=0,\quad \lim_{\eta\rightarrow \eta_0} K(\textbf{x},\textbf{y},\eta)=\delta(\textbf{x}-\textbf{y}),\quad \lim_{\eta \rightarrow -\infty}K(\textbf{x},\textbf{y},\eta)=0, \ \textrm{and}
\end{equation}
\begin{equation}
\mathcal{O}(\textbf{x}_1,\eta_1)G(\textbf{x}_1,\eta_1,\textbf{x}_2,\eta_2)=\delta(\eta_1-\eta_2)\delta^3(\textbf{x}_1-\textbf{x}_2),\quad \lim_{\eta_{1,2}\rightarrow \eta_0,-\infty} G(\textbf{x}_1,\eta_1,\textbf{x}_2,\eta_2)=0.
\end{equation}
From these we can formally construct the field $\phi_{\text{cl}}$ from
\begin{equation}
\phi_{\text{cl}}(\textbf{x},\eta)=\int d^3y K(\textbf{x},\textbf{y},\eta)\bar{\phi}(\textbf{y})+\int d^4x' \left.\left(-\frac{\partial }{\partial \eta }\frac{\delta \mathcal{L}_{\text{int}}}{\delta{\phi}'}+\frac{\delta\mathcal{L}_{\text{int}}}{\delta\phi}\right) \right\rvert_{\phi_{\text{cl}}(\textbf{x}',\eta')} G(\textbf{x},\eta,\textbf{x}',\eta').
\end{equation}
However, to make useful progress it is necessary to expand this second term perturbatively so we define
\begin{equation}
\phi_{\text{cl}}^{(0)}(\textbf{x},\eta)=\int d^3y K(\textbf{x},\textbf{y},\eta)\bar{\phi}(\textbf{y}),
\end{equation}
which lets us write,
\begin{align}
\phi_{\text{cl}}(\textbf{x},\eta)&\approx\phi_{\text{cl}}^{(0)}(\textbf{x},\eta)+\int d^4x' \left. \left(-\frac{\partial }{\partial \eta}\frac{\delta \mathcal{L}_{\text{int}}}{\delta{\phi}'}+\frac{\delta\mathcal{L}_{\text{int}}}{\delta\phi}\right)\right\rvert_{\phi^{(0)}_{\text{cl}}(\textbf{x}',\eta')} G(\textbf{x},\eta,\textbf{x}',\eta') \\
&=\phi_{\text{cl}}^{(0)}(\textbf{x},\eta)+\delta\phi(\textbf{x},\eta).
\end{align}
As we are ultimately interested in the Fourier coefficients of the action it is constructive to Fourier transform this expression, 
\begin{equation}
\phi_{\text{cl}}^{(0)}(\textbf{k},\eta)=K(\textbf{k},\eta)\bar{\phi}_{\textbf{k}},\quad \lim_{\eta \rightarrow \eta_0} K(\textbf{k},\eta)=1,\quad \lim_{\eta\rightarrow-\infty}K(\textbf{k},\eta)=0.
\end{equation}
We define $\phi^\pm(\textbf{k},\eta)$ to be the two linearly-independent solutions to
\begin{equation}
\mathcal{O}(\textbf{k},\eta)\phi(\textbf{k},\eta)=0 \,,
\end{equation}
where the $+$ corresponds to positive energy solutions and the $-$ to negative energy ones. The boundary conditions on $K(\textbf{k},\eta)$ fix it to be
\begin{equation}
K(\textbf{k},\eta)=\frac{\phi^+(\textbf{k},\eta)}{\phi^+(\textbf{k},\eta_0)}.
\end{equation}
If we then look at the Fourier transform of $\delta\phi$, after integrating the first term by parts, we find
\begin{equation}
\delta\phi_{\textbf{k}}(\eta)=\int d\eta'\left. \frac{\delta \mathcal{L}_{\text{int}}}{\delta{\phi}'}\right\rvert_{\phi_{\text{cl}}^{(0)}(\textbf{k},\eta')}\frac{\partial}{\partial \eta'}G(\textbf{k},\eta,\eta')+\left.\frac{\delta\mathcal{L}_{\text{int}}}{\delta\phi}\right\rvert_{\phi_{\text{cl}}^{(0)}(\textbf{k},\eta')} G(\textbf{k},\eta,\eta')\,.
\end{equation}
Where the Fourier Transform of the Greens function is defined to satisfy
\begin{align}
\mathcal{O}(\textbf{k},\eta)G(\textbf{k},\eta,\eta')&=\delta(\eta-\eta'), & \lim_{\eta_{1,2}\rightarrow \eta_0} G(\textbf{k},\eta,\eta')&=0,& \lim_{\eta_{1,2}\rightarrow -\infty} G(\textbf{k},\eta,\eta')&=0.
\end{align}
This has solutions given by
\begin{equation}
G(\textbf{k},\eta,\eta')=\begin{cases}\frac{\phi_1(\textbf{k},\eta')\phi_2(\textbf{k},\eta)}{W(\phi_1,\phi_2)(\eta')},\ \eta'\leq \eta \\\frac{\phi_1(\textbf{k},\eta)\phi_2(\textbf{k},\eta')}{W(\phi_1,\phi_2)(\eta')},\ \eta\leq \eta'.\end{cases}
\end{equation}
Where $\phi_1$ satisfies the boundary condition at $-\infty$ and $\phi_2$ satisfies the $\eta_0$ condition,
\begin{align}
\lim_{\eta\rightarrow -\infty(1-i\epsilon)} \phi_1(\textbf{k},\eta)&=0 && \Rightarrow &\quad \phi_1(\textbf{k},\eta)&=\phi^+(\textbf{k},\eta) \,,\\
\phi_2(\textbf{k},\eta_0)&=0&&\Rightarrow& \phi_2(\textbf{k},\eta)&=\phi^+(\textbf{k},\eta)-\frac{\phi^+(\textbf{k},\eta_0)}{\phi^-(\textbf{k},\eta_0)}\phi^-(\textbf{k},\eta).
\end{align}
The Wronskian is defined by
\begin{equation}
W(\phi_1,\phi_2)=a^3\phi_1\dot{\phi}_2-a^3\phi_2\dot{\phi}_1=-a^3 \frac{\phi^+(\textbf{k},\eta_0)}{\phi^-(\textbf{k},\eta_0)}\left( \phi^+(\textbf{k},\eta)\dot{\phi}^-(\textbf{k},\eta)-\dot{\phi}^+(\textbf{k},\eta)\phi^-(\textbf{k},\eta)\right).
\end{equation}
In order to fully specify the vacuum state we have already fixed the Wronskian of our positive and negative frequency solutions, $W(\phi^+,\phi^-)=-i$, (which is a constant as a consequence of the form of the background equation and the fact that $\phi^+$ and $\phi^-$ are a fundamental pair of solutions) so this expression simplifies to
\begin{equation}
W(\phi_1,\phi_2)=i\frac{\phi^+(\textbf{k},\eta_0)}{\phi^-(\textbf{k},\eta_0)}\,,
\end{equation}
The Greens function is
\begin{multline}
G(\textbf{k},\eta,\eta')=\\ i\left(\theta(\eta-\eta') \phi^+(\textbf{k},\eta')\phi^-(\textbf{k},\eta)+\theta(\eta'-\eta) \phi^+(\textbf{k},\eta) \phi^-(\textbf{k},\eta')- \frac{\phi^-(\textbf{k},\eta_0)}{\phi^+(\textbf{k},\eta_0)}\phi^+(\textbf{k},\eta') \phi^+(\textbf{k},\eta)\right).
\end{multline}
Having defined these terms, we can expand $iS[\phi_{\text{cl}}(\bar{\phi})]$ to linear order $\delta\phi$ (and hence second order in $\mathcal{L}_{\text{int}}$). Performing any $\eta$ integrals will then allow us to calculate the coefficients $\psi_n'$. The evaluation of $S$ in this way can equivalently be defined diagrammatically where the function $K(\textbf{k},\eta)$ is called the bulk-to-boundary propagator and $G(\textbf{k},\eta,\eta')$ is the bulk-to-bulk propagator because any lines connecting the vertex at $\eta$ to a boundary contribute $K(\textbf{k},\eta)$ whilst internal lines, which connect vertices at $\eta$ and $\eta'$ together, contribute a factor of $G(\textbf{k},\eta,\eta')$. 

We also note an important nuance in the evaluation of the action to second order. Whilst the quadratic action will, by definition, receive no correction at first order in $\delta\phi$ there is a second order correction that arises due to the introduction of the interaction term
\begin{equation}
iS[\phi_{\text{cl}}(\bar{\phi})]=iS_2[\phi_{\text{cl}}(\bar{\phi})]+iS_{\text{int}}[\phi_{\text{cl}}(\bar{\phi})]+iS_2[\delta\phi].
\end{equation}
Expanding this final term gives
\begin{equation}
S_2[\delta\phi]=\int d\eta \mathcal{L}_2[\delta\phi]=\int d\eta\frac{1}{2}\left.\frac{\delta\mathcal{L}_2}{\delta {\phi}'}\right\rvert_{\delta\phi_{\textbf{k}}}\delta{\phi}'_{-\textbf{k}}+\frac{1}{2}\left.\frac{\delta\mathcal{L}_2}{\delta\phi}\right\rvert_{\delta\phi_{\textbf{k}}}\delta\phi_{-\textbf{k}}=-\frac{1}{2}\int d\eta \delta\phi_{-\textbf{k}} \mathcal{O}(\textbf{k},\eta) \delta\phi_{\textbf{k}}
\end{equation}
Where the second equality follows due to $S_2$ being quadratic in the fields and the final equality results from integrating by parts and the definition of $\mathcal{O}$ in  \ref{eq:EL}. We can then use our equations of motion to expand this as
\begin{equation}
S_2[\delta\phi]=\frac{1}{2}\int d\eta\int d\eta' \left(\left.\frac{\partial}{\partial\eta}\frac{\delta\mathcal{L}_{\text{int}}}{\delta\phi'}-\frac{\delta\mathcal{L}_{\text{int}}}{\delta\phi}\right)\right\rvert_{\phi_{\text{cl}}^{(0)}(\textbf{k},\eta)}\delta\phi_{-\textbf{k}}
\end{equation}
This is precisely equal to minus half the quadratic order term that we find from the expansion of the interaction term,
\begin{equation}
S_{\text{int}}\left[\phi_{\text{cl}}(\bar{\phi})\right]=\int d\eta \mathcal{L}_{\text{int}}\left[\phi_{\text{cl}}^{(0)}\right]+\left(\left.-\frac{\partial}{\partial\eta}\frac{\delta\mathcal{L}_{\text{int}}}{\delta\phi'}+\frac{\delta\mathcal{L}_{\text{int}}}{\delta\phi}\right)\right\rvert_{\phi_{\text{cl}}^{(0)}(\textbf{k},\eta)} \delta\phi_{-\textbf{k}}
\end{equation}
Therefore, in our diagrammatic computations we can account for this additional term by introducing a factor of one half for exchange diagrams. 

\section{Details of the computation of $\psi_3^{\vpi\vpi\sigma}$}
\label{f3c}
In this appendix, we elaborate on the computation of the three point function $\psi_3^{\vpi\vpi\sigma}$ needed for  \eqref{f3}. Let us first explicitly write the associated time integral, 
\begin{align}
\psi_3^{\vpi\vpi\sigma} &= -2i \int a(\eta)^4\,d\eta\, K_\vpi(k_1,\eta)\, K_\vpi(k_2,\eta)\, K_\sigma(k_3,\eta)\,\\ \nonumber
&=\dfrac{-\pi\,2^{\frac{5}{2}-\Delta}}{H^4\,\Gamma[-\frac{3}{2}+\Delta]}\, (-\eta_0)^{-5+\Delta}\,\, k_3^{-2+\Delta} \,\,{\cal I}(p)\,, \\ \nonumber
{\cal I}(p)&\equiv \int _{-\infty}^0\,\dfrac{dx}{\sqrt{-x}}\,\exp(i p\,x) H^{(2)}_{\nu}(-x)
\end{align}
where it is seen that $k_3^{2-\Delta}\,\psi_3$ depends only on the following ratio, 
\be
p\equiv \dfrac{k_1+k_2}{k_3}\,.
\ee
By direct inspection, or by using the dS conformal symmetry \cite{Arkani-Hamed:2015bza}, one can show that 
\be
(p^2-1)\partial_p^2\,{\cal I}(p)+2p\partial_p\,{\cal I}(p)+\left( \frac{1}{4}-\nu^2 \right){\cal I}(p)=0\,.
\ee
The solution that is regular at $p=1$ (i.e. the collapsed limit) is given by, 
\be
{\cal I}(p)\propto  {}_2F_1 \left[\frac{1}{2}-\nu,\dfrac{1}{2}+\nu\,,1\,, \dfrac{1-p}{2}\right]\,.
\ee
The proportionally factor is crucial for checking the optical theorem we discovered for contact terms \eqref{unicont}. This factor can be easily found by looking at the total energy singularity of the three point function, i.e. 
\begin{align}
\lim_{p\to -1}\psi_3^{\vpi\vpi\sigma} &=-\dfrac{\sqrt{2}}{\sqrt{k_3}}\dfrac{1}{H^3\,\eta_0^2\,\sigma^+(k_3,\eta_0)}\ln (1+p)\,\\ \nonumber
&=\dfrac{\sqrt{\pi}\,2^{3-\Delta}}{H^4\, \Gamma[-\frac{3}{2}+\Delta]}\,k_3^{-2+\Delta}\, (-\eta_0)^{-5+\Delta}\, \exp\left[ \frac{i\,\pi}{2} \left( \nu+\frac{3}{2} \right) \right] \ln (1+p)\,.
\end{align}
On the other hand, 
\be
\lim_{p\to -1}\, _2F_1 \left[\frac{1}{2}-\nu,\dfrac{1}{2}+\nu\,,1\,, \dfrac{1-p}{2} \right]=-\dfrac{\ln (1+p)}{\Gamma[2-\Delta]\,\Gamma[-1+\Delta]}\,.
\ee
Finally arrived at the desired result, 
\begin{align}
\psi_3^{\vpi\vpi\sigma} &= -\dfrac{\sqrt{\pi}\,2^{3-\Delta}}{H^4}\dfrac{\Gamma[2-\Delta]\Gamma[-1+\Delta]}{\Gamma[-\frac{3}{2}+\Delta]} \exp\left[ \frac{i\pi}{2}\left( \nu+\frac{1}{2} \right)\right]\\ \nonumber 
&\times (-\eta_0)^{-5+\Delta}\,\, k_3^{-2+\Delta}\,_2F_1 \left[\frac{1}{2}-\nu,\dfrac{1}{2}+\nu\,,1\,, \dfrac{1-p}{2}\right]\,,
\end{align}
from which one infers   \eqref{f3}.


\section{Generalization of the COT Derivation} 
\label{gencopt}
In this Appendix, we relax some of the assumptions that we made for the derivation of the COT for exchange diagrams. First, we allow $\sigma$ to have any number of time derivatives, and second, we let the exchanging particle be identical to $\phi$. 
Instead of going through all the same steps taken in Section \ref{fourpoint}, i.e. starting with Eq. \eqref{unitexch} all the way to reach Eq. \eqref{exchCOPT}, here we rather short-circuit the complications of the original derivation by directly proving the conjectured COT. Regardless, the interpretation of the final result will stay the same---the COT is a direct implication of unitarity for the boundary observables.


\subsection*{$\sigma$ with one or more time derivatives}
Without loss of generality, we can assume that $\sigma$ carries only a single time derivative. This is because in perturbation theory, using the equations of motion, every operator with higher order time derivatives can be replaced with operators with at most one time and/or multiple spatial derivatives. For example, in flat space, the cubic operator $g\,\phi^2 \ddot{\sigma}$ can be traded for $g \partial_i^2 (\phi)^2 \sigma+g^2 \phi^2 \partial_t^2 (\phi^2)$. The first cubic term is among the vertices that we already accounted for in the proof of Eq. \eqref{exchCOPT}. The second term contributes to $\psi_4$, however, the COT derived for contact terms in Eq. \eqref{concop} enforces this part to cancel from the left hand side of Eq. \eqref{exchCOPT}. Therefore, in the following we assume a single derivative on the intermediate field, i.e. 
\begin{align}
S_{\text{int}} =& \int a^4(\eta)\,d\eta\, \left(\prod_{a=1}^3 \dfrac{d^3\bfk_a}{(2\pi)^3}\right)\,(2\pi)^3\,\delta^3(\bfk_1+\bfk_2+\bfk_3)\\ \nonumber
& \eta^{n+m+1}\,F\left[\dfrac{\bfk_a\cdot \bfk_b}{a^2(\eta)}\right]\,\partial^{n}_\eta \phi(\bfk_1,\eta)\,\partial^{m}_\eta \phi(\bfk_2,\eta)\sigma'(\bfk_3,\eta)\,.
\end{align}
The powers of $\eta$ are inserted to maintain the dilation symmetry, and the factor $F[\dots]$ accounts for possible spatial derivatives. Nevertheless, the following argument will straightforwardly generalize to flat space. Using integration by parts, the time derivative on $\sigma$ can be removed, at the cost of introducing the following boundary term, 
\begin{align}
S_{\text{boundary}}=& \int \, \left(\prod_{a=1}^3 \dfrac{d^3\bfk_a}{(2\pi)^3}\right)\,a^4(\eta_0)\,(2\pi)^3\,\delta^3(\bfk_1+\bfk_2+\bfk_3)\\ \nonumber
& \eta_0^{n+m+1}\,F_A\left[\dfrac{\bfk_a\cdot \bfk_b}{a^2(\eta_0)}\right]\, \partial^{n}_\eta \phi(\bfk_1,\eta)\rvert_{\eta_0}\,\partial^{m}_\eta \phi(\bfk_2,\eta)\rvert_{\eta_0}\,\sigma (\bfk_3,\eta_0)\,.
\end{align}
This boundary term does not influence the equations of motion but it can take part in the wavefunction of the universe $\Psi=\exp(i S_{\text{cl}}[\bar{\phi}(\bfk),\bar{\sigma}(\bfk)])$ through the three point function, i.e. 
\begin{align}
\psi'^{\phi\phi\sigma}_3(k_1,k_2,k_3) &\supset - i \, a^4(\eta_0)\,F_A\left[\dfrac{\bfk_a\cdot \bfk_b}{a^2(\eta_0)}\right]\,\eta_0^{n+m+1}\,\partial^{n}_\eta K_\phi(k_1,\eta)\rvert_{\eta_0}\,\partial^{m}_\eta K_\phi(k_2,\eta)\rvert_{\eta_0}+(1\leftrightarrow 2)\,,
\end{align}
where we have used $K_\sigma(k,\eta_0)=1$\footnote{Notice that in the de Sitter case, unless $n+m=1$ and $F=1$, this boundary term vanishes in the $\eta_0\to 0$ limit, so practically the only operator that the following discussion is needed for is $\phi \dot{\phi}\dot{\sigma}$. For this operator the total $\psi_3$ and $\psi_4$ are IR finite, although the boundary term above is IR infinite. Conversely, in flat space these boundary terms are generically non-vanishing and one needs the presented argument.}.
However, the contribution on the right hand side does not depend on the third momentum $k_3$ except through possible factors of  $\bfk_a\cdot\bfk_b$. Subsequently, these inner products can be written in terms of $k_1^2, k_2^2$ and $k_3^2$ using $\bfk_1+\bfk_2+\bfk_3=0$. As a result, the above boundary contribution cancels from the right hand side of the optical theorem Eq. \eqref{exchCOPT}, as it is invariant under the flipping of the sign of the third energy $k_3$. On the other hand, the appearance of this boundary term makes no difference in $\psi_4$. Consequently, our final COT equally applies to cubic interactions with arbitrary number of derivatives on $\sigma$.

\subsection*{$\sigma$ is the same field as $\phi$}
Without loss of generality, we again consider interactions with at most one time derivative per field, i.e. 
\begin{align}
\label{vertices}
S_{\text{int}} =g\,& \int a^4(\eta)\,d\eta\, \left(\prod_{a=1}^3 \dfrac{d^3\bfk_a}{(2\pi)^3}\right)\,(2\pi)^3\,\delta^3(\bfk_1+\bfk_2+\bfk_3)\\ \nonumber
& \eta^{s_1+s_2+s_3}\,F\left[\dfrac{\bfk_a\cdot \bfk_b}{a^2(\eta)}\right]\,\partial^{s_1}_\eta \phi(\bfk_1,\eta)\,\partial^{s_2}_\eta \phi(\bfk_2,\eta)\,\partial^{s_3}_\eta \phi(\bfk_3,\eta)\,,
\end{align}
with $s_a=0, 1$. For simplicity, henceforth we set $s_1=s_2=s_3=1$, but the computation will easily generalize to other cases. In this case we replace the factor $F\left[\dfrac{\bfk_a\cdot \bfk_b}{a^2(\eta)}\right]$ with the fully symmetric function, 
\be
\bar{F}[\bfk_1, \bfk_2,\bfk_3, \eta]\equiv \dfrac{1}{3!} (F\left[\dfrac{\bfk_a\cdot \bfk_b}{a^2(\eta)}\right]+\text{permutations})\,.
\ee
E.g. for $\dot{\phi}(\partial_i \phi)^2$ operator, 
\be
\bar{F}[\bfk_1, \bfk_2,\bfk_3, \eta]=\dfrac{1}{3 a^2}\left(\bfk_1\cdot\bfk_2+\bfk_2\cdot\bfk_3+\bfk_1\cdot\bfk_3\right)\,.
\ee

We wish to compute the four point function generated via the exchange of $\phi$ in a diagram where the two vertices are the same as \eqref{vertices}. 
In the wavefunction approach, up to linear order in $g$, the solution to $\phi_{\text{cl}}(\bfk,\eta)$ is given by, 
\begin{align}
\phi_{\text{cl}}(\bfk_1,\eta)&=K_\phi (k_1,\eta)\,\bar{\phi}(\bfk_1)\\ \nonumber
&+3\,g \int d\eta'\,G(k,\eta,\eta')\, (2\pi)^3\,\delta^3(\bfk_1-\bfk_2-\bfk_3)\,\\ \nonumber
& \partial_{\eta'}\left(a^4(\eta')\eta'^3\,\bar{F}(-\bfk_1,\bfk_2,\bfk_3,\eta')\,K'(k_2,\eta')\,K'(k_3,\eta')\right)\,\bar{\phi}(\bfk_2)\,\bar{\phi}(\bfk_3)\,.
\end{align}
Therefore, we find, 
\begin{align}
\psi'^s_4 (k_a,s) &=-36\,i\,g^2\int d\eta\,d\eta'\, G (s,\eta,\eta')\,\\ \nonumber
&  \partial_{\eta'}\left(a^4(\eta')\eta'^3\,\bar{F}(\bfk_1+\bfk_2,\bfk_3,\bfk_4,\eta')\,K'(k_3,\eta')\,K'(k_4,\eta')\right)\\ \nonumber
& \times  \partial_{\eta}\left(a^4(\eta)\eta^3\,\bar{F}(-\bfk_1-\bfk_2,\bfk_1,\bfk_2,\eta)\,K'(k_1,\eta)\,K'(k_2,\eta)\right)\,.
\end{align}
On the other hand, the cubic wavefunction coefficient is given by, 
\begin{align}
\psi'^{\phi}_3 &=-6\,i\,g\,\int d\eta\, a^4(\eta)\eta^3\,\bar{F}(\bfk_1,\bfk_2,\bfk_3,\eta)\,K'(k_1,\eta)\,K'(k_2,\eta)\,K'(k_3,\eta)\,.
\end{align}
Now, by employing the following property of the bulk-to-bulk propagator, 
\begin{align}
& G(s,\eta,\eta')+G^*(s,\eta,\eta')=\\ \nonumber
& i \left[\phi^+(s,\eta)\phi^-(k,\eta')+\phi^-(s,\eta)\phi^+(k,\eta')-\phi^-(s,\eta)\phi^-(k,\eta')-\phi^+(s,\eta)\phi^+(k,\eta')\right]
\end{align}
and making two integration by parts, we arrive at
\begin{align}
\label{exchp}
& \psi'^s_4 (k_a,s)+\psi'^s_4 (-k_a,s)=\\ \nonumber
& P(k)\, \left[\psi'^\phi_3(k_1,k_2,s)-\psi'^\phi_3(k_1,k_2,-s)\right]\,\left[\psi'^\phi_3(k_3,k_4,s)-\psi'^\phi_3(k_3,k_4,-s)\right]. 
\end{align}
Notice that in above the boundary terms that appear after integration by parts on $\eta$ and $\eta'$ cancel against each other due to\footnote{In this specific example, each of the individual boundary terms accidentally vanish in the $\eta_0\to 0$ limit. Yet, the argument is useful when this is not the case, e.g. in flat space.},
\be
\lim_{\eta_0\to 0}\sigma^+(s,\eta_0)=\lim_{\eta_0\to 0}\sigma^-(s,\eta_0)=\text{cst}.
\ee

\section{Additional Details on Polology}

In this appendix we lay out all the steps in the evaluation of the integrals in Section \ref{sec:polo} to allow the reader that is less familiar with performing these integrals to follow the logic required to reach our result and to make clear the methods used for anyone who wishes to check them. We also present a discussion of the generalisation of these results to diagrams with loops.


\subsection*{Amplitudes}
\label{amp}
Starting from the integral form of our rules for calculating amplitudes for diagrams without loops,
\begin{equation}
A_n=(2\pi)^3\delta^3\left(\sum_a^n\textbf{k}_a\right) i^{-V-1}\int\prod_\alpha^V g_\alpha F_\alpha(\textbf{k}) dt_\alpha \prod_\beta^I G_{\text{flat}}(K_\beta;t,t') \prod_a^nG_{B,\text{flat}}(k_a,t).
\end{equation}
As there are no loops, it will always be possible to choose a vertex that is connected to exactly one other vertex. This is because if there is no vertex with degree less than 2 then $\sum\limits_{v \in V}\textrm{deg}\  v \geq 2 V$ but by  Euler's formula, $V=I+1$ this means that $\sum\limits_{v \in V}\textrm{deg} \ v \geq 2I+2$ which violates the hand shaking lemma, $\sum\limits_{v\in V} \textrm{deg}\  v=2I$. Therefore, we can integrate out this vertex, which we can chose to label as vertex $V$ and which has some number, $m$, of external lines. To do this we first expand the integral over this vertex to give
\begin{multline}
A_n=(2\pi)^3\delta^3\left(\sum_a^n\textbf{k}_a\right)i^{-V-1}\int g_VF_V\prod_\alpha^{V-1} g_\alpha F_\alpha dt_\alpha\prod_\beta^{I-1} G_{\text{flat}}(K_\beta;t,t') \prod_a^{n-m}G_{B,\text{flat}}(k_a,t)\\ \prod_b^m f^*_{s_b}(k_b) \left(\int_{t_{V-1}}^\infty dt_V \frac{f_{s_{V-1}}^*(K_I)f_{s_V}(K_I)}{2K_I}e^{-iK(t_{V-1}-t_V)-i\sum\limits_b^m k_b t_V }+\right.\\\left.\int_{-\infty}^{t_{V-1}}dt_V \frac{f_{s_{V-1}}(K_I)f_{s_V}^*(K_I)}{2K_I}e^{iK(t_{V-1}-t_V)-i\sum\limits_b^m k_b t_V}	\right).
\end{multline}
Where $\textbf{K}_I=-\sum\limits_b^m\textbf{k}_b$ is the momentum of the internal line. These two integrals are straight forward and we find that
\begin{multline}
A_n=(2\pi)^3\delta^3\left(\sum_a^n\textbf{k}_a\right)i^{-V}\int g_VF_V\prod_\alpha^{V-1} g_\alpha F_\alpha dt_\alpha\prod_\beta^{I-1} G_{\text{flat}}(K_\beta;t,t') \prod_a^{n-m}G_{B,\text{flat}}(k_a,t) \\\prod_b^m f^*_{s_b}(k_b)\left(\frac{f_{s_{V-1}}^*(K_I)f_{s_{V}}(K_I)}{2K}\frac{1}{K_I-\sum\limits_b^m k_b}+ \frac{f_{s_{V-1}}(K_I)f_{s_V}^*(K_I)}{2K_I}\frac{1}{K_I+\sum\limits_b^m k_b}\right)e^{-i\sum\limits_b^m k_b t_{V-1}}.
\end{multline}
We can then reabsorb the new exponentials into bulk-to-boundary propagators at $t_{V-1}$ so that
\begin{multline}
A_n=\\ (2\pi)^3\delta^3\left(\sum_a^n\textbf{k}_a\right) i^{-(V-1)-1}\int g_VF_V\prod_\alpha^{V-1} g_\alpha F_\alpha dt_\alpha \prod_\beta^{I-1} G_{\text{flat}}(K_\beta;t,t') \prod_a^{n}G_{B,\text{flat}}(k_a,t)\mathcal{G}_I(K_I)
\end{multline}
where
\begin{equation}
\mathcal{G}_I(K_I)=\frac{f_{s_{V-1}}^*(K_I)f_{s_{V}}(K_I)}{2K_I}\frac{1}{K_I-\sum\limits_b^m k_b}+ \frac{f_{s_{V-1}}(K_I)f_{s_V}^*(K_I)}{2K_I}\frac{1}{K_I+\sum\limits_b^m k_b}.
\end{equation}
Therefore, it is possible to relate the amplitude for a diagram with $V$ vertices to one with $V-1$ vertices. In this way one can remove all internal lines, leaving just one vertex. This final time integral is just that from a contact term,
\begin{multline}
A_n=-(2\pi)^3\delta^3\left(\sum_a^n\textbf{k}_a\right)\prod_\alpha^V g_\alpha F_\alpha(\textbf{k})\prod_\beta^I \mathcal{G}_\beta(K_\beta)\prod_a^n f_{s_a}^*(k_a) \int_{-\infty}^\infty dt e^{-iE_T t}=\\-(2\pi)^4\delta^4\left(\sum_a^n k_a^\mu\right)\prod_\alpha^V g_\alpha F_\alpha (\textbf{k})\prod_\beta^I\mathcal{G}_\beta(K_\beta)\prod_a^n f_{s_a}^*(k_a),
\end{multline}
so we see that this final integral will enforce overall energy conservation, recovering the expected results from a standard set of Feynman rules. Using the primed notation defined previously we find that the amplitude is
\begin{equation}
A_n'=-\prod_\alpha^V g_\alpha F_\alpha (\textbf{k})\prod_\beta^I \mathcal{G}_\beta(K_\beta)\prod_a^n f_{s_a}^*(k_a).
\end{equation}


\subsection*{Correlators}
\label{corr}
Just as for the amplitude we start with the integral form of our rules in  \ref{eq:contactBN},
\begin{equation}
B_n^+= -i\int_{-\infty}^{\eta_0} \frac{H^{N+2n-4}}{\eta^4}(-\eta)^{R}g\tilde{F}(\textbf{k}) d\eta \prod_a^n\frac{f^*_{s_a}(k_a)}{2k_a^3}(-\eta)^{s_a}(1-s_a+ik_a\eta)(1-ik_a\eta_0)e^{-ik_a(\eta-\eta_0)}.
\end{equation}
Which is simplified by combining the product of $\eta$'s into a sum over powers of $\eta$,
\begin{equation}
(-\eta)^R\prod_a^n(-\eta)^{s_a}(1-s_a+ik_a\eta)=\sum_b^nC_b(k_a)\eta^{b+N},
\end{equation}
the integral is then
\begin{equation}
B_n^+=-iH^{N+2n-4}g\tilde{F}(\textbf{k})\prod_a^n\frac{f_{s_a}^*(k_a)}{2k_a^3}(1-ik_a\eta_0)\sum_b^n C_b(k_a)\sum_{c=0}^{N+b-4}\frac{(-1)^{c+b+N} \eta_0^c (N+b-4)!}{(-ik_T)^{N+b-3-c}c!}.
\end{equation}
The worst total energy pole is therefore when $b=n$ and $c=0$,
\begin{equation}
\lim_{k_T\rightarrow 0} B_n^+=-iH^{N+2n-4}g\tilde{F}(\textbf{k})\prod_a^n\frac{f_{s_a}^*(k_a)}{2k_a^3}(1-ik_a\eta_0) C_n(k_a)\frac{(-1)^{n+N} (N+n-4)!}{(-ik_T)^{N+n-3}},
\end{equation}
where the coefficient $C_n(k_a)$ is the contribution from the highest power of $\eta$ in  \ref{eq:contactBN}, so it comes from the product of the $ik_a\eta$ terms, i.e. $C_n=(-1)^{N}\prod_a^n ik_a$. Putting this into the expression for $B_n^+$ then gives  \ref{eq:BN+}
\begin{equation}
\lim_{k_T\rightarrow 0} B_n^+=\\(-1)^{n+1}ig\tilde{F}(\textbf{k})\prod_a^n\frac{(i+k_a\eta_0) f_{s_a}^*(k_a)}{2k_a^2} \frac{(N+n-4)!}{(-ik_T)^{N+n-3}},
\end{equation}
enabling us to relate contact correlators to amplitudes. Just as for the amplitude the hand shaking lemma guarantees that we will be able to find a vertex connected to only one other in order to connect tree-level diagrams so we can write the correlator as
\begin{multline}
B_n^+= (-1)^{s_{V-1}+s_V+\sum\limits_a^m s_a+R_V}i^{-V}H^{2I+2n+\sum\limits_\alpha^V(N_\alpha-4)}g_V\tilde{F}_V(\textbf{k})\int\prod_\alpha^{V-1}\frac{ g_\alpha \tilde{F}_\alpha (\textbf{k})}{a^{R_\alpha }}a^4 d\eta_a \\ \prod_\beta^{I-1} G_{++}(K_\beta;\eta,\eta') \prod_a^{n-m}G_{B+}(k_a,\eta)\prod_b^m\frac{f_{s_b}^*(k_b)}{2k_b^3}(1-ik_b\eta_0)\eta_V^{s_b}(1-s_b+ik_b\eta_V)\eta_{V-1}^{s_{V-1}}\\ \left(\int_{\eta_{V-1}}^{\eta_0} d\eta_V \frac{f_{s_{V-1}}^*(K_I)f_{s_V}(K_I)}{2K_I^3}\eta_V^{s_V}(1-s_{V-1}+iK_I\eta_{V-1})(1-s_V-iK_I\eta_V)e^{-iK_I(\eta_{V-1}-\eta_V)}\right.\\\left.+\int_{-\infty}^{\eta_{V-1}}d\eta_V\frac{f_{s_{V-1}}(K_I)f_{s_{V}}^*(K_I)}{2K_I^3}\eta_V^{s_V}(1-s_{V-1}-iK_I\eta_{V-1})(1-s_V+iK_I\eta_V)e^{iK_I(\eta_{V-1}-\eta_V)}\right)\\	\eta_V^{R_V-4}e^{-i\sum\limits_a^m k_a(\eta_V-\eta_0)}.
\end{multline}
In the same way as before the most singular term will be from the highest power of $\eta_V$. The integral $\int d\eta \eta^n e^{-i K\eta}$ cannot contribute any higher power of $\eta$ than $n$ and we have that in this limit
\begin{multline}\label{eq:BNtotal}
\lim_{k_T\rightarrow 0} B_n^+=(-1)^{N_V} i^{1-V}({N_V+m-3})!H^{2I+2n+\sum\limits_\alpha^V(N_\alpha-4)}  g_V\tilde{F}_V(\textbf{k})\sum_{k=0}^{N_V+m-3}\frac{(-1)^{N_V+m-3+k}}{i^{N_V+m-2-k}k!} \\ \int\prod_\alpha^{V-1}\frac{ g_\alpha \tilde{F}_\alpha(\textbf{k})}{a^{R_\alpha }}a^4 d\eta_\alpha  \prod_\beta^{I-1} G_{++}(K_\beta;\eta,\eta') \prod_a^{n-m}G_{B+}(k_a,\eta)\prod_b^m\frac{i(1-ik_b\eta_0)f_{s_b}^*(k_b)}{2k_b^2}(-\eta_{V-1})^{s_{V-1}}\\  \left( \frac{f_{s_{V-1}}^*(K_I)f_{s_V}(K_I)}{2K_I^2\left(K_I-\sum\limits_a^m k_a\right)^{N_V+m-2-k}} (1-s_{V-1}+iK_I\eta_{V-1}) \left(\eta_{V-1}^k-\eta_0^ke^{i\left(\sum\limits_a^m k_a-K_I \right)(\eta_{V-1}-\eta_0)}\right)\right.\\\left.+\frac{f_{s_{V-1}}(K_I)f_{s_V}^*(K_I)}{2K_I^2 \left(-\sum\limits_a^m k_a-K_I\right)^{N_V+m-2-k} }(1-s_{V-1}-iK_I\eta_{V-1})\eta_{V-1}^k\right)e^{-i\sum\limits_a^m k_a (\eta_{V-1}-\eta_0)}	.
\end{multline}
Whilst taking only the highest power of $\eta_{V-1}$ then gives
\begin{multline}
\lim_{k_T\rightarrow 0} B_n^+=(-1)^{N_V} i^{-V+1} H^{2I+2n+\sum\limits_\alpha^V(N_\alpha-4)} g_V \tilde{F}_V(\textbf{k})\int\prod_\alpha^{V-1}\frac{ g_\alpha \tilde{F}_\alpha (\textbf{k})}{a^{R_\alpha}}a^4 d\eta_\alpha \\\prod_\beta^{I-1} G_{++}(K_\beta;\eta,\eta') \prod_a^{n-m}G_{B+}(k_a,\eta)\prod_b^m\frac{(i+k_b\eta_0)f_{s_b}^*(k_b)}{2k_b^2}(-\eta_{V-1})^{s_{V-1}}e^{-i k_i (\eta_{V-1}-\eta_0)} \\ \eta_{V-1}^{m+N_V-2} \left( \frac{f_{s_{V-1}}^*(K_I)f_{s_V}(K_I)}{2K_I}\frac{1}{K_I-\sum\limits_i^m k_i}+\frac{f_{s_{V-1}}(K_I)f_{s_V}^*(K_I)}{2K_I}\frac{1}{\sum\limits_i^mk_i+K_I} \right).
\end{multline} 
Where we can recognise $\mathcal{G}_I(K_I)$ from  \ref{eq:P(K)}. The remaining steps to relate this to the amplitude are in the main text.


\subsection*{Loops}
\label{loop}
The polology of Amplitudes explores loop diagrams at the level of the loop integrand and so we will only attempt to compare correlators and amplitudes at this level. If we take a diagram and add a propagator connecting two already connected vertices to form a loop then we can remove any vertices that are not part of this loop in the same manner as was established previously, so that we are just left with external lines connected to the loop, this applies to both the amplitude and the correlator, although in the case of the correlator in order to perform this we must have taken only the highest power of time at each vertex, this is justified as we are only interested in the most singular terms which must come from this highest power of time. However, this still leaves the loop itself which we have not yet established how to deal with. Just as before, if this loop contains time and anti-time ordered vertices then the integral will be separable and we can break this loop simply, this leaves a tree level diagram that has both time and anti-time ordered vertices so it cannot contribute to the total energy pole, so we are only interested in totally time ordered diagrams. Therefore, to relate diagrams containing a single loop we must compare the integrals
\begin{align}
&I_A=\int_{-\infty}^{\infty}d\eta G_{\text{flat}}(K_1, \eta_1,\eta)G_{\text{flat}}(K_2,\eta,\eta_2) e^{-i\sum\limits_a^n k_a \eta}\\
&I=\int_{-\infty}^{\eta_0}d\eta \eta^{R+n-4} G_{++}(K_1, \eta_1,\eta)G_{++}(K_2,\eta,\eta_2)\eta^{\sum\limits_\alpha^VD_\alpha-4} e^{-i\sum\limits_a^n k_a (\eta-\eta_0)}.
\end{align}
Where the sum over $a$ is over all the external lines that are now associated with this vertex whilst the sum over $\alpha$ counts the removed vertices. We will also set $H$ to $1$ in this discussion as the proper inclusion of $H$ is just a matter of counting lines and vertices and will be exactly the same as for tree level diagrams. The most singular piece of $I$ will come from the limit that all times are taken to $\infty$ and we need to explore
\begin{equation}
\lim_{\eta_i\rightarrow-\infty}G_{++}(k,\eta_1,\eta_2)=(-1)^{s_1+s_2}\eta_1^{s_1+1}\eta_2^{s_2+1}G_{\text{flat}}(k,\eta_1,\eta_2).
\end{equation}
Therefore, we have that
\begin{multline}\label{eq:Ilim}
\lim_{\eta_i\rightarrow-\infty}I=\\\int_{-\infty}^{\eta_0}d\eta (-1)^{s+s'+s_1+s_2}\eta_1^{s_1+1}\eta_2^{s_2+1}\eta^{\sum\limits_\alpha^{V+1}D_\alpha-4}G_{\text{flat}}(K_1,\eta_1,\eta)G_{\text{flat}}(K_2,\eta,\eta_2)e^{-i\sum\limits_a^n k_a (\eta-\eta_0)}.
\end{multline}
If we then expand this product of propagators so that each term contains only one heaviside theta function that depends on $\eta$ we can see the how this integral would split up and the $k$ dependence on each integral, this applies to both the amplitude and the correlator in this limit, 
\begin{multline}
G_{\text{flat}}(K_1,\eta_1,\eta)G_{\text{flat}}(K_2,\eta,\eta_2)=\\\left[\left(\theta(\eta_1-\eta_2)\theta(\eta-\eta_1)+\theta(\eta_2-\eta_1)\theta(\eta-\eta_2)\right)\frac{f_{s_2}^*(K_2)f_{s'}(K_2)}{2K_2}e^{iK_2(\eta-\eta_2)}\frac{f_{s_1}^*(K_1)f_s(K_1)}{2K_1}e^{iK_1(\eta-\eta_1)}\right.\\\left. +\theta(\eta_1-\eta_2)\left(\theta(\eta-\eta_2)-\theta(\eta-\eta_1)\right)\frac{f_{s_2}^*(K_2)f_{s'}(K_2)}{2K_2}e^{-iK_2(\eta_2-\eta)}\frac{f_{s_1}(K_1)f_s^*(K_1)}{2K_1}e^{-iK_1(\eta-\eta_1)}+\right.\\\left. \theta(\eta_2-\eta_1)\left(\theta(\eta-\eta_1)-\theta(\eta-\eta_2)\right)\frac{f_{s_2}(K_2)f_{s'}^*(K_2)}{2K_2}e^{-iK_2(\eta-\eta_2)}\frac{f_{s_1}^*(K_1)f_s(K_1)}{2K_1}e^{-iK_1(\eta_1-\eta)}+\right.\\\left.\left(\theta(\eta_2-\eta_1)\theta(\eta_1-\eta)+\theta(\eta_1-\eta_2)\theta(\eta_2-\eta)\right)\frac{f_{s_2}(K_2)f_{s'}^*(K_2)}{2K_2}e^{iK_2(\eta_2-\eta)}\frac{f_{s_1}(K_1)f_s^*(K_1)}{2K_1}e^{iK_1(\eta_1-\eta)}\right]\\\!
\end{multline}
Each term in the correlator will look like
\begin{equation}
\int_{\eta_i}^{\eta_f} d\eta \eta^n Q(K_1,K_2)  e^{-i\sum\limits_a^n k_a (\eta-\eta_0)}e^{\pm iK_1(\eta-\eta_1)}e^{\pm iK_2(\eta-\eta_2)},
\end{equation}
where $Q$ is some function that depends on which integral we have taken and $\eta_i,\  \eta_f$ take the values $-\infty,\ \eta_1,\ \eta_2\textrm{ and } \eta_0$ and for each term in the correlator we have an equivalent term in the amplitude except we take  $n\rightarrow 0$  and $\eta_0\rightarrow \infty$ in the upper limit whilst it drops out of the exponential. Performing this integral gives
\begin{multline}
\sum_{k=0}^n Q(K_1,K_2)e^{i\left(\sum\limits_a^n k_a\eta_0\mp K_1 \eta_1\mp K_2\eta_2\right)}\frac{(-1)^nn!}{\left(i\left(\pm K_1 \pm K_2-\sum\limits_a^n k_a\right)\right)^{1+n-k}}\\ \left(e^{i\left(\pm K_1\pm K_2-\sum\limits_a^n k_a\right)\eta_f}\frac{\left(-\eta_f\right)^k}{k!}-e^{i\left(\pm K_1\pm K_2-\sum\limits_a^n k_a\right)\eta_i}\frac{\left(-\eta_i\right)^k}{k!}\right).
\end{multline}
When either $\eta_i$ or $\eta_f$ is $\eta_0$ we will drop this term as it contributes at a lower power of $\eta_{1,2}$. Likewise the regulation of this integral at $\pm\infty$ means that we will drop terms with $\eta_{i,f}=\pm\infty$ therefore in both amplitude and correlator the same terms remain,
\begin{equation}
\pm \sum_{k=0}^n Q(K_1,K_2)\frac{(-1)^nn!e^{i\left(\sum\limits_a^n k_a\eta_0\mp K_1 \eta_1\mp K_2\eta_2\right)} }{\left(i\left(\pm K_1 \pm K_2-\sum\limits_a^n k_a\right)\right)^{1+n-k}}e^{i\left(\pm K_1\pm K_2-\sum\limits_a^n k_a\right)\eta_{1,2}}\frac{\left(-\eta_{1,2}\right)^k}{k!}.
\end{equation}
When we look at the amplitude $n=0$ so there is only one term with $k=n=0$ whilst the highest power of $\eta_i$ in the correlator will come from the term $n=k$ so this is what we will keep and in both cases we have
\begin{equation}
\pm Q(K_1,K_2)\frac{e^{i\left(\sum\limits_a^n k_a\eta_0\mp K_1 \eta_1\mp K_2\eta_2\right)} }{i\left(\pm K_1 \pm K_2-\sum\limits_a^n k_a\right)}e^{i\left(\pm K_1\pm K_2-\sum\limits_a^n k_a\right)\eta_{1,2}}\eta_{1,2}^n.
\end{equation}
The only dependence on $n$ is the power of $\eta_i$ and so the difference between each term in the amplitude and the correlator after performing this integral is just $\eta_i^n$ but we saw in  \ref{eq:Ilim} that $n=\sum\limits_\alpha^{V+1}D_\alpha-4$ which is exactly the term that we expect as it results in the final pole in the correlator, therefore our relationship holds for diagrams with single loops. Furthermore, these arguments will hold for arbitrary products of internal propagators as in the limit of large $\eta$ the difference between the two is just that $G_{++}$ contains some additional multiplicative powers of $\eta$, therefore our relationship between amplitudes and correlators will hold, at the level of the integrand, for all diagrams, even those with loops.

\bibliographystyle{JHEP}
\bibliography{References}

\end{document}